\documentclass[aps, pra, 10pt, twocolumn, tightenlines, letterpaper, amsmath, amssymb, preprintnumbers, floatfix, longbibliography, nofootinbib]{revtex4-2}

\usepackage{dsfont}
\usepackage{graphicx}
\usepackage{amsmath}
\DeclareMathOperator*{\argmin}{argmin}
\usepackage{amssymb}
\usepackage{physics}
\usepackage{tabu}
\usepackage{tabularx}
\usepackage{bm}
\usepackage{tikz}
\usetikzlibrary{shapes.geometric}
\usepackage{graphicx}
\usepackage{placeins}
\usepackage{textgreek}
\usepackage{multirow}
\usepackage{makecell}
\usepackage{soul}
\usepackage{diagbox}
\usepackage{xcolor}
\usepackage[normalem]{ulem}
\usepackage[export]{adjustbox}
\usepackage{float}
\usepackage{wasysym}
\usepackage{fancyvrb}
\usepackage{listings}

\usepackage{algorithm}
\usepackage{algpseudocode}

\usepackage{quantikz}

\newcommand{\SubFigRef}[2]{\ref{#1}{\color{blue}{#2}}}

\usepackage[hypertexnames=false]{hyperref}
\hypersetup{
    colorlinks=true,       
    linkcolor=blue,          
    citecolor=blue,        
    filecolor=blue,      
    urlcolor=blue           
}

\newcolumntype{Y}{>{\centering\arraybackslash}X}

\usepackage{orcidlink}

\newlength{\emailcol}\newlength{\emailtmp}
\newcommand{\emailwidest}[1]{%
  \settowidth{\emailtmp}{#1}%
  \ifdim\emailtmp>\emailcol \setlength{\emailcol}{\emailtmp}\fi}

\newcommand{\emailtable}{%
  \begingroup
  \emailwidest{\textcolor{blue}{$^{\dagger}$\,grabow@uw.edu}}%
  \emailwidest{\textcolor{blue}{$^{\S}$\,jhartse@uw.edu}}%
  \emailwidest{\textcolor{blue}{$^{\ast\ast}$\,zhiyaol@uw.edu}}%
  \emailwidest{\textcolor{blue}{$^{\ddagger\ddagger}$\,spow9@uw.edu}}%
  \emailwidest{\textcolor{blue}{$^{\P\P}$\,xjyao@uw.edu}}%
  \textcolor{blue}{$^{*}$\,frolandh@uw.edu}; Corresponding author\par
  \begin{tabular}[t]{@{}p{\emailcol} l@{}}
    \textcolor{blue}{$^{\dagger}$\,grabow@uw.edu}      & \textcolor{blue}{$^{\ddagger}$\,segrie@uw.edu}    \\
    \textcolor{blue}{$^{\S}$\,jhartse@uw.edu} & \textcolor{blue}{$^{\P}$\,alash@uw.edu}     \\
    \textcolor{blue}{$^{\ast\ast}$\,zhiyaol@uw.edu}        & \textcolor{blue}{$^{\dagger\dagger}$\,ziyuanli@uw.edu} \\
    \textcolor{blue}{$^{\ddagger\ddagger}$\,spow9@uw.edu} & \\
  \end{tabular}\par
  \textcolor{blue}{$^{\S\S}$\,mjs5@uw.edu}; On leave from the Institute for Nuclear Theory.\par
  \begin{tabular}[t]{@{}p{\emailcol} l@{}}
   \textcolor{blue}{$^{\P\P}$\,xjyao@uw.edu} &\textcolor{blue}{$^{\ast\ast\ast}$\,zemlni@uw.edu}; Corresponding author\par\\
  \end{tabular}
  \endgroup}

\makeatletter
\newcommand\blfootnote[1]{
  \begingroup
  \renewcommand\thefootnote{}
  \renewcommand\@makefntext[1]{\noindent##1}
  \footnotetext{#1}
  \endgroup
  \addtocounter{footnote}{-1}
}
\makeatother

\makeatletter  
\renewcommand\onecolumngrid{
\do@columngrid{one}{\@ne}%
\def\set@footnotewidth{\onecolumngrid}
\def\footnoterule{\kern-6pt\hrule width 1.5in\kern6pt}%
}
\makeatother    

\let\oldaddcontentsline\addcontentsline

\begin{document}

\begin{figure}
  \vskip -1.cm
  \leftline{\includegraphics[width=0.15\textwidth]{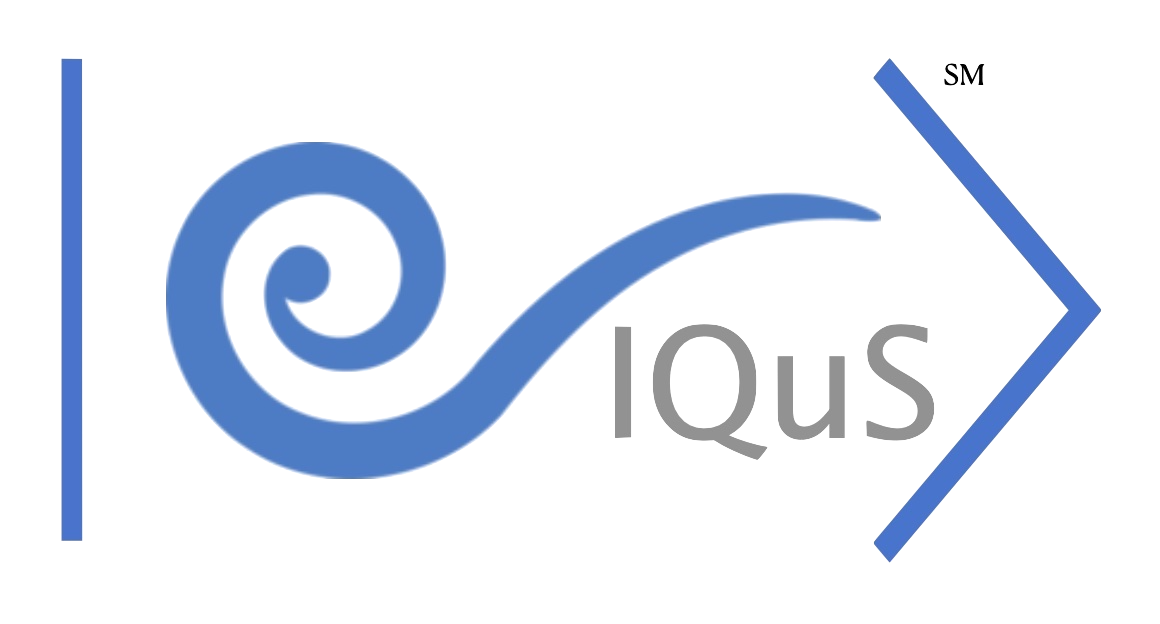}}
  \vskip -1.cm
\end{figure}

\title{Realizing Error Suppression in Partially Fault-Tolerant \\ Quantum Simulations with IBM Quantum Computers}

\author{Henry Froland$^{\ast}$\,\orcidlink{0009-0008-4356-0602}}
\author{Dorota M. Grabowska$^{\dagger}$\,\orcidlink{0000-0002-0760-4734}}
\author{Sebastian Grieninger$^{\ddagger}$\,\orcidlink{0000-0002-9523-5819}}
\author{Jeremy Hartse$^{\S}$\,\orcidlink{0009-0004-1943-421X}}
\author{Anne L. Lashbrook$^{\P}$\,\orcidlink{0009-0002-9120-7738}}
\author{Zhiyao Li$^{\ast\ast}$\,\orcidlink{0000-0002-7614-8496}}
\author{Ziyuan Li$^{\dagger\dagger}$\,\orcidlink{0009-0005-2034-1499}}
\author{Sarah J.~M.~Powell$^{\ddagger\ddagger}$\,\orcidlink{0000-0002-5228-8291}}
\author{Martin J.~Savage$^{\S\S}$\,\orcidlink{0000-0001-6502-7106}}
\author{Xiaojun Yao$^{\P\P}$\,\orcidlink{0000-0002-8377-2203}}
\author{Nikita A. Zemlevskiy$^{\ast\ast\ast}$\,\orcidlink{0000-0002-0794-2389}}

\affiliation{InQubator for Quantum Simulation (IQuS), Department of Physics, University of Washington, Seattle, WA 98195, USA.}

\preprint{IQuS@UW-21-130}
\date{\today}

\begin{abstract}
\noindent
Quantum error-detecting codes offer a near-term path for improving the performance of quantum simulations on noisy hardware.
Using IBM's superconducting quantum computer {\tt ibm\_boston}, we show that partially fault-tolerant encoded quantum simulations of the Ising model in 1+1D and 2+1D outperform their unencoded counterparts in estimating local observables.
To represent 42 logical qubits on the heavy-hex quantum processor, 21 blocks of the $[[4,2,2]]$ Iceberg code and up to 136 physical qubits are used.
By pairing fault-tolerant syndrome extraction with non-fault-tolerant logical operations, this scheme preserves many of the benefits of error detection while avoiding the overhead typically required for a fully fault-tolerant logical gate set.
The encoding's square logical connectivity, together with the freedom to place logical qubits within each block, enables simulations of a 2D spatial lattice with lower circuit depth than the unencoded implementation requires.
We introduce Observable-Ranked Postselection, a selective-filtering technique based on syndrome correlations that recovers reliable results without the prohibitive shot loss of full syndrome postselection.
Under the cumulative effect of device errors, this encoding improves local-observable accuracy over the unencoded baseline by $2$-$6\%$ at intermediate times in 1+1D simulations, growing with circuit depth to over $200\%$ in 2+1D at the latest times studied.
\end{abstract}

\maketitle
\blfootnote{\emailtable}

\renewcommand{\addcontentsline}[3]{}
\section{Introduction} 
\noindent
Performing quantum simulations with sufficient precision to compute many scientifically relevant observables requires levels of error suppression attainable only through quantum error correction (QEC)~\cite{Steane:1996ghp,Steane:1996va,Gottesman:1997zz,Gottesman:1997qd,Aliferis:2005ftz}.
The qubit and gate overhead of fault-tolerant (FT) quantum computation has so far kept large-scale QEC out of reach for practical quantum simulations.
Leveraging high gate fidelities and flexible connectivity, recent demonstrations on trapped ions~\cite{Dasu:2026dwm,Perlin:2026mph} and neutral atoms~\cite{Rodriguez:2024bhh,Reichardt:2024xfs,Bluvstein:2025ped} have achieved ``beyond break-even'' performance, with the encoded error rate falling below the unencoded rate~\cite{Gottesman:2016gef}. Solid-state platforms have seen similar success, albeit in more restricted settings~\cite{Urbanek:2020cza,Chen:2021num,GoogleQuantumAIandCollaborators:2024efv,Hetenyi:2024zvf,Caune:2024doa,Gupta:2023zei,Vigneau:2025avm,Abraham:2026slx}.
These results suggest that in certain cases, presently available hardware can benefit from FT implementations beyond what error mitigation alone can achieve. Realizing this benefit in quantum simulations requires carefully balancing the error-rate (ER) improvements from FT against its associated overhead.

An approach to computing with encoded qubits that is compatible with currently available hardware is to supplement FT components with non-FT (nFT) ones, 
allowing some physical errors to go undetected in exchange for lower circuit overhead~\cite{Preskill:2025cbl,Gerhard:2024peb,Reichardt:2026xbk,Akahoshi:2023xck,Yamamoto:2025iyx,Zhong:2025jox}.
In this partially FT approach, for a physical error rate $p$,\footnote{In practice many processes contribute to $p$. 
In this work $p$ characterizes the combined error rate of all of these processes, which is typically dominated by two-qubit gate errors.}
nFT components contribute a factor of $Ap$ to the logical error rate (LER), whereas FT components that remove $\mathcal O(p)$ errors provide a suppression of $Bp^2$, giving a total logical error rate of $p_L(p)=Ap+Bp^2$.
By carefully designing the nFT components such that $A\ll B\ll 1$, the computation can go below the ``pseudothreshold", i.e., $p$ such that $p_L(p)<p$.
The nFT components generally consist of logical rotations, for which a fully FT implementation would require magic state injection~\cite{Bravyi:2004isx} or code switching~\cite{Paetznick:2013kwr,Daguerre:2025boq}. 
Although topological codes such as the surface code~\cite{Kitaev:1997wr,Dennis:2001nw} and the heavy-hex code~\cite{Chamberland:2019zev} are naturally suited to limited-connectivity devices~\cite{Benito:2024mll}, the measurement and decoding overhead of their lattice-surgery operations places them out of reach for present-day simulations~\cite{Horsman:2011hyt}.

While the full power of QEC may still be beyond reach, immediate improvements can be achieved through the adoption of lightweight, low-overhead quantum error detection (QED) encodings~\cite{Linke:2017bvn,Vuillot:2018tqy,Harper:2019upm,Takita:2017blo,Corcoles:2014imo,Yamamoto:2023xan,Wang:2023qcn}. 
A motivation for this work is to use FT to advance quantum simulations of lattice gauge theories~\cite{Pato:2026wow,Rajput:2021trn,Spagnoli:2026qni,Turco:2026cte,Spagnoli:2024mib,Yao:2025cxs,Carena:2024dzu} describing the fundamental forces of Nature~\cite{Klco:2021lap,Bauer:2022hpo,Bauer:2023qgm,Davoudi:2022bnl,Beck:2023xhh,DiMeglio:2023nsa}. 
The mappings and dynamics of such simulations are organized by symmetries that group the degrees of freedom into repeating units.
As such, it is natural to study FT schemes whose code blocks coincide with these units.\footnote{In the Schwinger model (quantum electrodynamics in 1+1D), for example, a spatial site maps to two qubits, e.g. Refs.~\cite{Martinez:2016yna,Klco:2018kyo,Farrell:2023fgd,Farrell:2024fit}, suggesting the use of code blocks that implement two logical qubits.}
The recently studied Iceberg codes are a family of $[[n+2,n,2]]$ codes that use two extra physical qubits to encode $n$ logical qubits into a single block~\cite{Gottesman:1997zz,Linke:2017bvn,Roffe:2018oim,Chao:2017owu,Self:2022lsx,Chertkov:2025qzc,Vezvaee:2025yol,Dasu:2026dwm}.
Using multiple interacting code blocks of the $[[4,2,2]]$ Iceberg code, we perform Hamiltonian simulations on IBM's heavy-hexagonal quantum computers and achieve beyond break-even performance on the calculation of local observables.
This is done through the development of shallow syndrome extraction circuits designed for the device's native qubit connectivity.
We implement nFT logical rotations, minimizing their depth by selecting specific code block layouts and gate scheduling.
One unavoidable feature of QED codes is the exponential loss of ensemble size as the number of syndrome extraction rounds increases. 
For a computation with many code blocks, this loss prohibits running even a single round of syndrome extraction if all shots record an error and are discarded.
This work proposes a selective-filtering strategy,  Observable-Ranked Postselection (ORP), that discards detection events based on their probability of causing a logical error, determined through correlations between syndrome flips and logical observable values.

We demonstrate the benefits of these QED methods through simulations of false-vacuum decay in the 1+1D and 2+1D Mixed-Field Ising model (MFIM).
Despite a two to six times increase in circuit depth, encoded 1D simulations outperform their unencoded counterparts in estimating local observables by $2$-$6\%$ in the regimes considered in this work. 
As the logical degrees of freedom are spread across a block rather than being localized to a single physical qubit, the encoding relaxes the connectivity constraints of a logical computation. 
As a result, encoded simulations on a 2D square lattice are therefore possible at a 1.5 times lower overall circuit depth than a direct unencoded implementation. 
There, the encoding improvement grows with circuit depth, exceeding $200\%$ at late times.
Importantly, we compare the performance of encoded and unencoded circuits \emph{without any} error mitigation
schemes that utilize auxiliary circuits to ``learn'' the noise~\cite{Urbanek:2021oej,ARahman:2022tkr,Farrell:2023fgd,ZNE1,Froland:2026aff,Temme:2016vkz,PhysRevLett.120.210501,Klco:2018kyo,ZNE3,Berg:2022ugn}.
This choice isolates the improvements due to QED from those introduced by mitigation schemes.
In both 1D and 2D, we find that infrequent error detection in systems mapped to many Iceberg-code logical qubits occupies a crossover regime between NISQ-era heuristics and fully FT computation.

Sections~\ref{sec:422_code}--\ref{sec:discussion} constitute the main text of this paper and are designed to be read sequentially. 
Section~\ref{sec:422_code} describes the implementation of multiple blocks of the $[[4,2,2]]$ code on heavy-hex connectivity, as well as the details of our detector filtering technique, ORP.
Section~\ref{sec:results} presents results of encoded and unencoded simulations of the Ising model on {\tt ibm\_boston}.
Section~\ref{sec:discussion} concludes with a discussion of the results and the role of encoded computations in the early FT era of quantum simulation for scientific discovery.
The Methods sections contain information supporting the results in the main text.
\ref{sec:filtering} elaborates on ORP.
~\ref{sec:layouts} explains the procedure used to determine optimal device layouts.
Logical circuit scheduling is discussed in ~\ref{sec:scheduling}.
~\ref{sec:quantum_sim} and~\ref{sec:classical_sim}
provide the details and design choices behind our quantum and classical simulations, respectively.
The appendices provide additional information and expand on discussions in the Methods sections.

\section{Fault Tolerance in the \texorpdfstring{$[[4,2,2]]$}{[[4,2,2]]} Code}
\label{sec:422_code}
\noindent
The key consideration behind the FT design of logical circuits is that the spread of errors should be controlled and minimized. 
Namely, such circuits should not introduce more errors than they can detect, 
and they should not propagate detectable errors into undetectable ones. 
This work implements a version of FT that is inspired by level-1 FT (1-FT)~\cite{Aliferis:2005ftz} by choosing ``gadgets'', i.e. subcircuits that implement an encoded operation, that satisfy the following conditions:
\begin{enumerate}
    \item A gadget with no errors during operation detects one input error and outputs at most one error.
    
    \item A gadget with one error during operation and no input errors outputs at most one error.
\end{enumerate}
For modest-sized circuit gadgets, such as those used in this work, these conditions can be verified by classical simulations.
Two-qubit gates are the operations that typically suffer from the highest levels of noise and therefore set the dominant physical error rate. 
While faulty two-qubit gates can in principle produce undetectable errors at the same rate as single-qubit errors, this work finds that using these conditions as criterion for FT is sufficient for practical error suppression.

\begin{figure*}[ht]
    \centering
    \includegraphics[width=\linewidth]{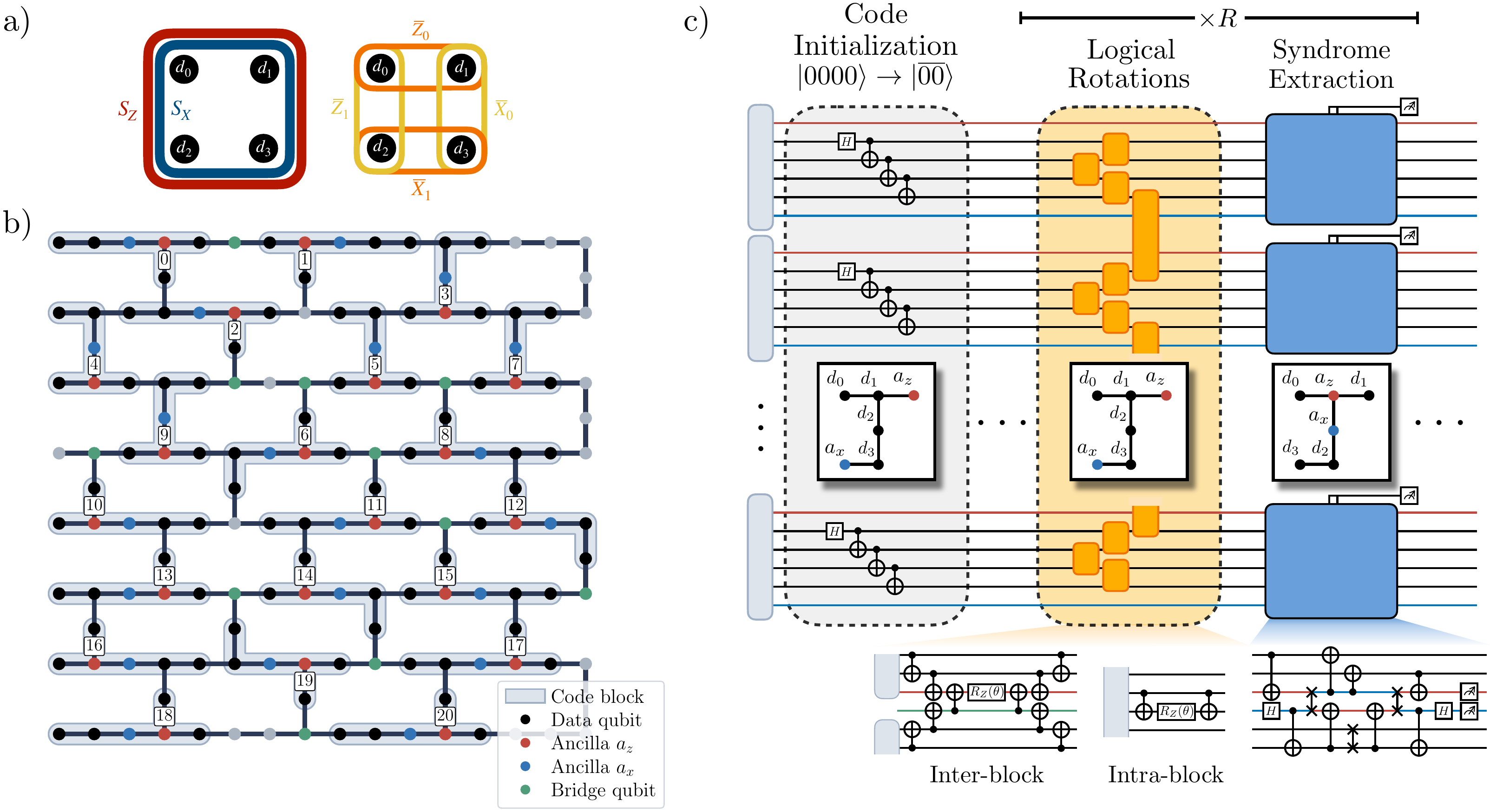}
    \caption{{\it Error Detection in Quantum Simulations on Heavy-Hex Topology.}
    a)~The stabilizers and logical operators that act on data qubits $d_0$-$d_3$ in the $[[4,2,2]]$ error-detecting code. 
    The $Z$-stabilizer is shown in red and the $X$-stabilizer is shown in blue. Logical operators are grouped into commuting sets represented by the color.
    b)~The placement of code blocks (grey) onto the {\tt ibm\_boston} quantum processor.
    Each code block has four data qubits (black) and two ancillas for $Z$- and $X$-stabilizer extraction (red and blue respectively).
    Bridge qubits between blocks used for routing are shown in green.
    c)~Circuit operation for encoded quantum simulations.
    Dashed outlines indicate nFT operations, while solid outlines indicate FT gadgets.
    Each code block starts in the ``compute" configuration (shown in the left box in the center).
    Within each code block, the logical $|\overline{00}\rangle$ state is prepared with a nFT circuit. 
    Non-FT logical operations (orange boxes) are performed in the compute configuration, example circuits of which are shown at the bottom.
    The blue boxes represent FT syndrome extraction done in the ``syndrome" configuration within each block (right two boxes), which is implemented with the circuit shown in the bottom.
    Applications of logical operations and syndrome extraction are repeated $R$ times in any given computation.
    }
    \label{fig:overview}
\end{figure*}

The $[[4,2,2]]$ code is a Calderbank-Shor-Steane (CSS)
code~\cite{Calderbank:1995dw} that forms two logical qubits from four physical qubits, and is the smallest code in the family of Iceberg codes.
It is a distance $d=2$ code and so is capable of detecting single-qubit errors.
The stabilizers of this code, shown on the left side of Fig.~\SubFigRef{fig:overview}{a}), are given by 
\begin{equation}
    S_X \ = \ X_0X_1X_2X_3,\quad S_Z \ = \ Z_0Z_1Z_2Z_3 \ .
\end{equation}
These operators define the codespace as the joint $+1$-eigenspace of $S_X$ and $S_Z$, i.e., the even-parity sector of the four-qubit Hilbert space. 
The logical $\ket{\overline{00}}$ state is a four-qubit GHZ state, ${\ket{GHZ(4)}=1/\sqrt{2}(|0000\rangle + |1111\rangle)}$,
and the logical operators on the codespace are given by 
\begin{equation}
\overline X_i \ = \ X_{i+1} X_3\ , \quad \overline Z_i \ = \ Z_0 Z_{i+1}
\label{eq:logical_ops}
\end{equation}
for $i=0,1$ and are shown on the right side of Fig.~\SubFigRef{fig:overview}{a}).
Any operator that contains only an even number of physical $X$s and $Z$s preserves the codespace, so errors of this form correspond to undetectable logical errors. 
Further, since operators on the logical space are equivalent up to multiplication by the stabilizers, the only errors this code fails to detect must contain two $X$s or two $Z$s.

\subsection{The \texorpdfstring{$[[4,2,2]]$}{[[4,2,2]]} Code on Heavy-Hex Quantum Computers}
\label{sec:422_device}
\noindent
To perform encoded computations on a quantum device, it is necessary to choose a hardware embedding that allows for efficient FT syndrome extraction and shallow implementation of logical operations. 
Six physical qubits are used to represent each $[[4,2,2]]$ code block: four data qubits $d_0$-$d_3$ and two ancilla qubits $a_x,\,a_z$ used for syndrome readout and logical operations. 
The embedding of a single code block into the heavy-hex connectivity requires one degree-3 vertex and two degree-2 vertices.
Within this single-block topology, logical operations and syndrome extraction have different optimal arrangements of qubits.
We call these the ``compute" and ``syndrome" configurations, which are shown in Fig.~\SubFigRef{fig:overview}{c}) in the center of the circuits. 
The FT syndrome extraction circuit shown at the bottom right of Fig.~\SubFigRef{fig:overview}{c}) measures $S_X$ and $S_Z$ simultaneously and requires all code blocks to have the same topology.
During syndrome extraction, the ancillas $a_x$ and $a_z$ also act as mutual error flags, controlling the proliferation of ``hook" errors that propagate from ancilla to data qubits~\cite{Chao:2017wck,Chao:2017owu,Chamberland:2018ken,Chao:2019leh}.
This circuit is derived from Ref.~\cite{Reichardt:2018kqi} and is verified to be 1-FT through classical simulations.
The circuits required to switch between the compute and syndrome configurations (and the gate overhead) are given in~\ref{sec:quantum_sim}.

To maintain scalability and locality of logical operations, we represent logical qubits with multiple $[[4,2,2]]$ code blocks.
The optimal block placement is chosen in several steps which are explained in~\ref{sec:layouts}.
Avoiding noisy two-qubit gates and qubits with high measurement error rates, the maximum number of code blocks we find is 21, corresponding to 42 logical qubits.
Logical qubits are labeled by their block and qubit number within the block: 
\begin{align}
    (b,l_b) \ 
     {\rm with}\ \  l_b\in\{0,1\}
     \ .
     \label{eq:blb}
\end{align}
Any given placement of code blocks can realize a number of distinct logical connectivity graphs.
This work selects the placement that has the most square-grid-like logical connectivity graph from all possible 21-block placements, which is relevant for the 2D simulations discussed in Sec.~\ref{sec:results}.
See \ref{sec:layouts} for more details on the layout selection process. 
The chosen block placement on {\tt ibm\_boston} has four missing internal edges and is shown in Fig.~\SubFigRef{fig:overview}{b}).
This block placement has both directly adjacent blocks, and blocks that have a ``bridge" qubit between them, shown in green.

\subsection{Observable-Ranked Postselection (ORP)}
\label{sec:filtering_main_text}
\noindent
Postselection is the task of identifying and discarding shots corrupted by errors. 
Syndrome measurements serve as imperfect proxies for identifying these affected shots, both due to the partially FT approach used in this work and because 
multiple errors can incorrectly return the logical state into the codespace.
As a result, errors affecting logical information will not always trigger a syndrome.
Further, a logical observable will be unaffected by errors that are causally disconnected from it or otherwise commute with it.
We introduce ORP to sidestep the exponential shot loss associated with complete postselection.\footnote{The form of the exponential shot loss is a constant plus an exponential in the number of rounds of stabilizer measurements.
The constant is determined by the ratio of the size of the codespace to the physical-qubit space.
It is typically small enough to be neglected for practical ensemble sizes.}
This method ranks error detection events by their probability of causing a logical error and uses this ranking to iteratively clean the measured dataset.

Active reset of the ancilla qubits between rounds consumes device coherence time, and so resets are omitted in this work.
Each measured outcome therefore reflects the accumulated parity of syndrome measurements in the computation up to that point~\cite{Geher:2024lkc}.
The measurement outcome for syndrome type-$g$ on block $b$ and round $r$ is denoted as $s_j=s_{(g,b,r)}$.
These are combined into ``detectors'' by comparing each stabilizer's outcome $s_j$ to its value in a previous round,
\begin{align}
    v_{j+1} \ = \ s_{j-1} \oplus s_{j+1} \ , 
    \label{eq:detectors}
\end{align}
where $j\pm1$ is shorthand for $(g,b,r\pm1)$ and $\oplus$ is the XOR operation implementing modulo-$2$ addition.
The definition~\eqref{eq:detectors} implies a detector only fires ($v_j=1$) when this comparison is violated. 
Crucially, in the absence of errors, all detectors will take the deterministic value $v_j=0$.

A detector $v_j$'s impact on a given observable $\overline O$ is inferred by conditioning $\langle\overline O\rangle$ on the value of $v_j$ by
\begin{align}\label{eq:delta_def}
    \Delta_j \ = \ \big|\langle \overline O \rangle_{v_j=0} - \langle \overline O\rangle_{v_j=1} \big| \ ,  
\end{align}
which measures the correlation between detection events and logical errors that bias $\langle \overline O\rangle$. 
For the $j$th detector, $\Delta_j$ can be interpreted as quantifying the effective probability that, given a detection event, a logical error occurred~\cite{Chen:2021num,Reichardt:2026xbk,Blume-Kohout:2025kvx}.\footnote{For a local observable $\overline O$, the detectors selected by ORP are found to coincide with the backwards lightcone of $\overline O$, such that detectors causally disconnected from $\overline O$ have values of $\Delta_j$ parametrically smaller than those within the lightcone.
This is discussed in~\ref{app:pp_lightcone}.}
Detectors are then ranked by $\Delta_j$, and ORP uses this ranking to postselect on detection events with high $\Delta_j$. 
A cut at level $k$ keeps only shots in which none of the top-$k$ detectors fire, resulting in a sequence of increasingly strict cuts with smaller ensemble sizes and progressively cleaner estimates of $\langle \overline O\rangle_k$.
Each time the cutoff $k$ is increased, the proportion of shots containing harmful errors relative to the remaining ensemble size decreases, and so the expectation value of a logical observable progressively approaches the noiseless value until it plateaus at a level set by the undetectable errors.
However, the ensemble size also decreases with increasing $k$, inflating the statistical uncertainty on $\langle O\rangle_k$. 
The optimal cut level therefore balances the bias from residual device noise against the variance from shot loss.

Since the residual bias is bounded below by the undetectable errors, once the shots carrying the dominant detectable errors have been removed $\langle \overline O\rangle_k$ stops moving and its distribution is consistent with statistical fluctuations alone.
We identify plateaus in $k$ over which this criterion holds, indicating that the estimate has converged. 
The optimal cut level $k^*$ is chosen to be at the lower edge of the most stable plateau, and ORP uses this value to compute $\langle \overline O\rangle_{k^*}$.
Technical details of ORP and the plateau-finding algorithm are given in~\ref{sec:filtering} and alternatives considered are discussed in~\ref{app:filtering_metric_details}.

\section{Error-Detected Quantum Simulations of The Ising Model}
\label{sec:results}

\begin{figure*}
    \centering
    \includegraphics[width=\linewidth]{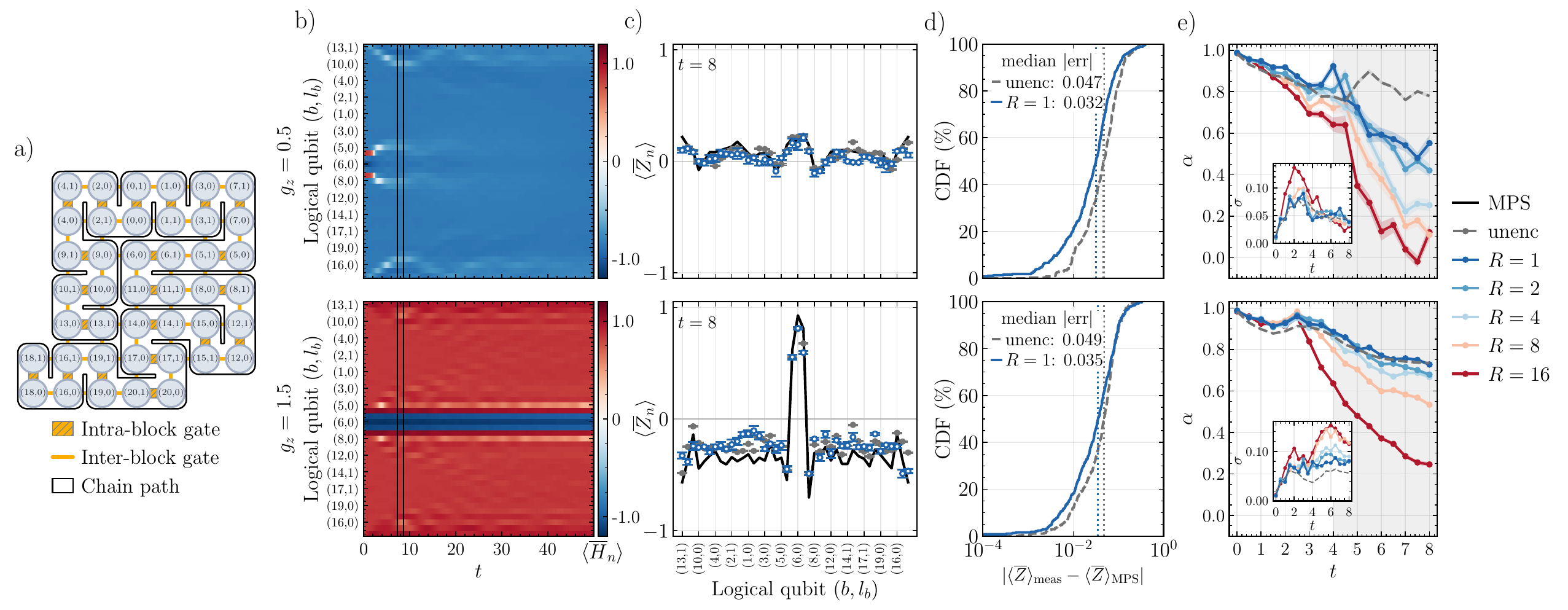}
    \caption{{\it Encoded Simulations of Dynamics in the 1D Ising Model.} 
    a)~The realized logical connectivity grid by the block placement in Fig.~\SubFigRef{fig:overview}{b}).
    Logical qubits $(b,l_b)$ (grey) are joined by intra-block gates (hatched orange) and inter-block gates (orange). 
    The black outline traces the gates used to simulate a 1D chain.
    b)~The energy density $\langle \overline{H}_n\rangle$ computed from MPS simulations as a function of time $t$ and logical qubit $n=(b,l_b)$ for $g_z=0.5$ (top) and $g_z=1.5$ (bottom).
    A size-three true-vacuum bubble located at the center is evolved with $n_T=500$ Trotter steps of size $\delta t=0.1$. 
    The black rectangle shows the $t$ for which device results are compared in c).
    c)~Comparison of encoded (blue) and unencoded (grey) magnetization $\langle \overline{Z}_n\rangle$ from {\tt ibm\_boston} to MPS simulations (black) at $t=8$, corresponding to 16 Trotter steps of $\delta t=0.5$ for $g_z=0.5$ (top) and $g_z=1.5$ (bottom).
    The encoded simulations shown have $R=1$ syndrome extraction rounds. 
    d)~The cumulative distribution function (CDF) of the per-qubit, per-time absolute error $|\langle \overline{Z}\rangle_\text{meas} - \langle \overline{Z}\rangle_\text{MPS}|$ for each value of $g_z$.
    The median error (dotted vertical lines) and CDFs corresponding to one layer ($R=1$) of syndrome measurements are shown compared to unencoded results (solid and dashed lines respectively).
    Results up to $t=4$ are used, where an encoding improvement is observed.
    e)~The signal-survival factor $\alpha$, defined in Eq.~(\ref{eq:Zslopedef}),
    as a function of $t$ for various $R$ (colored lines) and each value of $g_z$ compared to unencoded results (dashed grey line).
    The shading for $t=4$-$8$ represents times excluded for the CDFs shown in d). 
    The insets show the residual error $\sigma$ in the fit.
    The statistical uncertainty is determined through bootstrap resampling.
    }
    \label{fig:chain_results}
\end{figure*}
\noindent
The rich collective behavior emerging from simple nearest-neighbor spin interactions in the 
Ising model makes it an ideal platform for real-time quantum simulations.
Recent work has used it to examine the dynamics of thermalization~\cite{Jaschke:2019jka}, string breaking/confinement~\cite{De:2024smi,Surace:2020ycc}, scattering~\cite{Farrell:2025nkx,Zemlevskiy:2026kpc}, and Majorana edge modes~\cite{Mi:2022egw}.
This work focuses on simulations of false-vacuum decay under quench dynamics in the MFIM in 1D and 2D. 
False-vacuum decay has been well studied in both 1D and 2D using 
classical~\cite{Kormos:2016osj,Milsted:2020jmf,Lagnese:2021grb,Yin:2024hjm,Pavesic:2024ryc,Pavesic:2025nwm,Borla:2026fdb,Balducci:2022kvd,Pavesic:2024ryc,Pavesic:2026yiz,Lerose:2019jrs,Lagnese:2021hjt} and quantum methods~\cite{Darbha:2024srr,Chao:2025rhr,Luo:2025qlg,Humar:2026gbs}.
This work extends the tensor network study of Ref.~\cite{Pavesic:2024ryc} to nonzero longitudinal field.
The MFIM Hamiltonian with open boundary conditions (OBCs) is
\begin{align}
\overline{H} \ = \ -J\sum_{\langle ij\rangle} \overline{Z}_i \overline{Z}_j \ - \sum_i \left (g_x \overline{X}_i \ + \ g_z \overline{Z}_i \right ) \ .
\label{eq:h_ising}
\end{align}
The sum over $\langle ij\rangle$ denotes nearest-neighbor coupling in 1D and couples adjacent logical qubits on a 2D lattice.

False-vacuum decay phenomenology models the nucleation of true-vacuum bubbles from a false-vacuum background and their subsequent evolution~\cite{Coleman:1977py,Callan:1977pt}.
This work studies the quench dynamics of product-state bubbles, i.e., the evolution of bubbles after their nucleation.
Our simulations begin in a product state ${|\psi(t=0)\rangle = \prod_{i\in B}\overline X_i |\overline1\rangle^{\otimes N}}$ on $N$ qubits, where $B$ is the initial true-vacuum bubble region. 
Recent work has predicted that bubbles melt for $|g_z|/|g_x|<1$ and remain localized for $|g_z|/|g_x|>1$ in the limit of large $J$~\cite{Balducci:2022zym}, respectively indicating the presence and absence of false-vacuum decay.
We work with the couplings $J=1,\,g_x=0.75,$ and identify a bubble-melting regime at $g_z=0.5$ and a localized regime at $g_z=1.5$, where the bubble's extent is fixed by the quench dynamics.
This phenomenon is driven by the linear potential between bubble walls that causes the momentum of excitations to wrap around the Brillouin zone, resulting in bubble oscillations known as Bloch oscillations~\cite{Pomponio:2021ltz}.
In 1D, the oscillation of the bubble width (equivalently, Stark localization of the interface~\cite{Kormos:2016osj,Lerose:2019jrs}) is predicted to have an amplitude set by the ratio $r_\text{max} \sim |g_x|/|g_z|$.

\begin{figure*}
    \centering
    \includegraphics[width=\linewidth]{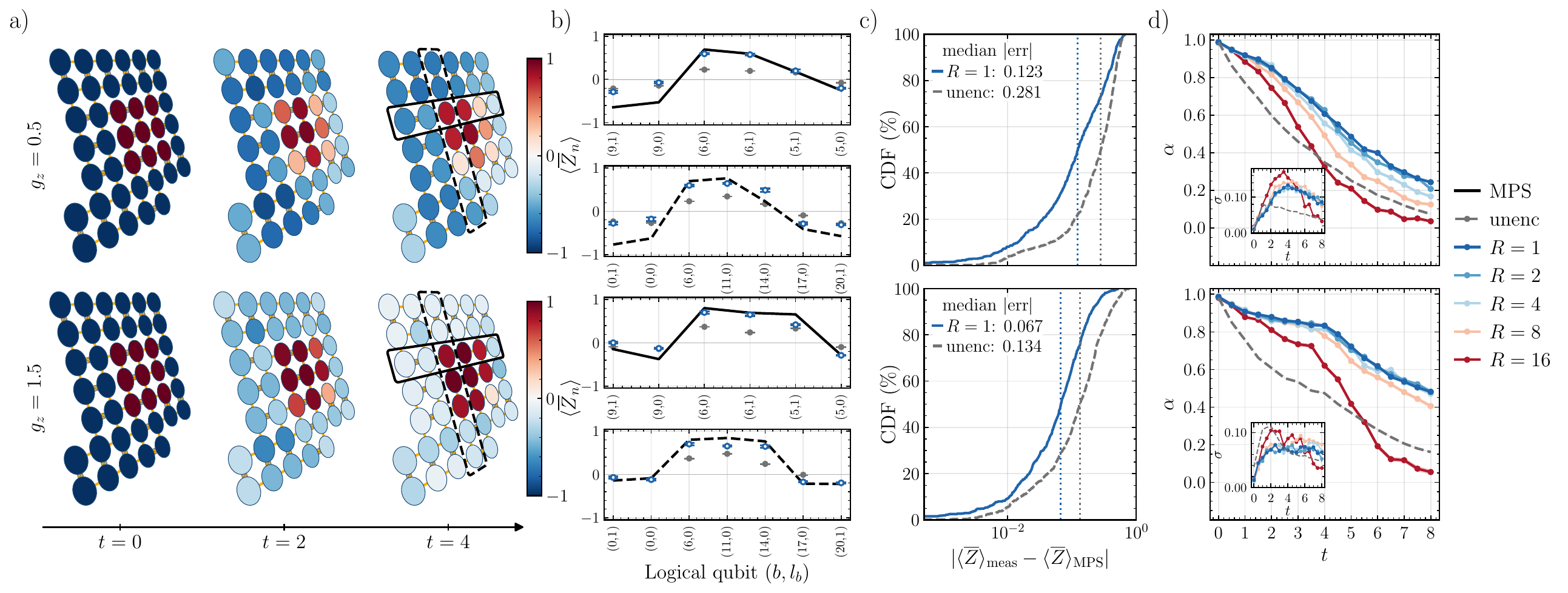}
    \caption{{\it Encoded Simulations of the 2D Ising Model.} 
    a)~The magnetization $\langle \overline{Z}_n\rangle$ of a $3\times3$ true-vacuum bubble as a function of time $t$ for $g_z=0.5$ (top row) and $g_z=1.5$ (bottom row),  computed from MPS simulations with $\delta t=0.1$.
    The black solid and dashed rectangles show cross-sections of the grid at $t=4$ for which data is plotted in b). 
    b)~Encoded (blue) and unencoded (grey) results from {\tt ibm\_boston} compared to MPS results (black) for the horizontal (solid) and vertical (dashed) cross-sections outlined in a), as a function of logical qubit number $n=(b,l_b)$.
    The encoded data has $R=1$ syndrome measurement rounds.
    c)~Cumulative distribution functions (CDFs) of the per-qubit, per-time absolute error $|\langle \overline{Z}\rangle_\text{meas} - \langle \overline{Z}\rangle_\text{MPS}|$ for both values of $g_z$.
    The median error and CDFs corresponding to  encoded runs with $R=1$ syndrome measurements and unencoded runs are shown compared to unencoded results (solid and dashed lines respectively).
    d)~The signal-survival factor $\alpha$ as a function of $t$ for various $R$ (colored lines) and each value of $g_z$, compared to unencoded results (dashed grey line).
    The insets show the residual error $\sigma$ in the fit.
    The statistical uncertainty is determined through bootstrap resampling.
    }
    \label{fig:grid_results}
\end{figure*}

In quantum simulations using the encoding discussed in the previous section, logical operations between qubits $(b,0)$ and $(b,1)$ within a block $b$ are straightforwardly implemented in the compute configuration, shown in Fig.~\SubFigRef{fig:overview}{c}). 
Depending on how adjacent blocks are placed relative to one another, some logical inter-block gates require routing through ancilla and bridge qubits.
These inter-block gates constitute the bulk of the depth in our time evolution circuits.
The MFIM Hamiltonian~\eqref{eq:h_ising} is realized by the operations $\overline Z_i,\, \overline X_i,\, \overline{Z}_i \overline{Z}_j$ available on our logical connectivity graph.

\subsection{Results from IBM's Quantum Computers}
\noindent
We simulate the dynamics of a 1D chain under the Hamiltonian in Eq.~(\ref{eq:h_ising}) using the logical qubit grid realized by the block placement shown in Fig.~\SubFigRef{fig:overview}{b}).
By retaining all intra-block connections and removing specific inter-block edges from the grid, we implement a chain of 42 logical qubits shown in Fig.~\SubFigRef{fig:chain_results}{a}).

The expectation value of the energy density ${ \overline{H}_n=-J\overline Z_n\overline Z_{n+1} -(g_x \overline X_n + g_z \overline Z_n)}$ as a function of time and logical qubit $n=(b,l_b)$ determined from MPS simulations is shown in Fig.~\SubFigRef{fig:chain_results}{b}).
For $g_z=0.5$ the bubble oscillates about its center, while for $g_z=1.5$ it remains localized.
This behavior is expected from the Bloch-oscillation picture of~\ref{app:physics}, in which $r_\text{max}$ falls from well above the bubble size to comparable with it, and is consistent with the melting/localization transition reported in Ref.~\cite{Balducci:2022zym}.
The magnetization $\langle \overline Z_n\rangle$ from encoded results determined through ORP and unencoded runs on {\tt ibm\_boston} are compared to MPS results for $t=8$ in Fig.~\SubFigRef{fig:chain_results}{c}).
Deviations from MPS expectations are quantified through the absolute error $|\langle \overline{Z}\rangle_\text{meas} - \langle \overline{Z}\rangle_\text{MPS}|$.
The cumulative distribution function (CDF) of the absolute errors over all qubits and $t\leq4$ is shown in Fig.~\SubFigRef{fig:chain_results}{d}) for both values of $g_z$. 
The largest encoding improvement in median absolute error is found to be $ 36.1 \% \pm3.3\%$ up to $t=4$ for $g_z=0.5$ and $R=1$ (corresponding to a single syndrome extraction round at the end of the circuit).
The median absolute error is lower for encoded runs and the improvement is consistent for both $g_z$. 
Further, their CDFs lie predominantly to the left of unencoded CDFs, overlapping only at large absolute errors.
Error detection is found to concentrate the bulk of the distribution at small errors but retains a tail of poorly-performing samples. 
This is consistent with most qubits and gates operating just below the pseudothreshold, with several outliers.

To quantify how faithfully the encoded circuit reproduces the ideal dynamics independent of the overall signal magnitude, 
$\langle \overline Z\rangle_\text{meas}$ is fit to $\langle \overline Z\rangle_\text{MPS}$
with a line through the origin.
The slope $\alpha$ is the signal-survival factor and $\sigma$ is the root-mean-square of the fit residuals 
\begin{align}
    \langle \overline Z\rangle_\text{meas}\ = \ \alpha \langle \overline Z\rangle_\text{MPS} + \epsilon \ , \ \sigma \ = \ \sqrt{\frac{1}{n}\sum_i\epsilon_i^2} 
    \ .
    \label{eq:Zslopedef}
\end{align}
Here the sums run over all $n$ samples of the observable (for a given qubit and time).
The signal-survival $\alpha=1$ indicates perfect retention, while $\alpha<1$ implies signal loss due to noise, so higher values of $\alpha$ at comparable $\sigma$ indicate more faithful results.
Figure~\SubFigRef{fig:chain_results}{e}) shows $\alpha$ and $\sigma$ as a function of $t$ for both couplings and various numbers of syndrome extraction rounds $R$.
Here $R$ denotes the application of up to $R$ syndrome extraction rounds, with at most one syndrome round per Trotter step. 
Encoded runs for all values of $R$ are found to outperform their unencoded counterparts up to intermediate times, after which only small-$R$ encoded results for $g_z=1.5$ show an improvement in $\alpha$.
As expected, $\alpha$ trends lower with $t$ due to the accumulation of undetected errors.
The typical encoding improvement in the localized regime compared to the melting regime over all $t$ is attributed to a narrower backwards lightcone at $g_z=1.5$, which restricts error propagation to a smaller spatial region and lets the encoding detect a greater fraction of errors. 
Several points in the melting regime break the monotonic decay of $\alpha$; this is again attributed to the faster propagation of errors at $g_z=0.5$.
Since $\alpha$ is similar for unencoded runs at $g_z=0.5$ and $1.5$, this suggests that encoding is more effective when dynamics are slow and error propagation is limited.

An unforeseen feature of the encoding improvement is its dependence on $R$, represented by the colors in Fig.~\SubFigRef{fig:chain_results}{e}). 
Early-time results show a modest advantage of increased syndrome extraction frequency (larger $R$), but beyond $t\sim4$ this reverses and fewer extraction rounds perform better.
This indicates that conserving coherence outweighs the benefit of additional syndrome extractions, and is due to the interplay between ORP and error behavior in the absence of ancilla resets. 
Without resets, an ancilla that detects an error remains in $|1\rangle$, and subsequent logical operations that couple to it inject effective errors that propagate through the system.
ORP removes some of this contamination, but the degradation with extraction frequency reveals its limitation: error detection events that ORP does not choose leave coherent contamination in the retained shots. 
Notably, adding resets does not significantly improve $\alpha$, indicating that the coherence-time cost of resets roughly cancels the benefit of eliminating syndrome-induced error propagation.
See~\ref{app:more_analysis} for a detailed analysis of this effect.

Two-dimensional lattice simulations are implemented by using all inter- and intra-block edges in the logical connectivity graph shown in Fig.~\SubFigRef{fig:chain_results}{a}).
Figure~\SubFigRef{fig:grid_results}{a}) shows the magnetization $\langle \overline Z_n\rangle$ on the grid for several selected times determined from MPS simulations.
Although the dynamics in 2D is found to be slower than in 1D, an initial $3\times3$ bubble is still found to spread for $g_z=0.5$ and remain localized for $g_z=1.5$. 

Figure~\SubFigRef{fig:grid_results}{b}) shows $\langle \overline Z_n\rangle$ for horizontal and vertical cross-sections on the grids in a) at $t=4$ (solid and dashed outlines respectively). 
While unencoded results show significant decoherence toward $\langle \overline Z_n\rangle=0$, encoded results computed with ORP are closer to MPS expectations. 
The encoding largely removes the geometric overhead associated with representing a square lattice on heavy-hex hardware, 
and only the depth overhead from Trotter step serialization is present. 
The retained geometric overhead is reflected in large unencoded circuit depths, and the encoding improvement is expected to be larger than in 1D simulations.
The CDFs of the absolute error over all logical qubits and times are shown in Fig.~\SubFigRef{fig:grid_results}{c}), with encoded CDFs lying entirely to the left of their unencoded counterparts. 
Further, the median absolute error for encoded runs is seen to be significantly smaller than unencoded runs. 
The largest encoding improvement in median absolute error is found to be $ 56.9 \% \pm0.8\%$ for $g_z=0.5$ and $R=1$.
Both the unencoded and encoded distributions show higher median error in Fig.~\SubFigRef{fig:grid_results}{c}) compared to Fig.~\SubFigRef{fig:chain_results}{d}) due to increased circuit depth.

The encoded signal-survival factor $\alpha$ is found to exceed unencoded values for all times and all $R$ with the exception of $R=16$ past $t\sim4$ (Fig.~\SubFigRef{fig:grid_results}{d})). 
Similar to the CDFs, the $\alpha$ values are observed to decrease faster in 2D than in 1D. 
Further, the improvement from encoding is significantly larger than in 1D.
~\ref{app:more_analysis} shows that $\alpha$ drops significantly and the improvement disappears for $R=4,8,16$ when postselection is removed.
This indicates that the improvement in $\alpha$ seen in Fig.~\SubFigRef{fig:grid_results}{} is largely due to postselection as opposed to reduced gate depth. 
In contrast to 1D, the decay of $\alpha$ is observed to be monotonic,
which we attribute to the slower proliferation of undetectable errors.
Runs with $R=1,2$ are found to produce comparable results,
indicating that the gain from more syndrome measurement rounds is balanced with the added errors from longer circuits.

\begin{figure}
    \centering
    \hspace*{.5cm}
    \includegraphics[width=0.9\linewidth]{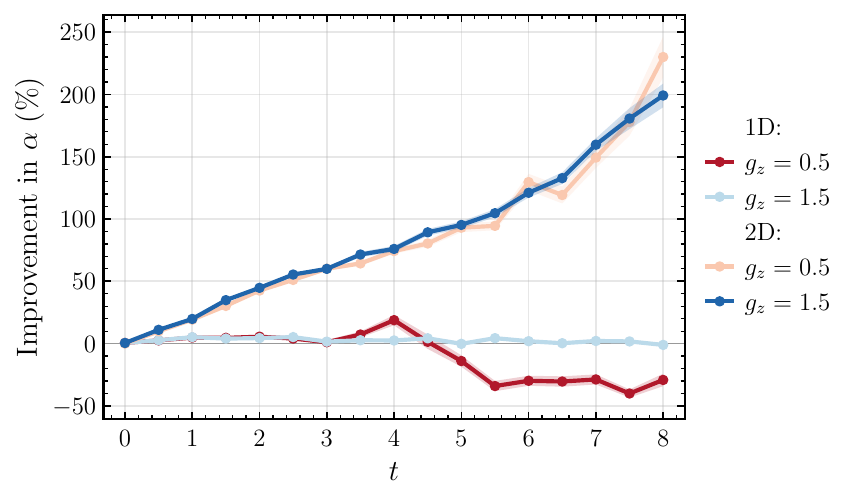}
    \caption{{\it Encoding Improvement in $\alpha$.} 
    Percent improvement in $\alpha$ of encoded over unencoded results as a function of time $t$ for 1D and 2D simulations. 
    Results with $R=1$ syndrome extraction rounds for the melting regime ($g_z=0.5$) are shown in shades of red, and the localized regime ($g_z=1.5$) is shown in shades of blue.
    The shading represents statistical uncertainty determined through bootstrap resampling.
    }
    \label{fig:improvement}
\end{figure}

Figure~\SubFigRef{fig:improvement}{} shows the percentage improvement in $\alpha$ over time for $R=1$ in 1D and 2D simulations.
The improvement in the results from runs with encoding grows approximately linearly in 2D, with an average slope of $23.5\pm0.8\%$ in the melting regime and $22.8\pm0.5\%$ in the localized regime.
Because each caught error prevents corruption over a lightcone volume scaling as $\mathcal{O}(t^2)$ in 2D compared to $\mathcal{O}(t)$ in 1D, detection has greater leverage in 2D and the encoding improvement grows with $t$.
The greater than $200\%$ improvement in 2D is also driven by rapidly decohering unencoded results, as is similarly seen in Fig.~\SubFigRef{fig:grid_results}{b}). 
Dynamics at $g_z=0.5$ and $g_z=1.5$ are more similar in 2D than in 1D, explaining their close trajectories in Fig.~\SubFigRef{fig:improvement}{}.
Beyond $t\sim4.5$, the melting regime in 1D shows a disadvantage from encoding because of the proliferation of errors enabled by fast dynamics. 
In 1D, the encoding improvement is approximately constant: for $g_z=0.5$ an improvement is seen up to to $t=4.5$ (average improvement of $5.99\pm0.99\%$), and for $g_z=1.5$ the average improvement is $2.5\pm0.23\%$ over all $t$.
Additional analysis of these results is shown in~\ref{app:more_analysis} and data, including acceptance rates, is given in~\ref{app:tables}.
Together, these results show that both the density of nFT gates and the dynamics being simulated affect the encoding improvement.

\section{Discussion}
\label{sec:discussion}
\noindent
The partially FT quantum simulations performed in this work using IBM's quantum computers provide a concrete demonstration of the benefit of encoded error detection on contemporary hardware.
Despite significant overhead associated with encoding and the shot loss resulting from postselection, results obtained using our scheme show improvements for both 1D and 1D simulations. 
These results reside in a crossover regime between NISQ heuristics and fully FT device operation~\cite{Preskill:2025cbl}, which shapes our circuit design.
Our circuits combine FT syndrome extraction with nFT logical operations, resulting in jointly optimized circuit overhead and error resilience properties.
Increased frequency of syndrome extraction is seen to improve results at early times, as expected from a fully-FT computation. 
Late-time performance instead benefits from the nFT reduction in overhead provided by sparser error detection.
This balance reflects a central tradeoff in the crossover regime: more syndrome measurements aid filtering but consume some of the coherence time of the quantum computer. 
A similar tradeoff exists for the addition of ancilla resets following syndrome readout. 
Designed to operate inside this tradeoff, ORP postselects on a ranked subset of observable-relevant checks and recovers much of the benefit of full postselection without an exponential loss in statistics.
The implementation of a QEC code would avoid the filtering machinery built to manage the postselection overhead.
Our circuit design choices and error-detection scheme are tailored to the specific simulation and observables of interest.
We expect this kind of customization to be a persistent feature of the crossover regime.

The potential for encoded advantage in quantum simulation is underscored by its sensitivity to device performance. 
The absence of an advantage on devices with slightly worse error rates ($\mathcal{O}(10^{-3})$) and coherence times ($\mathcal{O}(100\mu \text s)$) than {\tt ibm\_boston} shows that even modest improvement in device characteristics is enough to cross the pseudothreshold.
Since our figure of merit is local observable estimation, our results do not directly imply a favorable logical error-rate scaling.
However, the improvement in observable estimation suggests that state-of-the-art devices are in the vicinity of the pseudothreshold for encoding procedures with a small footprint.
This reinforces the understanding that any improvement in the crossover regime is tied to specifics of code choice, overhead, and hardware operating point rather than being a generic feature of the encoding, and should become more robust and widespread as physical fidelities improve.

While the encoded improvements found in this work hold across dynamics and dimensionality, both encoded and unencoded results would likely benefit further from error mitigation.
Studying FT constructions that remove leading-order errors, alongside mitigation techniques that suppress higher-order contributions, is a very promising direction and should be pursued. 
This includes both standard error mitigation techniques based on physical-noise learning~\cite{Wallman:2015uzh,Farrell:2023fgd,Urbanek:2021oej,Froland:2026aff,ARahman:2022tkr,Berg:2020ibi,Temme:2016vkz,Berg:2022ugn}, as well as potential techniques operating directly on syndrome information.
The incorporation of such techniques involves circuit and observable-dependent design choices, and the implications for the encoded advantage remain to be seen.

Beyond the direct advantage in observable estimation, encoding expands the versatility of logical computations where no shallow unencoded analog exists. 
It enables simulations whose direct unencoded implementation carries substantially higher overhead from connectivity constraints alone. 
The 2D lattice simulations in this work illustrate this: the realized connectivity graph is costly to embed directly on heavy-hex hardware. 
This approach extends to more complex logical connectivities, such as three-dimensional lattices required for lattice gauge theories, e.g. Refs.~\cite{Klco:2021lap,Bauer:2022hpo,Beck:2023xhh,Bauer:2023qgm,DiMeglio:2023nsa} and the Fermi-Hubbard model, e.g. Ref.~\cite{Esslinger:2010rek}.
As these constraints relax, we expect a more diverse range of simulations to become feasible in the crossover regime, where the dominant constraints shift to coherence time and circuit depth.
The benefit of encoding is therefore not limited to solely improving the accuracy of a given simulation, but also extends to making entirely new classes of simulations accessible.

\section*{Acknowledgements}
\noindent
We would like to thank John Preskill for helpful discussions.
This work was supported, in part, by U.S. Department of Energy, Office of Science, Office of Nuclear Physics, InQubator for Quantum Simulation (IQuS) under Award Number DOE (NP) Award DE-SC0020970 via the program on Quantum Horizons: QIS Research and Innovation for Nuclear Science.
Support is also acknowledged from the U.S. Department of Energy, Office of Science, National Quantum Information Science Research Centers, Quantum Systems Accelerator (Award No. DE-SCL0000121).
This work was also supported, in part, by the Department of Physics and the College of Arts and Sciences at the University of Washington,
and enabled, in part, by the use of advanced computational, storage and networking infrastructure provided by the Hyak supercomputer system at the University of Washington.
We acknowledge the use of {\tt Claude} for code development.
We acknowledge the use of IBM Quantum Credits for this work. 
The views expressed are those of the authors, and do not reflect the official policy or position of IBM or the IBM Quantum team.
The Quimb~\cite{gray2018quimb} library was used for tensor network simulations, and QuSpin~\cite{WeinbergED} was used for exact diagonalization computations.

\let\addcontentsline\oldaddcontentsline

\setcounter{section}{0}
\renewcommand{\thesection}{Methods~\Alph{section}}
\renewcommand{\thesubsection}{\arabic{subsection}}

\section{ORP Details}
\label{sec:filtering}

\subsection{Detector Construction From Syndromes}
\label{sec:detector_construction}
\noindent
Syndrome measurements record the eigenvalue of a stabilizer at the time of measurement, and connecting this value to the underlying error mechanisms is done through the construction of detectors.
A detector $v_j$ is a parity of various syndrome measurements that has a deterministic value for a noiseless circuit~\cite{Derks:2024jyw}, where $j=(g,b,r)$ is a ``space-time'' label denoting the stabilizer $g$, block $b$, and round $r$ that the detector measures.
The same construction applies to both $X$-type and $Z$-type detectors: the generator index $g$ runs over the $X$ stabilizers to build the $X$ detectors and over the $Z$ stabilizers to build the $Z$ detectors, with each detector formed only from syndrome bits of a single stabilizer within a single block.
The two stabilizer types are therefore structurally identical, and differ only through the ``temporal boundaries'', or how detectors are constructed at initialization and when terminal measurements on the data qubits are made.
In this case detectors are set by the logical state in which each block is prepared and the basis in which it is measured out.

\medskip
\noindent\textit{No-reset circuits.} For the no-reset circuits used in this work, the ancilla measuring a given stabilizer is not reinitialized between cycles, so its outcome records the running parity of that stabilizer's eigenvalue history rather than its instantaneous eigenvalue.
The deterministic combination is then the parity of two syndrome measurements in the same code block two cycles apart~\cite{Geher:2024lkc},
\begin{equation}
    v_{j+1} \ = \ s_{j+1}\oplus s_{j-1} \ ,
\end{equation}
where $j\pm 1$ is shorthand for $(g,b,r\pm 1)$, so that the intervening cycle is skipped.
Taking the parity across two cycles cancels the accumulated history carried by the unreset ancilla and isolates the change in the stabilizer eigenvalue, ensuring that an isolated error flips only a small, bounded set of detectors. 
For the first detector with $j=1$, the syndrome measurements are padded with a reference value of $0$ (so no error has occurred at the beginning of the circuit). 
This padding renders the first detector deterministic for every stabilizer whose eigenvalue is fixed by the state preparation. 
In this work, each block of the $[[4,2,2]]$ code is initialized in $\ket{\overline{00}}$, which is a simultaneous $+1$ eigenstate of both $S_X$ and $S_Z$.
Both stabilizer eigenvalues are therefore fixed at preparation, and the padding yields a deterministic first detector for the $X$-type and $Z$-type stabilizers alike.
Errors during initialization will go undetected unless the the preparation circuits is itself FT, however since $\ket{GHZ(4)}$ circuits are exceptionally shallow errors during this stage do not contribute very much to the overall logical error rate.
Had the block instead been prepared in a bare computational-basis product state, only the $Z$-type stabilizers would be fixed by the preparation. 
The $X$-type eigenvalue on the other hand would be random and its first detector would form only once two in-circuit syndrome measurements are available and their relative parities are deterministic.
At the final cycle the data-qubit readout reconstructs the stabilizers compatible with the measurement basis, and this reconstructed value plays the role of $s_{j+1}$ in the boundary detector, closing the space-time volume.
For $R$ rounds, the final detector in the measurement basis reads
\begin{equation}
    v_{R+1} \ = \ s_{R-1}\oplus s_{R}\oplus\left(\bigoplus_{d\in\text{supp}(g)}d\right) \ ,
\end{equation}
where $\text{supp}(g)$ are the data qubits on which stabilizer $g$ is defined; in this work it is always the $S_Z$ stabilizer as measurements are always be made in the $Z$ basis.
For the simulations with only a single stabilizer-measurement round, the two-cycle rule has no earlier in-circuit round to reference, and the $0$-padding convention makes each detector coincide with its stabilizer measurement, $v_{1}=s_{1}$. 
Because the GHZ initialization fixes both eigenvalues, this holds for the $X$-type and $Z$-type stabilizers alike, supplemented by the measurement-basis boundary detector,
\begin{align}
    v_{2} \ & = \  s_{1}\oplus\left(\bigoplus_{d\in\text{supp}(S_Z)}d\right)
    \ .
\end{align}

\medskip
\noindent\textit{Reset circuits.}
When the ancilla is instead reset to $\ket{0}$ after each syndrome measurement, its outcome reports the instantaneous stabilizer eigenvalue rather than a running parity. 
The detector then reduces to the parity of syndrome measurements in consecutive cycles,
\begin{equation}
    v_j \ = \ s_{j-1} \oplus s_{j} \ ,
\end{equation}
where the index convention is the same as in the no-reset case and the same padding convention supplies the reference value for the first detector. Similarly, the construction of the $X$ and $Z$ detectors and the initialization boundary are unchanged, and both stabilizer types again acquire a deterministic first detector from the GHZ preparation. 
The only modification at the terminal boundary is that the reconstructed data value is compared against a single syndrome measurement round,
\begin{equation}
    v_{R+1} \ = \ s_R\oplus \left( \bigoplus_{d\in\text{supp}(g)}d \right) \ ,
\end{equation}
since $s_{R}$ is already the instantaneous eigenvalue. The remaining practical difference is in which detectors are flipped by an error: a measurement fault flips a pair of \emph{adjacent} detectors, $v_j$ and $v_{j+1}$, rather than the next-nearest pair $v_j$ and $v_{j+2}$ produced in the no-reset case, so the time-like edges connect neighboring rather than next-neighboring cycles.

\subsection{Detector Selection}\label{sec:det_selec}
\begin{figure}
    \centering
    \includegraphics[width=0.8\linewidth]{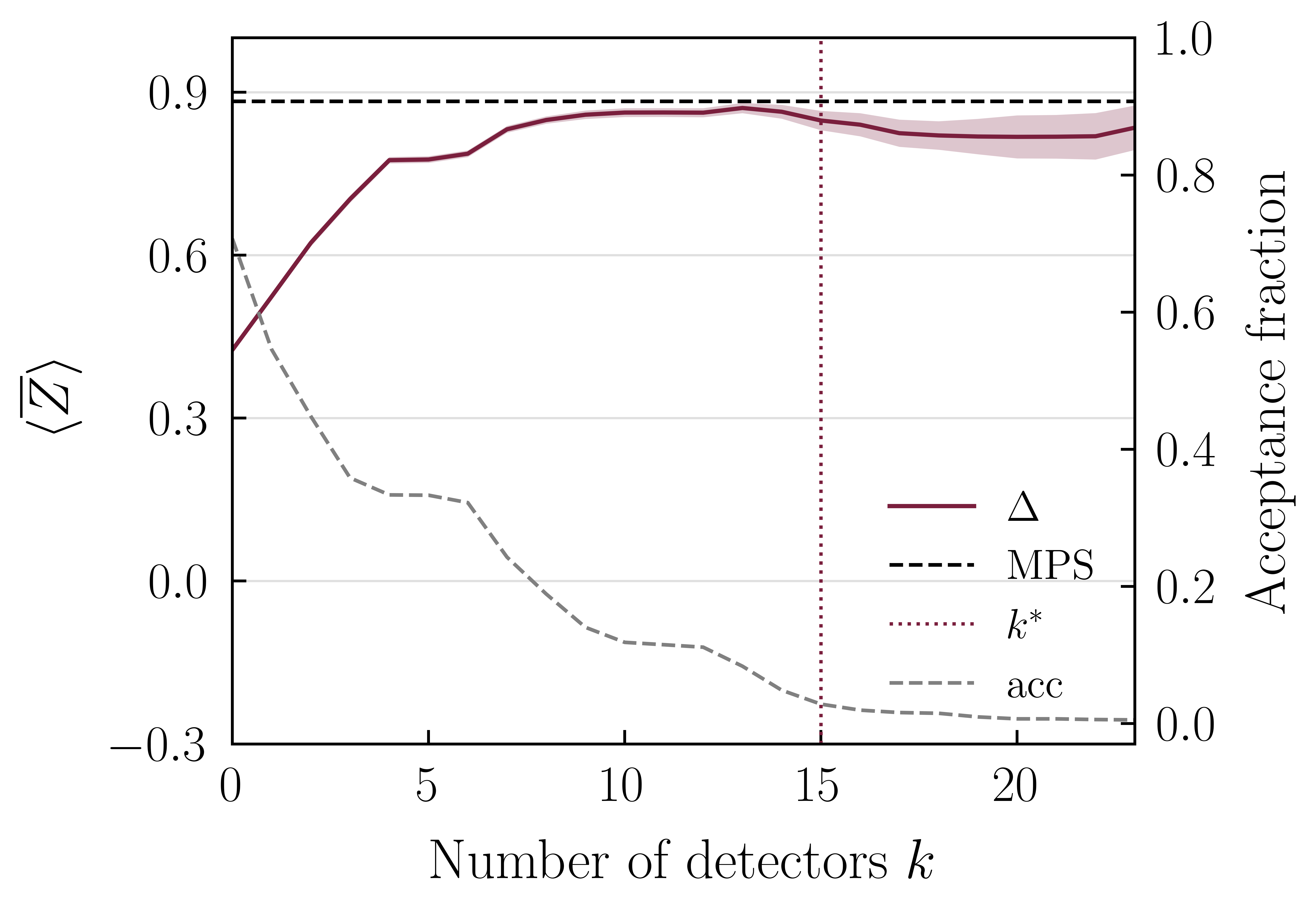}
    \caption{\textit{Convergence of ORP on {\tt ibm\_boston}}. 
    Four Trotter steps of size $\delta t=0.4$ are applied to four code blocks, each initialized in the $|\overline{00}\rangle$ state, with couplings $g_x=g_z=1$.
    The solid line shows $\langle \overline Z_{(0,0)}\rangle$ computed through ORP as a function of the number of detectors $k$ used in the postselection criterion, with the noiseless MPS prediction given by the horizontal dashed line. 
    The corresponding acceptance fraction is shown by the dotted line (right axis). 
    The vertical red dotted line marks the value of $k^*$ chosen by the plateau-finding algorithm.
    \label{fig:nkeep_methods}}
\end{figure}

\noindent
The following plateau-finding algorithm is used with ORP to determine which detectors should be used for postselection.
\begin{enumerate}
    \item Given an observable $\overline{O}$, detectors $v_j$ are ranked in order of decreasing ${\Delta_j = |\langle \overline{O} \rangle_{v_j=0} - \langle \overline{O}\rangle_{v_j=1} |}$.\footnote{Since $\overline O$ is defined up to conjugation by $S_X$ and $S_Z$, the calculation of $\Delta_j$ depends on which data qubits are used to compute $\langle \overline O\rangle$.}
    \item A sweep over cutoff levels $k$ is done, where each level $k$ only retains shots for which none of the top-$k$ detectors record an error.
    Final blockwise parity postselection is always applied regardless of $k$.
    This produces a sequence of postselected expectation values $\langle \overline{O}\rangle_k$ with bootstrapped uncertainties $\sigma_k$.
    The acceptance fraction among all blockwise parity-even shots $f_k$ is recorded for each $k$.
    \item A reference value is built by combining the most heavily filtered levels using an inverse-variance weighted average, 
    \begin{align}
        O_\text{ref} \ = \ \frac{\sum_{k\in\{k_\text{ref}\}}\langle \overline{O}\rangle_k/\sigma_k^2}{\sum_{k\in\{k_\text{ref}\}}1/\sigma_k^2} \ ,
    \end{align}
    where $\{k_\text{ref}\}$ includes levels $k$ that satisfy ${f_\text{min}\leq f_k \leq f_\text{ref}}$.
    \item Starting from large $k$ and moving down, a plateau is found by looking for the longest run of $k$ where ${|\langle \overline{O}\rangle_k-O_\text{ref}|\leq z\sigma_k}$. 
    \item The selected value $k^*$ is the most permissive $k$ in this run.
    If a sufficiently long plateau isn't found, the cut level is set to be ${k^* = \argmin\limits_k \, [(\langle \overline{O}\rangle_k - O_\text{ref})^2 + \sigma_k^2]}$.
\end{enumerate}
This procedure is carried out on the held-out half of the results, and the resulting $k^*$ is applied to the complementary half to report $\langle O\rangle_{k^*}$.
This work uses the parameters $f_\text{min}=0.005,\,f_\text{ref}=0.05,\,z=1$, although the quality of $\langle \overline{O}\rangle_k$ is observed to be similar for a range of parameters.
Observables for all times $t$ and logical qubits $(b,l_b)$ are calculated using this algorithm. An example of the convergence of local expectation values for four blocks on {\tt ibm\_boston} as a function of the number of detectors used in the postselection criteria is shown in Fig.~\SubFigRef{fig:nkeep_methods}{}. The plateau-finding algorithm successfully chooses a point within the vicinity of the noiseless MPS prediction, predicting $\langle \overline Z\rangle = 0.85(2)$ compared to an ideal value of $0.88$. While the associated acceptance fraction decreases exponentially, choosing to filter only on $k^*$ detectors does not leave the total ensemble exponentially small in system size.

\section{Block Placement Selection}
\label{sec:layouts}
\noindent
The goal of layout selection is to arrange the logical qubits so that they closely resemble some target geometry, in our case a regular square lattice that we call a ``logical grid".
The heavy-hex quantum processor's physical connectivity, combined with the choice of logical connectivity, creates a large search space of candidate logical grids. 
This can be formulated as a rectangle placement problem, since each $[[4,2,2]]$ block encodes two logical qubits, we represent it as a $2\times1$ rectangle occupying two sites of a square lattice.
Two logical qubits couple when their blocks occupy adjacent lattice sites such that they are directly connected by a physical gate.
We apply this formulation to the heavy-hex connectivity, though it generalizes to arbitrary device geometries.

We first enumerate all placements $P_i$ of blocks onto the device that maximize the number of blocks used, subject to the heavy-hex connectivity and excluding a set of qubits with poor performance. 
For {\tt ibm\_boston} (avoiding four noisy qubits), we find 42 different embeddings $P_i$ supporting 21 code blocks each, i.e., 42 logical qubits per embedding.

\begin{figure}
    \centering
    \includegraphics[width=0.75\linewidth]{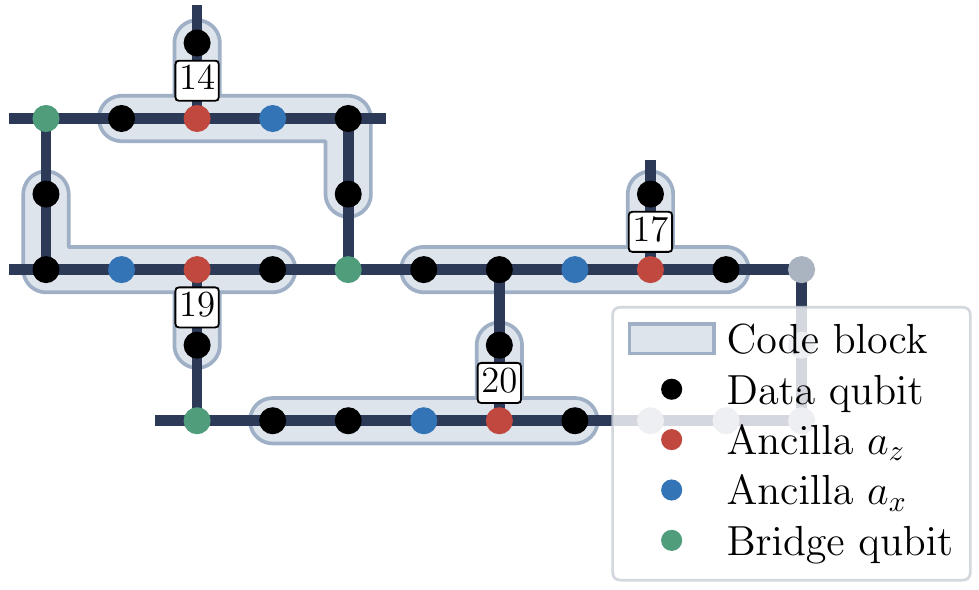}
    \caption{{\it Example $[[4,2,2]]$ Code Block Placement onto Heavy-Hex Connectivity}.
    Several blocks (grey) are shown placed onto a heavy-hex connected quantum processor.
    Data qubits are given in black, $Z$- and $X$-type ancillas are shown in red and blue respectively, and a bridge qubit is shown in green. }
    \label{fig:block_placement_example}
\end{figure}

Within a given $P_i$, logical qubits within two blocks are allowed to couple if their blocks are adjacent or share a single ``bridge'' qubit on the device.
An example of this is shown in Fig.~\SubFigRef{fig:block_placement_example}{}, where blocks 20 and 17 are directly adjacent, while blocks 17 and 14 are considered adjacent through the green bridge qubit.
Relabeling logical qubits $(b,0)\leftrightarrow(b,1)$ or ancillas $a_x\leftrightarrow a_z$ within a block leaves the intra-block topology unchanged, so we treat these configurations as equivalent.
For a given placement $P_i$, the realized edge set is
\begin{align}
    E(P_i) \ =\ &\underbrace{\{(b,0),(b,1)\}_{b\in P_i}}_{\text{intra-block, always present}} \nonumber\\
    &\cup\ \bigl\{\{(a,l_a),(b,l_b)\} : \nonumber\\
    &\qquad (a,l_a)\ \text{adjacent to}\ (b,l_b) \;\forall b\in P_i\bigr\}
    \ ,
\end{align}
i.e., the set of intra-block edges and available adjacent inter-block edges.
The degree of a given vertex $v$ is $\text{deg}(v) = 1+\#\{\text{inter-block edges}\}$, and we require $2\leq \text{deg}(v) \leq 4$ for all $v$.

Finding the best logical grid is a combinatorial optimization over the space of edge configurations, which is too large to search exhaustively.
Instead, we use simulated annealing~\cite{Kirkpatrick:1983zz} to find the approximate solution.
Each candidate configuration is scored as ${S(E) = |E| - \lambda (\kappa(E)-1)}$.
Here $\kappa(E)$ is the number of connected components and $\lambda$ is a large positive coefficient that penalizes disconnected graphs.
This form allows the algorithm to briefly pass through disconnected configurations to reach higher-scoring connected ones.
In the annealing search, modifications $E'$ to the current layout $E$ are accepted with the probability $p_\text{accept} = \min(1, e^{(S(E') - S(E))/T})$, where ${T(t) = T_0 \left(T_\text{end}/T_0\right)^{t/(N_\text{it}-1)}}$ is the cooling schedule which specifies the search rate.
A periodic reheating every $R_h$ steps, $T\leftarrow \max(T,T_\text{reheat})$ to escape local minima is included. 
We use the parameters ${T_0=4},{T_\text{end}=0.01},{T_\text{reheat}=2},{R_h=6000},{N_\text{it}=25000}$.

The set of allowed updates are single-rectangle relocations, single rectangle rotations, two-rectangle swaps, and three-rectangle cyclic swaps.
In addition, with probability $0.15$ an under-realized block (one that is not yet coupled to all its available neighbors) is relocated to a spot that maximizes its realized edges. 
This prevents the algorithm from getting stuck on grids with thin chains attached.
This process is repeated 500 times for every placement $P_i$ with random seeds.
The parameters for the simulated annealing are heuristically chosen to determine a suitable logical grid.
With this algorithm, we find the grid shown in Fig.~\SubFigRef{fig:overview}{b}), which has 66 logical edges (45 inter-block and 21 intra-block).\footnote{In principle, the Trotter step depth could be optimized jointly with the layout. 
This work does not do so and instead fixes it in advance. Trotter step optimization is discussed in~\ref{sec:scheduling}.} 
\ref{app:other_layouts} discusses other block placements that may be more favorable on similar heavy-hex quantum processors.

\section{Circuit Scheduling}
\label{sec:scheduling}
\noindent
Once the optimal logical grid is determined, a Trotter step of the Hamiltonian in Eq.~\eqref{eq:h_ising} must be built. 
Naive placement of inter-block $ZZ$ rotations results in deep circuits and long idle times.
As shown later in this section, finding the optimal Trotter step for a given grid reduces to determining the best placement of inter-block gates $U_e(\delta t) = e^{-i \delta t/2 \overline{Z}_{(a,l_a)} \overline{Z}_{(b,l_b)}}$.
The logical operations within a single block are defined in Eq.~\eqref{eq:logical_ops}.
It is assumed that blocks start in the compute configuration shown in the center of Fig.~\SubFigRef{fig:overview}{c}).
In each block $b$, 
$\overline{Z}_{(b,0)}$ and $\overline{X}_{(b,1)}$ can be implemented directly in this configuration while $\overline{Z}_{(b,1)}$ and $\overline{X}_{(b,0)}$ require SWAP($d_1,d_2$). 
A block's frame $f\in\{0,1\}$ records whether $d_1$ and $d_2$ have been swapped relative to their starting position.
Together with intra-block gates, inter-block gates on both logical qubits within a block require at least one frame toggle per Trotter step.

\begin{figure*}
    \centering
    \includegraphics[width=0.9\linewidth]{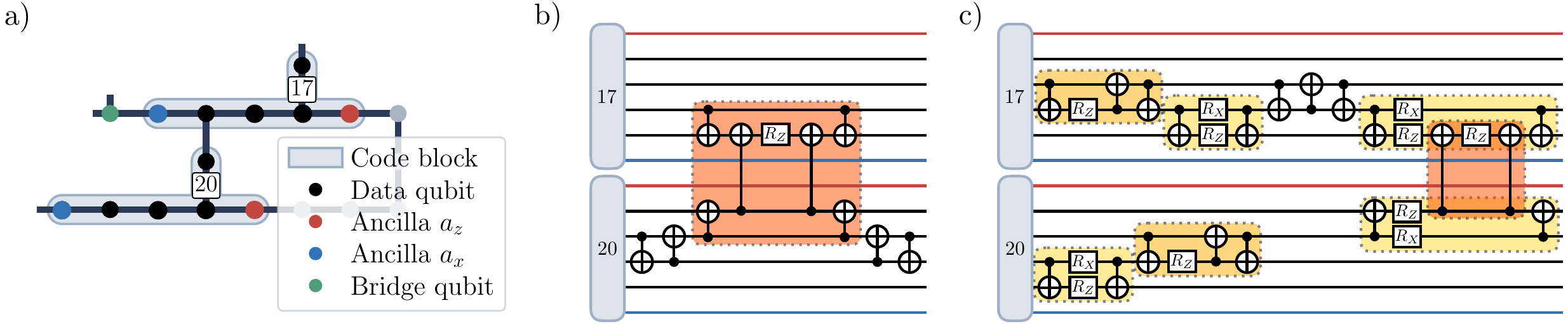}
    \caption{{\it Encoded Trotter Step Circuits.}
    a)~An example placement of two blocks in the compute configuration. 
    Data qubits are given in black, $Z$- and $X$-type ancillas are shown in red and blue respectively, and a bridge qubit is shown in green.
    b)~The CNOT fanout necessary for implementing a $\overline R_{ZZ}$ between logical qubits $(17,0)$ and $(20,1)$ (dark orange), with the associated SWAP gate converted to CNOTs.
    c)~A Trotter step involving logical qubits in blocks 17 and 20, with the inter-block gate between qubits $(17,0)$ and $(20,1)$.
    Single-logical-qubit gates (yellow) and intra-block $\overline R_{ZZ}$ (light orange) are folded into the CNOT fanout required for the inter-block $\overline R_{ZZ}$ gate, the last part of which is shown in dark orange. 
    Block 20 exits the Trotter step in frame $f=1$.}
    \label{fig:trot_step_circs}
\end{figure*}

Given two blocks $a$ and $b$, and a single connection point between them (a direct connection or through a bridge qubit), the shallowest implementation of the unitary $U_e$ applies CNOT fanout on both blocks (to accumulate the required parity), applies a $Z$ rotation on the central qubit, and uncomputes the accumulated parities by applying the fanouts in reverse.\footnote{In the case where two adjacent blocks have two connections, such as blocks 14 and 19 in Fig.~\SubFigRef{fig:block_placement_example}{}, the algorithm selects a schedule that may use both connections to minimize overall circuit depth.} 
An example of such a circuit between qubits $(17,0)$ and $(20,1)$ is shown in Fig.~\SubFigRef{fig:trot_step_circs}{b}). 
The depths of the inter-block $ZZ$ gadgets between all neighboring pairs $(a,l_a),(b,l_b)$ are shown in Table~\ref{tab:cost}.
\begin{table*}
\centering
\renewcommand{\arraystretch}{1.4}
\begin{tabularx}{\linewidth}{|c||Y|Y|Y|Y||c||Y|Y|Y|Y|}
\hline
\multirow{3}{*}{\makecell{Blocks\\$(a,b)$}} & \multicolumn{4}{c||}{Two-qubit gate depth} & \multirow{3}{*}{\makecell{Blocks\\$(a,b)$}} & \multicolumn{4}{c|}{Two-qubit gate depth}  \\ \cline{2-5}\cline{7-10}
 & \multicolumn{4}{c||}{$(l_a,l_b)$} &  & \multicolumn{4}{c|}{$(l_a,l_b)$} \\\cline{2-5}\cline{7-10}

 & (0,0) & (0,1) & (1,0) & (1,1) &  & (0,0) & (0,1) & (1,0) & (1,1) \\
\hline\hline
(0, 1) & 6 & 8 & 8 & 10 & (10, 13) & 4 & 8 & 6 & 8 \\
\hline
(0, 2) & 4 & 8 & 6 & 8 & (11, 12) & 6 & 8 & 8 & 10 \\
\hline
(1, 3) & 6 & 8 & 8 & 10 & (11, 14) & 4 & 8 & 6 & 8 \\
\hline
(2, 4) & 6 & 8 & 8 & 10 & (11, 15) & 6 & 8 & 8 & 10 \\
\hline
(2, 9) & 6 & 10 & 8 & 10 & (12, 15) & 6 & 8 & 8 & 10 \\
\hline
(3, 5) & 4 & 8 & 6 & 8 & (13, 14) & 6 & 8 & 8 & 10 \\
\hline
(3, 7) & 4 & 8 & 6 & 8 & (13, 16) & 4 & 8 & 6 & 8 \\
\hline
(4, 9) & 4 & 8 & 6 & 8 & (13, 19) & 6 & 10 & 8 & 10 \\
\hline
(5, 6) & 6 & 8 & 8 & 10 & (14, 15) & 4 & 8 & 6 & 8 \\
\hline
(5, 7) & 6 & 8 & 8 & 10 & (14, 17) & 8 & 10 & 10 & 12 \\
\hline
(5, 8) & 6 & 8 & 8 & 10 & (14, 19) & 6 & 10 & 8 & 10 \\
\hline
(6, 8) & 6 & 8 & 8 & 10 & (15, 17) & 4 & 8 & 6 & 8 \\
\hline
(6, 9) & 4 & 8 & 6 & 8 & (16, 18) & 4 & 8 & 6 & 8 \\
\hline
(6, 11) & 6 & 8 & 8 & 10 & (16, 19) & 6 & 8 & 8 & 10 \\
\hline
(7, 8) & 6 & 8 & 8 & 10 & (17, 19) & 6 & 10 & 8 & 10 \\
\hline
(8, 11) & 6 & 8 & 8 & 10 & (17, 20) & 4 & 8 & 6 & 8 \\
\hline
(8, 12) & 4 & 8 & 6 & 8 & (19, 20) & 6 & 10 & 8 & 10 \\
\hline
(9, 10) & 6 & 8 & 8 & 10 &  &  &  &  &  \\
\hline
    \end{tabularx}
    \caption{The inter-block $R_{ZZ}$ CNOT fanout depth for pairs of logical qubits $(l_a,l_b)$ between pairs of code blocks $(a,b)$ corresponding to the block placement shown in Fig.~\SubFigRef{fig:overview}{b}).
    The depth includes CNOTs used for SWAPs and bridge qubits where necessary.}
    \label{tab:cost}
\end{table*}
Three choices leave the logical circuit invariant, 
but may lower the depth: the orientation of block $o_b\in\{0,1\}$ (whether $a_x\leftrightarrow a_z$ is applied), $l_b\in \{0,1\}$ which specifies whether the $(b,0)\leftrightarrow (b,1)$ relabeling is applied, and $w_e$ controlling which connection to use between two blocks when multiple exist. 
We parameterize the schedule with $A = \{o,l,w\}$ for all blocks and edges in the logical grid.
Within a block, couplings in the same frame share the same accumulated $Z$-parity, so this parity is computed once per frame segment via a single fanout.

Scheduling for a fixed $A$ is itself a combinatorial optimization problem: even with the gate placements fixed, the ordering and layering of two-qubit gates into a minimum-depth circuit has a search space too large to explore exhaustively. 
The schedule for a given $A$ is determined using the constrained-satisfaction solver CP-SAT from the package {\tt ortools}~\cite{cpsatlp,ortools}.
This is possible because of commutation between the terms of each Trotter step. 
Together with fixing $f=0$ at the start of each Trotter step and placement of single qubit gates described below, this constrains the Trotter step to a single consistent ordering on all blocks, leaving the depth as the only free objective.
The objective minimizes the total depth with the additional degree of freedom that the exit frame need not be $f=0$. 
The parameters $A$ are determined by coordinate descent.

A full Trotter step of the Hamiltonian~\eqref{eq:h_ising} can be implemented at no extra depth using the schedule of inter-block $\overline R_{ZZ}$ gates, provided the logical grid satisfies the conditions in~\ref{sec:layouts}.
The single-logical qubit gates and the intra-block two-logical qubit gates can be folded into the inter-block $\overline R_{ZZ}$ gates.  
These fanouts already collect the $XX$ and $ZZ$ parities required for single qubit $\overline X$- and $\overline Z$-rotations.
Since each vertex $v$ on the selected lattice has $2\leq \text{deg}(v) \leq 4$, all four single-logical-qubit rotations can be implemented in all blocks with the addition of only single-physical-qubit rotations.
The parity required for the intra-block $\overline R_{ZZ}$ is also collected in each fanout at the time of the SWAP($d_1,d_2$) application, so the addition of this interaction also carries no extra two-qubit depth cost.
The specific location at which each single-logical-qubit gate and the intra-block $\overline R_{ZZ}$ is inserted depends on both the block's current frame and the structure of the particular inter-block $ZZ$ gadget being used.
An example of a Trotter step spanning two adjacent blocks is shown in Fig.~\SubFigRef{fig:trot_step_circs}{c}), with the inter-qubit and intra-qubit gates highlighted.
Since a Trotter step does not necessarily return every block to $f=0$, consecutive odd-numbered Trotter steps are appended in reverse order, so that frames of all blocks are restored to $f=0$ after even $n_T$.

\section{Quantum Simulation Details}
\label{sec:quantum_sim}
\noindent
While more efficient time evolution techniques and Trotter step implementations are possible for unencoded circuits and MPS simulations,
we execute circuits with the same Trotter order to hold observable expectation values fixed.
Further, results from encoded and unencoded runs with the same number of shots are compared, resulting in biased unencoded results with small error bars.
The ``extra" shots in the unencoded runs could be instead used to efficiently mitigate errors in the results through noise-learning methods~\cite{Wallman:2015uzh,Farrell:2023fgd,Urbanek:2021oej,ARahman:2022tkr,Berg:2020ibi,Temme:2016vkz,Berg:2022ugn}.
Ideally, syndromes would be measured after every operation that could spread errors to multiple qubits.
In practice this is prohibitively costly due to limited device coherence time.
Tables~\ref{tab:chain_circ_nums} and~\ref{tab:grid_circ_nums} show the overhead in terms of depth, gate counts, and coherence time for encoded and unencoded runs in 1D and 2D, respectively.
\begin{table*}
\renewcommand{\arraystretch}{1.4}
\begin{tabularx}{\linewidth}{|c||Y|Y|Y|Y|Y|Y|}
\hline
& Unencoded & $R=1$ & $R=2$ & $R=4$ & $R=8$ & $R=16$ \\
\hline\hline
\# measurements & 42 & 126 & 168 & 252 & 420 & 756\\\hline
Measurement duration ($dt$) & 446 & 446 & 671 & 1121 & 2021 & 3821\\\hline
Total duration ($dt$) & 6017 & 10171 & 11139 & 13075 & 16947 & 26451\\\hline
Depth & 116 & 272 & 297 & 347 & 447 & 719 \\\hline
\# 2-qubit gates & 1126 & 3266 & 4112 & 5804 & 9188 & 16404\\\hline
\# shots & 32000 & 32000 & 32000 & 32000 & 32000 & 32000 \\\hline
\end{tabularx}
\caption{{\it Details of 1D Chain Simulations on {\tt ibm\_boston} at $t=8$}.
The columns compare unencoded counts to counts in encoded runs with various numbers of syndrome extraction rounds $R$.
The second row shows the total number of measurements (both mid-circuit and terminal).
The maximum coherence time used by measurements on any qubit in units of $dt=4\times10^{-9}$s is given in the third row.
The fourth row gives the total duration of the circuit.
Depths, total numbers of two-qubit gates, and numbers of shots executed are given in the fifth, sixth and seventh rows respectively}
\label{tab:chain_circ_nums}
\renewcommand{\arraystretch}{1.0}
\end{table*}
\begin{table*}
\renewcommand{\arraystretch}{1.4}
\begin{tabularx}{\linewidth}{|c||Y|Y|Y|Y|Y|Y|}
\hline
& Unencoded & $R=1$ & $R=2$ & $R=4$ & $R=8$ & $R=16$ \\
\hline\hline
\# measurements & 42 & 126 & 168 & 252 & 420 & 756\\\hline
Measurement duration ($dt$) & 446 & 446 & 671 & 1121 & 2021 & 3821\\\hline
Total duration ($dt$) & 19135 & 13931 & 14935 & 16943 & 20959 & 30591 \\\hline
Depth & 626 & 443 & 470 & 524 & 632 & 912 \\\hline
\# 2-qubit gates & 4286 & 5540 & 6414 & 8162 & 11658 & 19418 \\\hline
\# shots & 32000 & 32000 & 32000 & 32000 & 32000 & 32000 \\\hline
\end{tabularx}
\caption{{\it Details of 2D Grid Simulations on {\tt ibm\_boston} at $t=8$}.
The same rows and columns as Table~\ref{tab:chain_circ_nums} but for the 2D simulations presented in Sec.~\ref{sec:results}.}
\label{tab:grid_circ_nums}
\renewcommand{\arraystretch}{1.0}
\end{table*}
Given that the two-qubit gate duration is 
$17\ dt$ to $22\ dt$ on {\tt ibm\_boston}, 
both the encoded and unencoded circuits have significant idle time. 
The circuits in this work do not exceed the median coherence times on {\tt ibm\_boston}.\footnote{The median T1 and T2 times on {\tt ibm\_boston} were $234\mu$s and $275\mu$s when the data in this work was taken.}
However, degradation with depth is still observed, dominated by qubits with low T1/T2 times, high measurement error rates, or regions with faulty two-qubit gates.
The 1D encoded circuits are two to six times deeper than their unencoded counterparts.
This is similarly reflected in the durations and total gate counts. 
Table~\ref{tab:grid_circ_nums} shows the geometric overhead of embedding a square lattice onto heavy-hex connectivity for the unencoded circuits, which are $\sim1.5$ times deeper than the shallowest encoded circuit. 
This is one of the driving factors of their degrading performance as seen in Fig.~\SubFigRef{fig:grid_results}{}, and is also reflected in inflated gate counts and durations. 

\begin{figure}
    \centering
    \includegraphics[width=\linewidth]{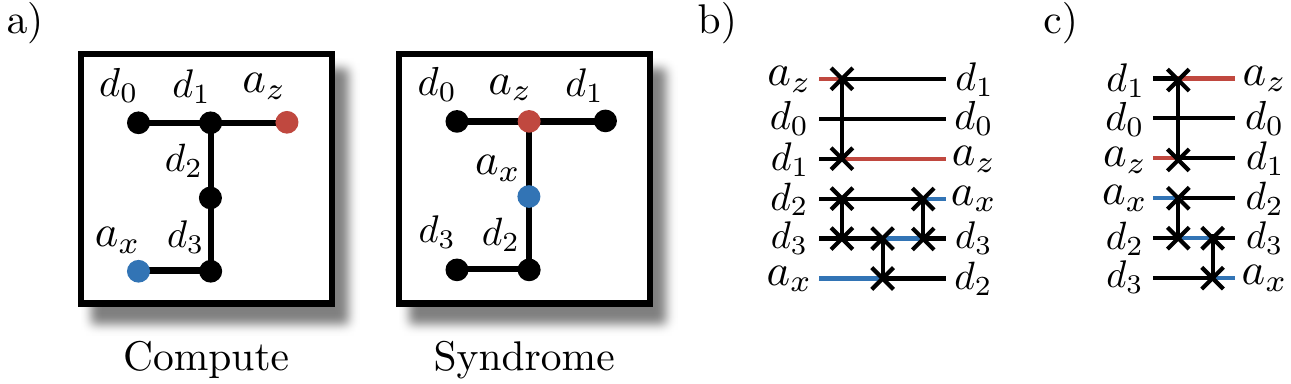}
    \caption{{\it Circuits for Switching Between Compute and Syndrome Block Configurations.}
    a)~The compute and syndrome block configurations as shown in the center of Fig.~\SubFigRef{fig:overview}{c}). 
    Data qubits are shown in black, $Z$-type ancillas are shown in red, and $X$-type ancillas are shown in blue.
    b)~The circuit to switch from the compute to the syndrome configuration.
    c)~The corresponding circuit to switch from the syndrome to the compute configuration.
    The syndrome extraction circuit applies a SWAP$(d_2,d_3)$, which is taken into account in the configuration switching circuit.
    }
    \label{fig:switch_layouts_circs}
\end{figure}

Switching between the compute and syndrome block configuration introduces a depth $\sim15$ overhead at each syndrome measurement.\footnote{Some of this depth is reduced by CNOT cancellations with the Trotter steps in practice.}
Figure~\SubFigRef{fig:switch_layouts_circs}{} shows the SWAP circuits necessary to switch between the compute and syndrome configuration necessary for the syndrome extraction circuit , which are shown at the bottom right of Fig.~\SubFigRef{fig:overview}{c}).

\section{Classical Simulation Details}
\label{sec:classical_sim}
\noindent
Matrix product state simulations are used as a classical comparison to the results from {\tt ibm\_boston}. 
Despite the quench dynamics, for the 1D chain this is efficient because of the small system size, while In 2D the cost grows sharply.
Mapping an $L\times L$ grid onto a 1D MPS introduces long-range couplings, and the entanglement across a cut scales with the linear cut length.
Faithfully representing a general state on a 2D lattice requires a {\tt max\_bond} growing roughly exponentially in $L$, with per-step cost $\mathcal{O}(L^2\ {\tt max\_bond}^3)$. 
The required {\tt max\_bond} is set by the entanglement, which grows in time as the quench spreads correlations.
While more efficient classical simulation techniques for 2D lattices exist, 
such as tree tensor networks, belief propagation, and sparse Pauli dynamics, MPS is used for simplicity.

Time evolution is simulated by applying Trotter steps with ${\tt max\_bond}=1500$ and singular-value cutoff of ${\tt cutoff}=10^{-8}$.
Two sets of simulations are run. 
Late-time simulations with $\delta t=0.1$ with up to $n_T=500$ (corresponding to Fig.~\SubFigRef{fig:mps_2d_gz_comparison}{}) are used to confirm the presence of physical effects absent at early times and washed out by large $\delta t$. 
While truncation errors in 1D simulations are negligible, significant error accumulates for a 2D lattice.
The missing links of the logical connectivity grid shown in Fig.~\SubFigRef{fig:chain_results}{a}) modify the conventional ``snake" MPS ordering, which likely modifies the convergence.
The error, quantified by the norm of the MPS wavefunction at $n_T=100$ (the latest time in Fig.~\SubFigRef{fig:mps_2d_gz_comparison}{}) is $\mathcal{O}(10^{-3})$.
While this value is relatively large, changing {\tt max\_bond} from 1200 to 1500 only modifies observables at $\mathcal{O}(10^{-2})$ which is sufficient for this qualitative study.
The latter simulations execute the exact unencoded circuits run on ${\tt ibm\_boston}$ with $\delta t=0.5$ and up to $n_T=20$ Trotter steps for a noiseless comparison to device results.
The truncation errors in these simulations, both for 1D and 2D lattices, are negligible.
Larger values of $\delta t$ deform the dispersion relation and add an additional quench to the dynamics~\cite{Yang:2023nak,Milsted:2020jmf,Farrell:2025nkx}.
The localized regime ($g_z=1.5$) is observed to spread entanglement less than the melting regime ($g_z=0.5$).
The results of the $5\times5$ simulations displayed in Fig.~\SubFigRef{fig:mps_2d_gz_uniform_comparison}{} 
were computed by {\tt qiskit} statevector simulation.

\clearpage
\onecolumngrid
\setcounter{section}{0}
\renewcommand{\thesection}{Appendix~\Alph{section}}
\renewcommand{\thesubsection}{\arabic{subsection}}

\section{Details on Filtering Metrics}
\label{app:filtering_metric_details}
\noindent
This appendix presents a detailed analysis of the filtering metric $\Delta_j$ used in ORP. 
It also introduces $\beta_j$, an alternative metric based on linear regression.
First, both of these quantities are defined in~\ref{app:metric_defs}. 
Then, a worked example on a small system of one code block and two rounds of syndrome measurements is given in~\ref{app:worked_example} to provide a concrete implementation of these metrics.
After,~\ref{app:correlation_description} introduces a general, high-level description of errors in terms of the correlations among detector measurements and correlations between detector and observable measurements. 
The quantities $\Delta_j$ and $\beta_j$ are then expressed in terms of the general model parameters in~\ref{app:metric_details}, demonstrating the types of errors that they filter. 
These expressions simplify substantially when considered in a weak-noise limit, and can be seen to directly connect to the probability of an error detection event propagating to a logical error on an observable.
Finally,~\ref{app:pp_lightcone} connects the detectors chosen based on the $\Delta_j$ ranking, as done in ORP, to the backwards lightcone of a target local observable using Pauli Propagation.

In what follows, the observables considered are Pauli strings and are denoted by $\overline{O}$ when considered as operators, with expectation values denoted as $\langle \overline{O} \rangle$. 
The result of a single measurement of such an observable will be denoted as $Q$ ($Q_i$ when the measurement is a member of an ensemble, where $i$ is the shot index), which can take values $Q\in\{-1,1\}$.
Expectation values and conditional expectation values of $Q$ over an ensemble are denoted as $\mathbb{E}[\cdot]$ and $\mathbb{E}[\cdot|\cdot]$. 
More general observables can be decomposed into a weighted sum of Pauli strings, and many of the following results can be extended to this context.

\subsection{Definitions of \texorpdfstring{$\Delta_j$}{Delta\_j} and \texorpdfstring{$\beta_j$}{Beta\_j}}
\label{app:metric_defs}
\noindent
Recall the definition of $\Delta_j$ from Eq.~\eqref{eq:delta_def}, rewritten as an expectation value over shots
\begin{align}\label{eq:delta_exp}
    \Delta_j \ = \ \big|\mathbb{E}[Q | v_j=0] - \mathbb{E}[Q | v_j=1] \big| \ .
\end{align}
This quantity is calculated by taking the difference in the expectation values of observable $Q$ over shots postselected on the detector $v_j$ not detecting/detecting an error respectively. 
To define $\beta_j$, let $X_i$ be a length $N_d$ vector of detector outcomes for a single shot, so that $X$ is an $N_{s}\times (N_d+1)$ matrix where $N_d$ is the number of detectors and $N_s$ is the number of shots.
The first column of $X$ is all $1$s, and captures detector-independent behavior.
The relationship between a particular detector outcome and the value of the logical observable can be modeled as
\begin{equation}
    Q_i \ = \ \beta_0 \ +\ \sum_{j=1}^{N_{d}}\ \beta_j\ X_{ij} \ + \ \varepsilon_i
    \ ,
\end{equation}
where $\beta_0$ is a shot-independent bias term 
and $\varepsilon_i$ is an error term that accounts for observable behavior that is not explainable by the detector values. 
This relation is represented as a matrix equation $\vec{Q}=X\vec\beta+\vec \varepsilon$ where $\vec{Q}$ collects all $Q_i$ into a vector.
To minimize the cumulative error across all shots, we seek $\beta^*$ which is the solution of the following minimization
\begin{equation}\label{eq:beta_min}
    \beta^* \ = \ \operatorname*{arg\,min}_{\beta} L(\vec{Q},X,\beta)
    \ ,
\end{equation}
where 
$L(\vec{Q},X,\beta)=||\vec{Q}-X\beta||^2_2$ is the $2$-norm of the error vector $\varepsilon$. 
The solution to Eq.~\eqref{eq:beta_min} satisfies the \emph{normal equation}
\begin{equation}\label{eq:normal_equations}
    X^TX\beta^* \ = \ X^T\vec{Q} \ .
\end{equation}
The metric $\beta_j$ is then the magnitude of the $j$-th element of the vector $\beta^*=(X^TX)^{-1}X^T\vec{Q}$.

\subsection{A Worked Example of Calculating \texorpdfstring{$\Delta_j$}{Delta\_j} and \texorpdfstring{$\beta_j$}{Beta\_j}}
\label{app:worked_example}
\noindent
As an explicit example, consider a circuit with one $[[4,2,2]]$ code block and two rounds of $S_Z$ syndrome measurements and no $S_X$ measurements (using one ancilla without resets, with measurements recorded to classical bits) corresponding to three detectors. 
The initial state is $\ket{\overline{00}}$ and the observable of interest is $\overline{O}=\overline{Z}_{0}$. 
For each shot, we consider the different times during syndrome extraction that an $X$ error can occur. 
For the set of $N_s=4$ measurements in this example, the outcomes on codeblock $A$ are:
\begin{equation}
\Bigl\{\ [\text{syndrome}\  |\  \text{data}]\ \Bigr\}
\ = \ 
\Bigl\{\ 
    [\,0\;0 \mid\;0\;0\;0\;0]\ ,\quad
    [\,1\;0 \mid\;1\;0\;0\;0]\ ,\quad
    [\,0\;1 \mid\;0\;0\;0\;1]\ ,\quad
    [\,0\;0 \mid\;1\;0\;0\;0]
    \ \Bigr\}
    \ ,
\end{equation}
where the notation $[\text{syndrome}\  |\  \text{data}]$ describes a measurement outcome. 
In the first shot, no error occurred. 
In the second shot, an error occurred before the first syndrome measurement. 
In the third shot, an error occurred between the first and second round of extraction. 
In the fourth shot, an error occurred after the second round of extraction. 
The detectors and observable outcomes (with notation $[v_1\;v_2\;v_3\;|\;Q_i]$) using the construction in~\ref{sec:detector_construction} are then given by,
\begin{equation}
    [\,0\;0\;0 \mid +1\,]\ ,\quad
    [\,1\;0\;0 \mid -1\,]\ ,\quad
    [\,0\;1\;0 \mid +1\,]\ ,\quad
    [\,0\;0\;1 \mid -1\,]
    \ .
\end{equation}
The values of $\Delta_j$ can then be calculated as,\footnote{This example is a pure memory experiment on a single block, and so each detector heralds errors equally impacting the logical information causing all $\Delta_j$ to be equal. Adding more blocks or nFT logical rotations would introduce a hierarchy in the $\Delta_j$, which is the case for the simulations presented in the main text.}
\begin{equation}
    \Delta_1 \ = \ 4/3\ , \quad\Delta_2 \ = \ 4/3\ , \quad\Delta_3 \ = \ 4/3
    \ .
\end{equation}
To calculate the $\beta_j$, the detector matrix $X$ and observable vector $\vec{Q}$ are constructed as,
\begin{align}
    X \ & = \
    \begin{bmatrix}
        1 & 0 & 0 & 0 \\
1 & 1 & 0 & 0 \\
1 & 0 & 1 & 0 \\
1 & 0 & 0 & 1
    \end{bmatrix}
    \ \ ,\ \ 
\vec{Q} \ = \
\begin{bmatrix}
    +1 \\ -1 \\ +1 \\ -1
\end{bmatrix}
\ ,
\end{align}
and the vector of $\beta_j$ is given by $\beta^*=\left[(X^TX)^{-1}X^TQ\right]_{j>0}=-[2,0,2]$.

\subsection{Model of Error Mechanisms}
\label{app:correlation_description}
\noindent
This section presents a higher-level description of the noise on a quantum processor, inspired by the treatments in Ref.~\cite{Chen:2021num, Kishony:2026qtg}.
This model stands in contrast to a circuit-level noise model, which describes the probabilities of a specific error occurring, where the probabilities of flipping a detector or causing a logical error are derived quantities.
Instead, this model treats error mechanisms that flip detectors as the fundamental objects and directly assigns probabilities to these events, which is favorable when compared to a circuit-level noise model because the probability of a detector flip can be directly calculated from the raw device data.
We will refer to the underlying causes of detector flips as `mechanisms' to emphasize that detector flips are due to the aggregate effect of a number of different underlying physical errors~\cite{Blume-Kohout:2025kvx}. 
In general there is no single underlying process that can be identified with the flip of a detector on a quantum computer. 

Let $G=(V,E)$ be a graph describing the errors, where $V$ is a set of nodes and $E$ are the edges. 
Each node $v_j\in V$ corresponds to a detector and can be flipped by error mechanisms $e_j,e_{ij}\in E$ with $e_j/e_{ij}=\text{Bernoulli}(p_j/p_{ij})$ respectively. 
A single index $j$ corresponds to a flip of detector $v_j$ if $e_j=1$ and the two indices $(i,j)$ correspond to different ``pair" processes flipping two detectors $v_i,v_j$ if $e_{ij}=1$. 
Triplet processes $(i,j,k)$ and higher could be included to enhance the expressivity of this model, adding hyperedges to the error graph $G$, but analyzing these higher-order structures is beyond the scope of the current work. 
For brevity, both single node $j$ and pair processes $(i,j)$ will be denoted by a single index $m$. 
The support $\text{supp}(e_m)$ of an error mechanism $e_m$ is defined by the set of detectors that are flipped when that mechanism occurs.

If an error mechanism occurs, it flips the value of an observable $Q$ with probability $q_m$.
For an individual physical error in a Clifford circuit with Pauli noise, the observable flip probability is deterministic, $q_m \in\{0,1\}$, corresponding to whether the resulting error flips the operator.
The mechanisms of the present model are cumulative: each $(p_m, q_m)$ aggregates all microscopic faults sharing the same detector signature, with $p_m$ the combined rate of the class and $q_m$ its occurrence-weighted average flip probability. 
An intermediate value of $q_m$ therefore arises whenever the errors identified by a detector induce faults with different logical actions, and is expected even in Clifford experiments, for instance when a Pauli error coincides in detector signature with a leakage or crosstalk process. 
Intermediate values likewise arise for individual faults under non-Clifford gates.

The error mechanisms are taken to be mutually independent. 
The measured value of a logical observable $\overline{O}$ for a single shot is then represented as
\begin{equation}\label{eq:O}
    Q \ = \ Q_I\prod_{m}(-1)^{f_m} \ ,
\end{equation}
where $Q_I\in\{-1,1\}$ is the noiseless value of $Q$\footnote{$\mathbb{E}[Q_I]$ approaches the noiseless value of the observable $\langle \overline{O}\rangle$ for a large ensemble size.} and $f_m\in\{0,1\}$ correspond to parity flips of the measured observable so that $Q\in\{-1,1\}$. 
The noiseless value $Q_I=-1$ with probability $\lambda$ depends on the encoded wavefunction and $f_m=1$ indicates process $m$ has occurred and flipped the logical observable, i.e. $P(f_m=1)=q_mp_m$. 
Separating the measured $Q$ into its ideal part $Q_I$ and associated parity flips $f_m$ allows one to treat the effect of noise independently of the encoded information in the state. 
This is an approximation that is only strictly true for Clifford circuits and restricted types of noise, e.g. stochastic Pauli noise. Therefore the circuits presented in the main text will exhibit correlations between detector and observable values that are not captured by this model.
However, its simplicity affords both analytical tractability and insight, and it is empirically found to capture certain features of device noise like dependence on the magnitude of $\langle\overline{O}\rangle$.

\subsection{Expressing \texorpdfstring{$\Delta_j$}{Delta\_j} and \texorpdfstring{$\beta_j$}{Beta\_j} in Terms of Detector Probabilities}
\label{app:metric_details}
\noindent
\subsubsection{Calculations for \texorpdfstring{$\Delta_j$}{Delta\_j}}
\noindent
This section expresses Eq.~\eqref{eq:delta_exp} in terms of the probabilities $p_m,q_m$.
A detector $v_j$ can be written in terms of the error mechanisms $e_m$ that it detects as
\begin{equation}
    v_j \ = \ \bigoplus_{e_m\in E | j\in \text{supp}(e_m)} e_m
    \ ,
\end{equation}
where the condition $e_m\in E \,|\, j\in \text{supp}(e_m)$ will be denoted as $m\ni j$. 
In words, $v_j$ is defined the relative parity of the error mechanisms whose support contains it. 
First, we seek to solve for the expectation value $\mathbb{E}[Q | v_j=b]$ where $b\in\{0,1\}$. 
We begin by separating the product over $m$ in Eq.~\eqref{eq:O}
into those whose support $\text{supp}(e_m)$ contains $j$ and those where it does not
\begin{align}\label{eq:Qexpt}
    \mathbb{E}[Q | v_j=b] \ &= \ \mathbb{E}\left[Q_I\prod_{m\not\ni j}(-1)^{f_m}\right]\mathbb{E}\left[\prod_{m\ni j}(-1)^{f_m} \bigg|\ v_j=b\right]
    \nonumber\\
    \ &= \ \langle \overline{O}\rangle B_j\times\mathbb{E}\left[\prod_{m\ni j}(-1)^{f_m}\bigg|\ v_j=b\right] \ .
\end{align}
Here, $\langle \overline{O}\rangle=1-2\lambda$ and $B_j=\prod_{m\not\ni j}(1-2p_mq_m)$ denotes the background effect of the error mechanisms that do not touch detector $j$.
Thus, we must simplify the remaining expectation value in Eq.~\eqref{eq:Qexpt}. Consider a single mechanism $e_n$ whose support contains $j$. 
The expectation value is written in terms of the conditional outcomes of $e_n$,
\begin{align}\label{eq:remove_n}
    \mathbb{E}\!\left[(-1)^{\sum\limits_{m\ni j}f_m}\big|\ v_j=b\right] \ 
    &= \ \sum_{c\in\{0,1\}} P(e_n=c \mid v_j=b)\,(1-2q_n)^{c}\,
    \mathbb{E}\!\left[(-1)^{\sum\limits_{m\ni j,\, m\neq n}f_m}\Big|\ 
    \bigoplus\limits_{m\ni j,\, m\neq n}e_m=b\oplus c\right]
    \nonumber\\[4pt]
    &= \ P_0(e_n\mid v_j=b)\,
    \mathbb{E}\!\left[(-1)^{\sum\limits_{m\ni j,\, m\neq n}f_m}\Big|\ 
    \bigoplus\limits_{m\ni j,\, m\neq n}e_m=b\right]
    \nonumber\\
    &\hspace{30pt}
    + P_1(e_n\mid v_j=b)\,(1-2q_n)\,
    \mathbb{E}\!\left[(-1)^{\sum\limits_{m\ni j,\, m\neq n}f_m}\Big|\ 
    \bigoplus\limits_{m\ni j,\, m\neq n}e_m=\bar{b}\right] \ ,
\end{align}
where $P_b(v)=P(v=b)$. In the first line, the total parity has been conditioned on the value $e_n=c$, which shifts the target parity to $b\oplus c$ and introduces a factor $(1-2q_n)^c$; here $\oplus$ denotes addition modulo $2$. In the second line, $\bar{b}$ denotes the flipped value of $b$. The conditional probabilities are solved for by using
\begin{equation}
    P_c(e_n|v_j=b) \ = \ \frac{P(e_n=c,v_j=b)}{P(v_j=b)} \ = \ (1-c-p_n+2cp_n)\frac{P_{b\oplus c}(v_j')}{P_b(v_j)}
    \ ,
\end{equation}
where the notation $v'_j$ denotes the detector with mechanism $n$ removed. 
Both these quantities can be solved for by observing that
\begin{align}
    P_b(v_j) + P_{\bar{b}}(v_j) \ &= \ 1
    \ \ , \ \  
    P_b(v_j) - P_{\bar{b}}(v_j) \ = \ (-1)^b\ \mathbb{E}[(-1)^{v_j}]
    \ ,
\end{align}
with
\begin{equation}
    \mathbb{E}[(-1)^{v_j}] \ = \ \prod_{m\ni j}\mathbb{E}[(-1)^{e_m}] \ =\ \prod_{m\ni j}(1-2p_m)
    \ ,
\end{equation}
implying that
\begin{equation}
    P_b(v_j) \ = \ \frac{1}{2}\left(1+(-1)^b\prod_{m\ni j} (1-2p_m)\right)
    \ .
\end{equation}

Returning to evaluating Eq.~\eqref{eq:remove_n}, the process of removing a node is now repeated. 
Denote the detector with $k$ mechanisms removed as $v_j^{(k)}=\bigoplus\limits_{m\ni j-k}e_m$, with $v_j^{(0)}=v_j$, and the expectation value with $k$ mechanisms removed as
\begin{equation}
    E_b^{(k)} \ = \ \mathbb{E}\left[(-1)^{\sum\limits_{m\ni j-k}f_m} \big|\ v_j^{(k)}=b\right]
    \ ,
\end{equation}
where Eq.~\eqref{eq:remove_n} denotes the first step in this iteration. 
Generalizing to all nodes, the iterative equation reads
\begin{equation}\label{eq:iteration}
    P_{b}(v_j^{(k)}) \ E_b^{(k)} \ = \ \left(1-p_{n_k}\right)\ 
    P_b\left(v_j^{(k+1)}\right)\ 
    E^{(k+1)}_b+p_{n_k}\left(1-2q_{n_k}\right)\ 
    P_{\bar{b}}\left(v_j^{(k+1)}\right)\ E_{\bar{b}}^{(k+1)}
    \qquad {\rm for}\ \ \ k\geq 0
    \ ,
\end{equation}
where $n_k$ denotes the removed mechanism. 
Note that the R.H.S of this equation only refers to the mechanisms that have $j$ in their support, without the removed mechanism $n_k$.
For both values of $b$, Eq.~\eqref{eq:iteration} can be written as a matrix equation,
\begin{equation}
    V^{(k)} \ = \ M_{n_k}V^{(k+1)},\quad 
    V^{(k)} \ = \  
    \begin{bmatrix}
    P_0(v_j^{(k)})E_0^{(k)} \\
    P_1(v_j^{(k)})E_1^{(k)}
    \end{bmatrix},\quad
    M_{n_k}
     \ = \
    \begin{bmatrix}
    (1-p_{n_k}) & (1-2q_{n_k})p_{n_k} \\
    (1-2q_{n_k})p_{n_k} & (1-p_{n_k})
    \end{bmatrix}
    \ .
\end{equation}
After all $M$ mechanisms containing $j$ have been removed, the parity is even and the sum vanishes, giving the base case $V^{(M)}=(1,0)^{T}$. 
The matrix $M_{n_k}$ can be diagonalized by the basis 
$E_{\pm}^{(k)}=P_0(v_j^{(k)})\ E_0^{(k)}\pm P_1(v_j^{(k)})\ E_1^{(k)}$, 
with $E_{\pm}^{(M)}=1$, allowing the iteration relation to be solved exactly,
\begin{equation}
    E_{\pm}^{(0)} \ = \ \prod_{m\ni j} \Big[(1-p_m)\pm(1-2q_m)p_m\Big]
    \ .
\end{equation}
Rewriting this in terms of the original $E_{0}^{(0)}$ and $E_{1}^{(0)}$, the desired expectation value is then given by,
\begin{equation}
    \mathbb{E}[Q|v_j=b] \ = \ 
    \langle \overline{O}\rangle \ B_j \ 
    \frac{\prod\limits_{m\ni j}(1-2p_mq_m)+(-1)^b\prod\limits_{m\ni j}(1-2p_m(1-q_m))}{1+(-1)^b\prod\limits_{m\ni j}1-2p_m}
    \ .
\end{equation}
Finally, the filtering metric is obtained by subtracting the two conditional expectation values. Denote the three products over mechanisms containing $j$ as,
\begin{align}
    A_j \ &= \ \prod_{m\ni j}(1-2p_mq_m)\ ,\ \ 
    C_j \ = \ \prod_{m\ni j}\big(1-2p_m(1-q_m)\big)\ ,\ \ 
    D_j \ = \ \prod_{m\ni j}(1-2p_m)\ ,
\end{align}
so that 
\begin{align}
\mathbb{E}[Q|v_j=b] \ & = \ \langle \overline{O}\rangle\  B_j\ 
\frac{(A_j+(-1)^bC_j)}{(1+(-1)^bD_j)}
\ .
\end{align}
Subtracting the $b=0$ case from the $b=1$ case gives 
\begin{align}
    \Delta_j
     \ &= \ \Bigg|2\langle \overline{O}\rangle B_j\,\frac{A_jD_j-C_j}{1-D_j^2}\Bigg|
     \ .
\end{align}
The denominator is directly measurable: comparing with the parity probability derived above, ${D_j = 1-2P_1(v_j)}$, so that $1-D_j^2 = 4P_1(v_j)\big(1-P_1(v_j)\big)$. 
Expanding to leading order in the mechanism probabilities, ${A_jD_j-C_j\approx -4\sum_{m\ni j}p_mq_m}$ and ${1-D_j^2\approx 4\sum_{m\ni j}p_m}$, giving
\begin{equation}
    \Delta_j \ \approx \ \Bigg|2\langle \overline{O}\rangle \,\left(\sum\limits_{m\ni j}p_mq_m\right)\Bigg/\left(\sum\limits_{m\ni j}p_m\right)\Bigg|
    \label{eq:DeltajestA}
    \ .
\end{equation}
The denominator can be viewed as the cumulative ``firing'' rate for all processes whose support contains detector $v_j$, while the numerator is this same quantity weighted by the probability of that process causing a logical error. 
Thus, $\Delta_j$ takes on the interpretation of an effective probability of all processes affecting $v_j$ that cause a logical error, weighted by $\langle \overline{O}\rangle$. 
As such, detectors that take on higher values of $\Delta_j$ should be postselected on first, as these are the most detrimental to the encoded information. This forms the basis of ORP.

This analysis reveals that the scale of $\Delta_j$ is set by the value of the logical observable and the background factor $B_j$. 
If $\langle \overline{O} \rangle\ll 1$ or if the background attenuation is large, i.e., $B_j\ll 1$, and all the individual firing rates $p_m$ are close to $1$, then every value of $\Delta_j$ will take on parametrically small values. 
Due to finite measurement statistics, it can become challenging to resolve differences at this scale, reducing the effectiveness of $\Delta_j$ as a quality metric distinguishing different detectors.

\subsubsection{Calculations for \texorpdfstring{$\beta_j$}{Beta\_j}}
\noindent
In the limit of many shots, $\frac{1}{N_s}X^TX$ and $\frac{1}{N_s}X^T\vec{Q}$ converge to matrices of moments.
The normal equation (Eq.~\eqref{eq:normal_equations})
can be expressed component-wise explicitly in terms of the moments of detectors $v_j$ and observable $Q$
\begin{equation}
    \begin{bmatrix}
        1 & \mathbb{E}[v_k]^{T} \\
        \mathbb{E}[v_j] & \mathbb{E}[v_jv_k]
    \end{bmatrix}
    \begin{bmatrix}
        \beta_0^* \\ \beta_k^*
    \end{bmatrix}
    \ = \ 
    \begin{bmatrix}
        \mathbb{E}[Q] \\ \mathbb{E}[Qv_j]
    \end{bmatrix}\ ,
    \qquad j,k \ = \ 1,\dots,N_d
    \ .
\end{equation}
Eliminating $\beta_0^* = \mathbb{E}[Q]-\sum_k\mathbb{E}[v_k]\beta_k^*$ from
the remaining rows reduces this to
\begin{equation}
    G\beta^* \ = \ a
    \ ,
\end{equation}
for the connected two point function $G_{jk}=\mathbb{E}[v_jv_k]-\mathbb{E}[v_j]\mathbb{E}[v_k]$ and $a_j=\mathbb{E}[Qv_j]-\mathbb{E}[Q]\mathbb{E}[v_j]$, with the bias recovered from the eliminated row. 
Each moment is evaluated using the products derived above. 
Rewriting the detectors in terms of parities, $v_j=\tfrac{1}{2}\big(1-(-1)^{v_j}\big)$, the one-point functions are
\begin{equation}
    \mathbb{E}[v_j] \ = \ \frac{1-D_j}{2} \ ,\qquad
    \mathbb{E}[Q] \ = \ \langle \overline{O}\rangle B_j A_j
    \ ,
\end{equation}
where the second expression is independent of the choice of $j$. 
For the two-point function, the shared mechanism $e_{jk}$ cancels from the parity $v_j\oplus v_k$, so that $\mathbb{E}[(-1)^{v_j\oplus v_k}]=D_jD_k/(1-2p_{jk})^2$, and
\begin{equation}
    \mathbb{E}[v_jv_k] \ = \ \frac{1}{4}\left(1 - D_j - D_k + \frac{D_jD_k}{(1-2p_{jk})^2}\right)\ ,\qquad
    \mathbb{E}[v_j^2] \ = \ \mathbb{E}[v_j] \ = \ \frac{1-D_j}{2}
    \ ,
\end{equation}
where the diagonal follows from $v_j^2=v_j$ for a detector. 
The cross moment follows from the conditional expectation derived above, or directly from $\mathbb{E}[Q(-1)^{v_j}]=\langle \overline{O}\rangle B_jC_j$,
\begin{equation}
    \mathbb{E}[Qv_j] \ = \ \frac{\langle \overline{O}\rangle B_j}{2}\left(A_j-C_j\right)
    \ .
\end{equation}
The connected components are then given as,
\begin{equation}
    G_{jk} \ = \ \frac{D_jD_k}{4}\left(\frac{1}{(1-2p_{jk})^2}-1\right)\quad(j\neq k)\ , \qquad 
    G_{jj} \ = \ \frac{1-D_j^2}{4}\ , \qquad
    a_j \ = \ \frac{\langle \overline{O}\rangle B_j}{2}\left(A_jD_j-C_j\right)
    \ .
\end{equation}
The limit of weak noise, $p_m\ll 1$, is used to simplify the expressions. 
Expanding to leading order, $D_j\approx 1-2\sum_{m\ni j}p_m$
and $(1-2p_{jk})^{-2}-1\approx 4p_{jk}$, so that
\begin{equation}
    G_{jk} \ \approx \ p_{jk}\quad(j\neq k)\ ,\qquad
    G_{jj} \ \approx \ p_j + \sum_{k\neq j}p_{jk}\ ,\qquad
    a_j \ \approx \ -2\langle \overline{O}\rangle\Big(p_jq_j+\sum_{k\neq j}p_{jk}q_{jk}\Big)
    \ ,
\end{equation}
where in $a_j$ the terms carrying rates alone cancel between $A_jD_j$ and
$C_j$, leaving the flip-weighted sums. 
The normal equation then reads
\begin{equation}
    \Big(p_j+\sum_{k\neq j}p_{jk}\Big)\beta^*_j
    + \sum_{k\neq j}p_{jk}\,\beta^*_k
    \ = \ -2\langle \overline{O}\rangle\Big(p_jq_j+\sum_{k\neq j}p_{jk}q_{jk}\Big)
    \ .
\end{equation}
Here, the RHS is the same magnitude as the numerator of $\Delta_j$ for the same detector.
Furthermore, omitting the second term on the LHS yields $\beta_j^*=\Delta_j$, showing that the correlations that cause these two metrics to differ are the pair processes $p_{ij}$ that connect two detectors. 
Specifically, $\Delta_j$ attributes the logical error caused by pair mechanisms $e_{ij}$ to \emph{both} detectors $v_i$ \emph{and} $v_j$. 
The regression, on the other hand, splits the contribution of $e_{ij}$ between $v_i$ and $v_j$ in proportion to all the other rates of mechanisms that have $v_j$ in their support. 
\begin{figure}
    \centering
    \includegraphics[scale=.5]{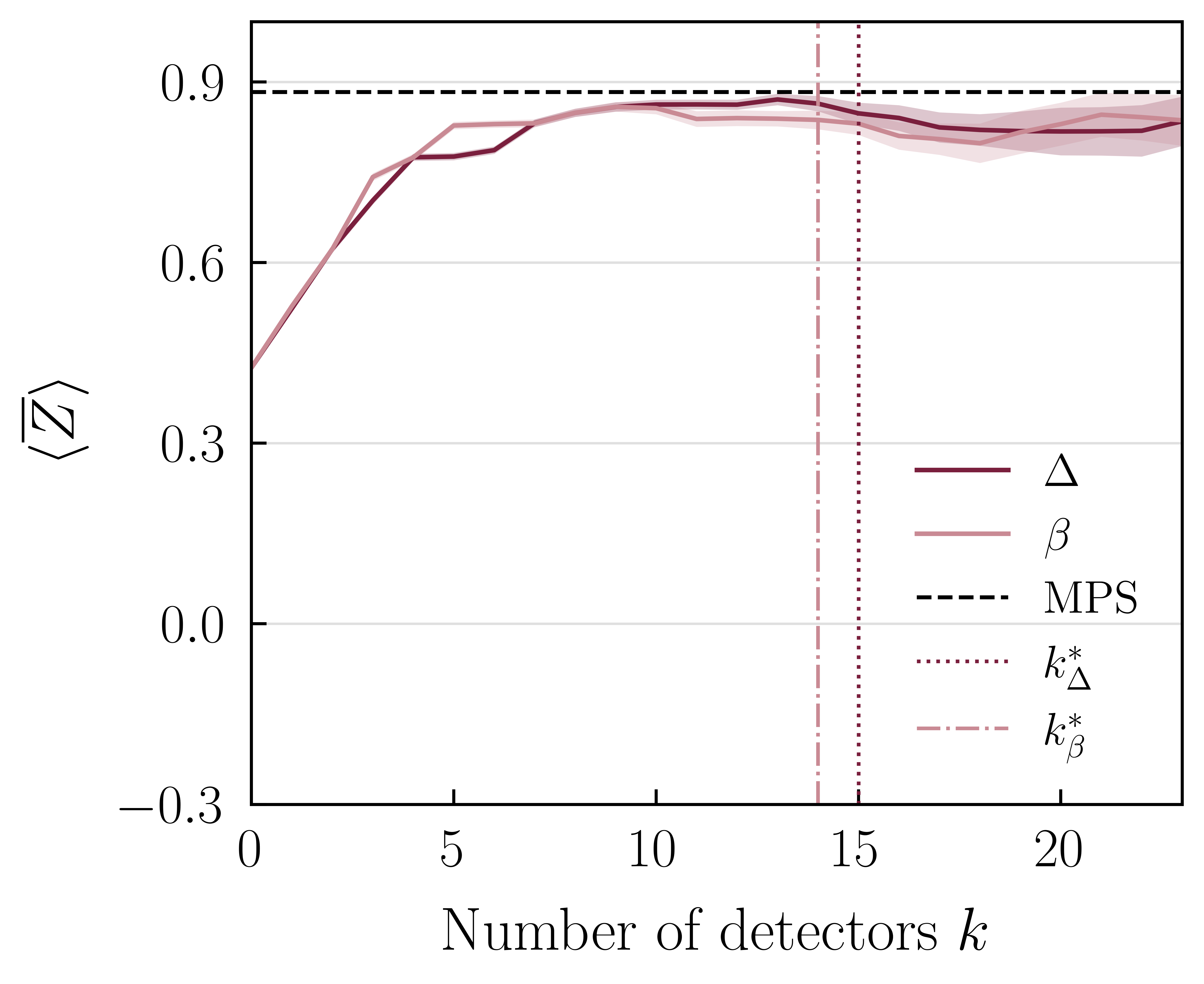}
    \caption{\textit{Comparison of $\Delta_j$ and $\beta_j$ Predictions on }{\tt ibm\_boston}. The metrics $\Delta_j$ and $\beta_j$ are calculated using Eq.~(\ref{eq:delta_exp}) and Eq.~(\ref{eq:normal_equations}) respectively.
    The two solid lines show the predicted value of $\langle \overline{Z}\rangle$ as a function of the number $k$ of detectors used in postselection for $\Delta_j$ (dark) and $\beta_j$ (light). The horizontal dashed line is the noiseless MPS prediction. The vertical dashed lines correspond to the value of $k^*$ returned from the plateau finding algorithm using $\Delta_j$ (dark) and $\beta_j$ (light). The expectation value is from four Trotter steps on four code blocks.}
    \label{fig:nkeep_appendices}
\end{figure}
Therefore, a ranking based on $\Delta_j$ quantifies the probability that detection events at $v_j$ result in a logical error, whereas a ranking based on $\beta_j$ quantifies the degree to which the logical error rate is uniquely attributed to the detector $v_j$. 
This distinction suggests that $\Delta_j$ is a more effective measure for identifying the noisiest shots,  and thus $\Delta_j$ is used in ORP to produce the results in the main text.
However, in practice, filtering based on $\beta_j$ or $\Delta_j$ is found to give comparable results, see e.g. Fig.~\SubFigRef{fig:nkeep_appendices}{}.

\subsection{The Lightcone of the Filtered Hierarchy}
\label{app:pp_lightcone}
\noindent
To connect the detector ranking chosen by $\Delta_j$ to the properties of a partially FT circuit $U$, we Pauli-propagate the observable $\overline{O}$ on which $\Delta_j$ is conditioned backward through $U$ to the location of each detector $v_j$. 
The propagated observable can be represented as
\begin{align}\label{eq:pp_obs}
    \tilde{O} \ &= \ U\overline{O}U^{\dagger}\nonumber\\
     \ &= \ \sum_P c_p P \ ,
\end{align}
where the sum is over all Pauli strings $P$. 
For a detector $v_{j=(g,b,r)}$, located at code block $b$ in round $r$, its ranking, as determined by Pauli-propagation, is given by
\begin{equation}\label{pp_ranking}
    \gamma_j \ = \ \frac{1}{\mathcal{N}}\sum_{P : \text{supp}(P)\in b} |c_P|
\end{equation}
where the sum is over all Pauli strings that have support (i.e. non-identity components) contained in code block $b$, and $\mathcal{N} = \sum_P |c_P|$ is a normalization that ensures $0\leq \gamma_j \leq 1$.
The ranking based on $\Delta_j$ is set by the magnitude of $\Delta_j$.

\begin{figure}
    \centering
    \includegraphics[scale=0.5]{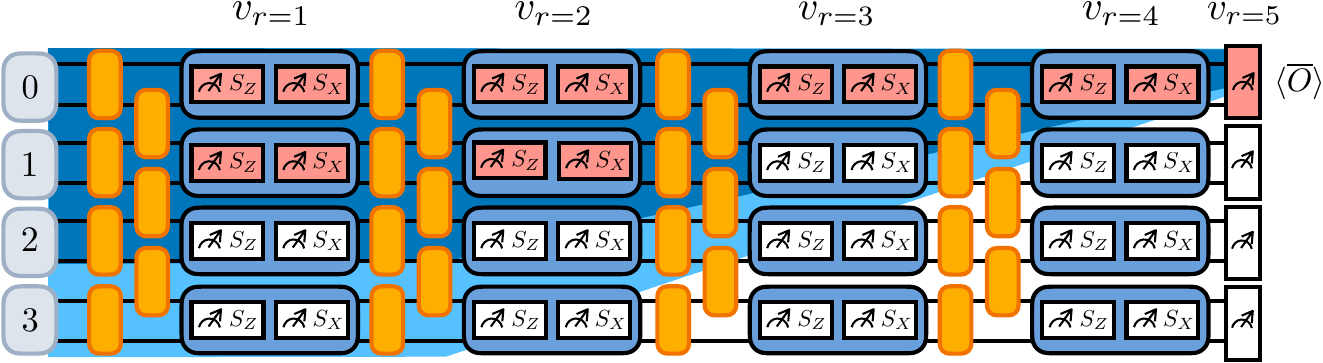}
    \caption{{\it Depiction of Most Important Detectors} Calculated for the logical observable $\overline{O}=\overline{Z}_{(b=0,0)}$, as determined by the rankings calculated in Table~\ref{tab:filtering_comparison}. 
    Highlighted in red are the locations of the seven most important detectors, as determined by both ORP and Pauli-propagation. 
    The grey region shows the observable's circuit backwards lightcone, containing all of the detectors that are connected to the observable through gates. 
    The dark blue region captures the physical revserse-lightcone of $\overline O$, showing the detector locations where the back-propagated observable has the largest Pauli weights.
    }
    \label{fig:detector_lightcone_ranking}
\end{figure}

As an example, Table~\ref{tab:filtering_comparison} shows a comparison between the detector ranking determined by $\gamma_j$ and the detector ranking determined by $\Delta_j$ using simulated results.
\begin{table}
\renewcommand{\arraystretch}{1.4}
\begin{tabularx}{\linewidth}{|c||Y|Y|Y|Y|Y|Y|Y|Y|Y|Y|}
\hline
Detector ranking & 1 & 2 & 3 & 4 & 5 & 6 & 7 & 8 & 9 & 10 \\
\hline\hline
$\Delta_j$ & $v_{(0,1)}$ & $v_{(0,2)}$ & $v_{(0,4)}$ & $v_{(0,3)}$ & $v_{(0,5)}$ & $v_{(1,1)}$ & $v_{(1,2)}$ & $v_{(3,1)}$ & $v_{(3, 3)}$ & $v_{(3,4)}$ \\\hline
$\gamma_j$ & $v_{(0,5)}$ & $v_{(0,4)}$ & $v_{(0,3)}$ & $v_{(0,2)}$ & $v_{(0,1)}$ & $v_{(1,1)}$ & $v_{(1,2)}$ & $v_{(1,3)}$ & $v_{(2,1)}$ & $v_{(3,5)}$\\\hline
\end{tabularx}
\caption{{\it Comparison Between Detector Ranking Using Pauli-propagation and ORP.} 
The top row shows the position in the ranking, where $1$ is the highest ranking and $10$ is the lowest ranking among the detectors shown.
The middle row shows the detectors $v_{(b,r)}$ ranked by their $\Delta_j$ value and the bottom row shows them ranked by their $\gamma_j$ value.
The rankings are computed for the observable $\overline{Z}_{(b=0,0)}$ using a four-block circuit consisting of $n_T=4$ Trotter steps with step size $\delta t=0.1$, and one syndrome extraction round per Trotter step, starting from an all $\ket{\overline{00}}$ initial state. The rankings are computed for $g_x=g_z=1$.
}
\label{tab:filtering_comparison}
\renewcommand{\arraystretch}{1.0}
\end{table}
The rankings are computed for the observable $\overline{Z}_{(b=0,0)}$ on four code blocks $b=0,1,2,3$ with $n_T = 4$ Trotter steps and a syndrome extraction rate of one round per Trotter step, as illustrated in Fig.~\SubFigRef{fig:detector_lightcone_ranking}{}.
\noindent
The $\Delta$-rankings are computed using classical simulations subject to two-qubit depolarizing noise with error rate $p=0.003$, which is comparable with error rates on current IBM hardware. 
These rankings are found to be consistent with those computed from hardware results.
Although the two orderings do not match exactly, ORP is consistent with Pauli-propagation, since the set of the top-$k$ detectors selected typically are the same, with the most important detectors residing in the code block containing the observable being measured. 
Figure~\SubFigRef{fig:detector_lightcone_ranking}{} shows the locations of the most important detectors chosen by the two methods and illustrates how the highest ranked detectors selected by ORP coincide with the local observable's backwards lightcone.

\section{Fault Tolerant Benchmarking of Circuit Elements}
\label{app:circuit_elts_benchmark}

\subsection{Logical Error Rates of Circuit Gadgets}
\label{sec:ler_cg}
\noindent
This appendix numerically studies the logical performance in the simulations by calculating the logical error rate (LER) as a function of the device error rate.
The noise model assumes a two-qubit depolarizing channel, defined as
\begin{equation}
    \mathcal{E}(\rho) = (1-p)\rho + \frac{p}{15}\sum_{P\neq I} P\rho P
\end{equation}
where $p\in[0,1]$ is the physical error rate and the sum is over all two-qubit Paulis $P$, excluding the identity. 
Additionally, single-qubit gates are followed by a single-qubit depolarizing channel with error rate $p/10$.
In the numerical experiments, the ideal state is a logical $\ket{\overline{00}}$, which is a GHZ state on four qubits. 
A logical error occurs when there is a physical that keeps the state in the codespace $\mathcal{C}$ and goes undetected by the syndrome measurements. Removing shots where a nontrivial syndrome is detected is equivalent to projecting the noisy state back onto the codespace, defined by the projector
\begin{equation}\label{eq:code_space_proj}
    \Pi_{\mathcal{C}} = \frac{1}{4}(1+S_Z)(1+S_X)
    \ ,
\end{equation}
where undetectable errors take the state out of the simultaneous $+1$ eigenspace of $S_Z$ and $S_X$.
For a state $\rho$ whose noisy evolution is given by $\tilde{\rho}$, the LER is defined by
\begin{equation}
    p_L \ = \ 1-F_L,\qquad F_L \ = \ \text{Tr}\,\rho\tilde{\rho}\, \Pi_{\mathcal{C}}
    \ .
\end{equation}

\begin{figure}
    \centering
    \includegraphics[scale=.5, trim = 0 120 0 120]{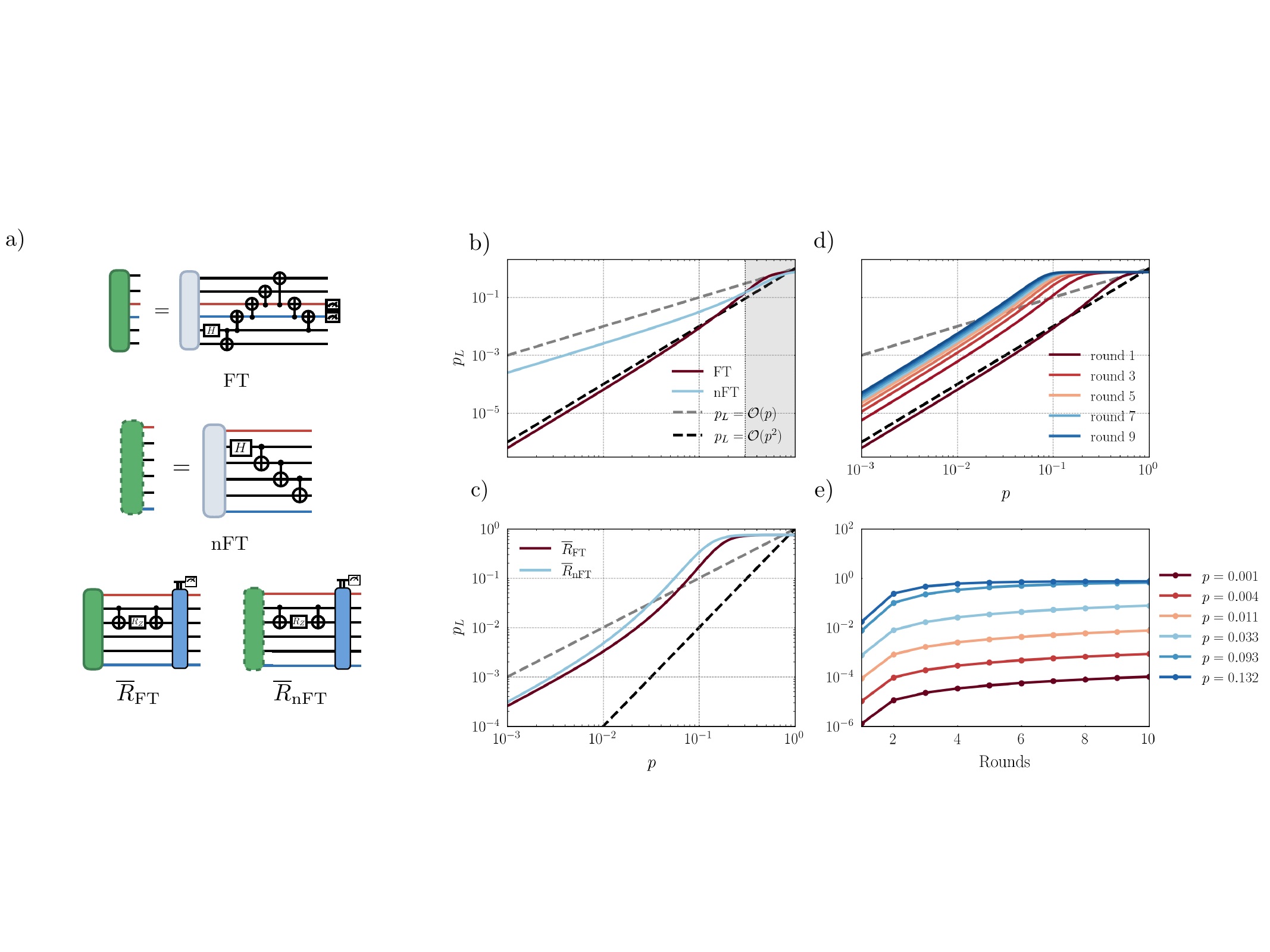}
    \caption{\textit{Logical Error Rate of Different Circuit Components} 
    a)~The different FT and nFT circuit elements whose logical error rates are considered in ~\ref{sec:ler_cg}.
    The green elements denote initialization gadgets, labeled FT (solid) and nFT (dashed). 
    The bottom two circuits $\overline R_\text{FT}/\overline R_\text{nFT}$ are FT/nFT initialization gadgets followed by a nFT logical rotation $\overline{R}_{ZZ}$.
    b)~The LER of the two different initialization gadgets. 
    The grey region is where nFT crosses to a lower LER than the FT procedure. The grey and black dashed lines represent $\mathcal{O}(p)$ and $\mathcal{O}(p^2)$ scaling for the logical error rate $p_L$.
    c)~ The LER for the FT and nFT initialization followed by a logical rotation.
    d)~The LER of the syndrome extraction circuits for multiple rounds.
    e)~The LER of syndrome extraction as a function of number of rounds for different $p$.
    }
    \label{fig:p_vs_pl_gadget}
\end{figure}

The LER $p_L$ can be expressed as a function of $p$. For $p$ small, the leading order contribution is a good approximation for how $p_L$ scales with $p$.
For level-$0$ FT (no encoding) the leading order scaling is $\mathcal{O}(p)$.
Gadgets that are 1-FT have leading order behavior $\mathcal{O}(p^2)$, and so mixing both types results in an LER of the form $p_L=Ap+Bp^2$, where the constants $A,B$ encode the geometric features of a circuit. 
In particular, $A$ represents nFT circuit features that can cause undetectable errors while $B$ represents the extra circuit overhead required to detect errors. 
The relative magnitudes of $A$ and $B$ can have meaningful impacts on the error rates, and in some cases imply that a nFT circuit gadget can give superior performance to its encoded counterpart.
An example of this is seen by comparing the nFT rotations used in this work to a FT approach like magic state injection.

The clearest examples of the improved performance of including nFT elements are the $\ket{\overline{00}}$ initialization circuits that prepare a four-qubit GHZ state.
A simple nFT way of preparing this state requires a CNOT depth of four, whereas a 1-FT technique for preparing such a state on heavy-hex topology requires a CNOT depth of seven followed by two measurements.
Both of these circuits are shown in Fig.~\SubFigRef{fig:p_vs_pl_gadget}{a}), with their respective LERs determined from classical simulations shown in Fig.~\SubFigRef{fig:p_vs_pl_gadget}{b}) over a range of $p$.
When considering the LER of both of these circuit gadgets, the FT circuit has error suppression quadratic in $p$ compared to the lack of suppression in the nFT case.
For devices with long measurement times, idle errors can occur on the data qubits being measured, amplifying the FT overhead $B$ by almost an order of magnitude and increasing the number of errors.
Together with a two-qubit gate depth almost twice that of the nFT gadget, these extra circuit elements degrade the performance at large enough $p$.
This is reflected in the LER curves, where the FT $p_L$ passes the nFT value at a physical error rate of $p\sim 0.2$.

Considering both FT and nFT elements together in the same circuit qualitatively modifies this picture. 
In particular, examining the leading-order expansion for $p_L$ reveals a crossover point at $p^*\approx A/B$ such that for $p>p^*$ there is a region of quadratic suppression of errors, which then degrades to a linear decay with no suppression for $p\lesssim p^*$.
Examples of such behavior are shown in Fig.~\SubFigRef{fig:p_vs_pl_gadget}{c}), which compares the effect of both FT and nFT state preparation, followed by a logical rotation (nFT) and syndrome extraction (FT). 
The extra circuit depth pushes the LER high enough such that the nFT state prep is no longer able to outperform its FT counterpart. 
Further, even though the asymptotic error dependence is linear, the addition of FT components can push the LER below that of nFT rotations. 
The suppression at intermediate $p>p^*$ indicates that for physical error rates that are small but not yet below threshold, there are still advantages in error suppression gained by incorporating FT elements into circuit design. 
Finally, Fig.~\SubFigRef{fig:p_vs_pl_gadget}{d}) analyzes the LER of the syndrome extraction circuits with FT initialization. 
Here, all of the circuit elements are 1-FT and so $p_L = \mathcal{O}(p^2)$. 
The overhead factor $B$ increases with the number of syndrome measurement rounds, but then quickly plateaus to an $\mathcal{O}(1)$ constant as shown in Fig.~\SubFigRef{fig:p_vs_pl_gadget}{e}).

\subsection{Memory Experiments}
\label{sec:memory}
\begin{figure}
    \centering
    \includegraphics[width=1\linewidth, trim = 0 250 0 200, clip]{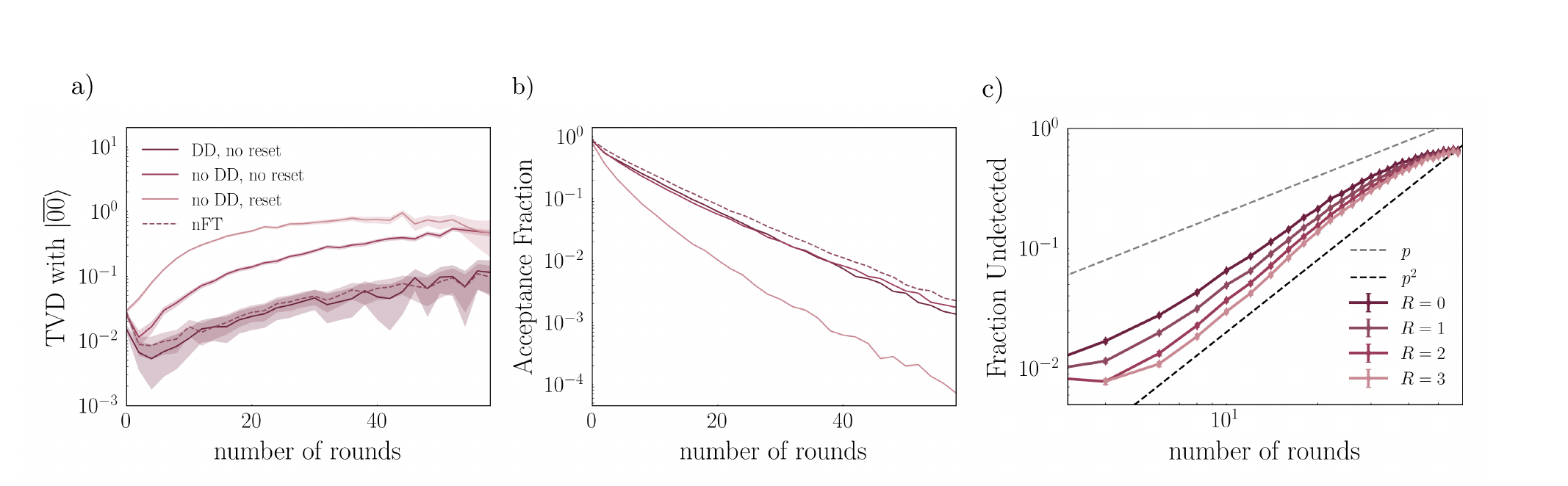}
    \caption{\textit{Memory Experiment Performance of Multiple Rounds on {\tt ibm\_boston}} 
    a)~The Total Variation Distance (TVD) from the logical $\ket{\overline{00}}$ state 
    over $60$ rounds of syndrome extraction.
    b)~The acceptance fraction over $60$ rounds. 
    c)~Suppression of errors at leading order as a function of number of syndromes used for postselection.}
    \label{fig:memory_figs}
\end{figure}

\noindent
To assess the effectiveness of the syndrome extraction circuits used in this work, memory experiments are performed on {\tt ibm\_boston} where both FT and nFT state prep are followed by many rounds of syndrome extraction. 
Assuming noiseless circuit operation, the system will remain in the logical $\ket{\overline{00}}=\frac{1}{\sqrt{2}}(\ket{0000}+\ket{1111})$ state, and measurements in the $Z$-basis result in either ``0000" or ``1111" bitstrings.
Noise populates other bitstrings with a non-zero probability, with some being detected by syndrome extraction and others causing undetectable errors. 
The undetected errors will appear as even parity bitstrings with Hamming-weight two in the final bitstring distribution. 
After postselecting all shots whose syndrome measurements flagged an error, the filtered probability distribution is compared to that of an ideal GHZ state using the Total Variation Distance (TVD) defined by
\begin{equation}
    \text{TVD}(q,q') \ = \ \frac{1}{2}\sum_s|q(s)-q'(s)|
    \ ,
\end{equation}
where the sum is over all bit strings, $q$ is the measured distribution and $q'$ is the GHZ distribution, with $q'(\text{``0000"})=q'(\text{``1111"})=0.5$.
Figure~\SubFigRef{fig:memory_figs}{a}) shows the TVD of the postselected distribution as a function of the number of syndrome extraction rounds.
The four curves compare 
(1) no DD and reset, 
(2) no DD and no reset, 
(3) DD and no reset and 
(4) DD and no reset 
with nFT state prep over the course of 60 rounds of syndrome extraction (all with a circuit depth of 761 CNOTs).
Incrementally adding in DD and removing resets have dramatic impacts on the error, lowering it by over an order of magnitude. 
The nFT state preparation contains more errors for a few rounds of syndrome extraction, but remains comparable to the FT state over many rounds.
Figure~\SubFigRef{fig:memory_figs}{b}) shows the corresponding acceptance fraction for these cases, defined as the number of measurements that record no errors. 
This fraction decreases exponentially with the number of rounds, with DD having the most profound effect on preserving shot lifetime.

The $[[4,2,2]]$ code is a distance-2 code, and so all events contributing to the LER at first order in the physical error rate $p$ are be flagged, and $p_L$ will scale as $\mathcal{O}(p^2)$.
Observing this scaling in quantum hardware runs requires tuning the physical error rate, which is not possible to do directly. 
We leverage the fact that syndrome extraction acts as the logical identity in the codespace, and so multiple rounds of extraction can serve to amplify the effective physical error rate.
As such, the physical error rate can is increased by modifying which rounds of syndrome extraction are postselected against. 
Specifically, instead of considering the results of syndrome measurements on every round, we only consider postselecting on the final round of extraction. To account for the errors from initializing the code state $\ket{\overline{00}}$, a constant number of syndromes are additionally postselected on.
As a proxy for the LER $p_L$, we count the fraction of even parity Hamming-weight 2 bitstrings that survive the postselection, which at leading order is proportional to the number of undetectable errors that occurred. 
Figure~\SubFigRef{fig:memory_figs}{c}) shows the fraction of undetected errors as a function of the number of rounds with an increasing number of syndromes postselected at the beginning. 
This fraction grows quadratically with the number of rounds, and this number is interpreted as the scaling factor for the physical error rate. 
At low amplification, 
this fraction plateaus due to nFT state prep errors and finite measurement statistics, whereas at high amplification there is a plateau due to exceeding the pseudothreshold.
Therefore, the scaling should be understood as appearing in an intermediate regime, in the plot approximately between rounds $8$-$30$.

\subsection{Rotations with Ancilla Qubits}
\label{app:analog_ancilla_rots}
\begin{figure}
    \centering
    \includegraphics[width=0.9\linewidth]{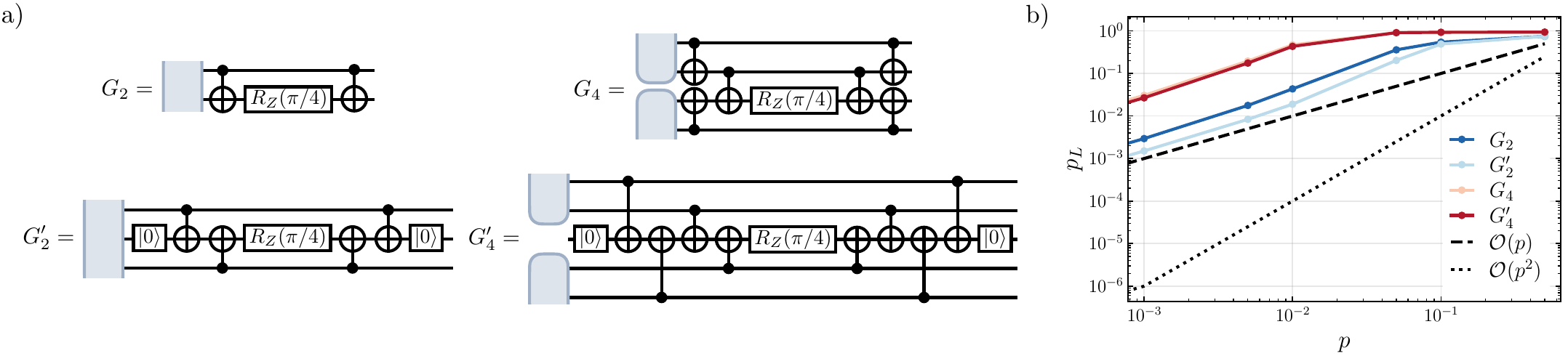}
    \caption{{\it Ancilla vs. No-Ancilla Rotations.}
    a)~The top row shows ancilla-free circuits $G_2$ and $G_4$ implementing $e^{-i\pi/8 \overline{Z}_{(b,0)}\overline{Z}_{(b,1)}}$ within a single code block (left) $b$ and $e^{-i\pi/8 \overline{Z}_{(a,0)}\overline{Z}_{(b,0)}}$ between two code blocks $a$ and $b$ (right).
    The bottom row shows circuits $G_2'$,\, $G_4'$ with ancilla qubits that implement the same rotations.
    b)~The LER $p_L$ as a function of the physical error rate $p$ for the circuits in a). 
    The logical error rate is computed by taking the average over Haar-random states under repeated rotations with depolarizing noise.}
    \label{fig:ancilla_rotations}
\end{figure}

\noindent
A method to reduce the error spread by nFT rotations, at the cost of ancillas and measurements, was studied in Refs.~\cite{Zhong:2026jps,Gerhard:2024peb}.
This method couples an ancilla qubit prepared in $|0\rangle$ to the rotation circuit in such a way that the bulk of single-qubit $X$ errors are concentrated on the ancilla qubit.
As an example, let $P=ZZ$ and $P'=ZZZ$ where the first two qubits are data and the last one is an ancilla.
Then $e^{-i\theta P'}=e^{-i\theta P}\ket{0}\bra{0}+e^{i\theta P}\ket{1}\bra{1}$ and the correct rotation is recovered by postselection on ancilla state $|0\rangle$.
Figure~\SubFigRef{fig:ancilla_rotations}{a}) shows such circuit pairs $G_2/G_2'$ and $G_4/G_4'$, implementing $ZZ$-rotations within a block and between logical qubits of neighboring blocks.
In these circuits, the only errors that spread to become high-weight errors are $Z$-errors indistinguishable from logical rotations.
Ancilla measurements can be deferred to the end of the circuit, provided later errors do not affect them.
Since $X$ errors are accumulated on the ancilla qubit, 
an undetectable (logical) error can occur when two physical errors cancel
\footnote{This cancellation can be studied using an error model with $X$ errors only on the ancilla qubit.}.
Note that applying this gadget with an ancilla qubit in state $|1\rangle$ implements a rotation with the opposite angle, which is a logical error that is detectable via postselection on ancilla measurement.
The implications of this are discussed in~\ref{app:more_analysis}.

While this method enables detecting more errors, its gate and measurement overhead introduces additional errors.
To determine which approach is best-suited for currently available hardware, 
we compare implementations of $e^{-i\pi/8 \overline{Z}_{(b,0)}\overline{Z}_{(b,1)}}$ within a single code block $b$ and $e^{-i\pi/8 \overline{Z}_{(a,0)}\overline{Z}_{(b,0)}}$ between two code blocks $a$ and $b$.
The data qubits are positioned within code blocks so that SWAP gates are not required to implement these rotations.
The circuits in  Fig.~\SubFigRef{fig:ancilla_rotations}{a}) are repeatedly applied such that a net $\theta=2\pi$ rotation is implemented.
The logical error rate $p_L$ is determined from classical simulations with depolarizing noise by computing the fidelity of the initial state $\rho$ with the noisy final state $\tilde\rho$, $p_L = 1-\text{Tr}(\rho\tilde\rho\Pi_{\mathcal{C}})$.
Since the effect of rotations depends on the initial state, we compute the Haar average of $p_L$ over all initial states.
A two-design is necessary to compute the Haar average of the fidelity~\cite{Dankert:2009yux}.
The set of stabilizer states forms a three-design~\cite{Zhu:2017psv}, so one- and two-qubit stabilizer states are used as initial states for the circuits in Fig.~\SubFigRef{fig:ancilla_rotations}{a}). 
The LER as a function of physical error rate $p$ for eight repeated $\pi/4$ rotations of $G_2,\,G_2',\,G_4,\,G_4'$ is shown in Fig.\SubFigRef{fig:ancilla_rotations}{b}).
Both implementations exhibit $p_L=\mathcal{O}(p)$ scaling. 
The ancilla method $G_2'$ is found to have a two to three times smaller prefactor than $G_2$, and $G_4$ is found to be roughly equivalent to $G_4'$ on average when accounting for errors detected by the $[[4,2,2]]$ code.
This suggests that although the ancilla method can offer an advantage in specific cases, consistent with Refs.~\cite{Zhong:2026jps,Gerhard:2024peb}, the two approaches perform similarly on average under depolarizing noise.
Further, this examination assumes noiseless (and instantaneous) measurements.
On realistic hardware, a measurement consumes significant coherence time and introduces measurement errors. 
For these reasons the simulations in this work are carried out with the standard NISQ logical rotation gadget.\footnote{Ancillas are used to implement rotations between blocks through bridge qubits where necessary, see~\ref{sec:scheduling}.}

\section{Additional Analysis of Encoding Improvement}
\label{app:more_analysis}

\subsection{Qubit Dependence}
\begin{figure}
    \centering
    \includegraphics[width=\linewidth]{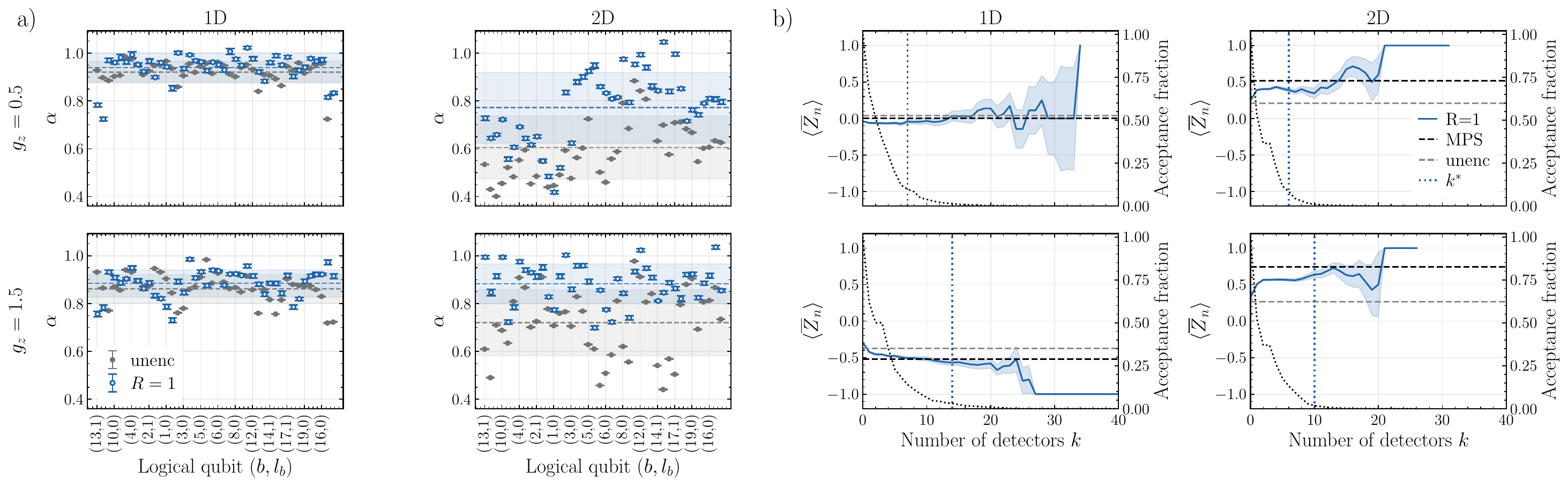}
    \caption{{\it Postselection Performance and Signal Survival in Encoded Simulations.}
    a)~The signal-survival factor $\alpha$ as a function of logical qubit over all $t$ for $g_z=0.5$ (top) and $g_z=1.5$ (bottom).
    Encoded results with $R=1$ are given in blue, and unencoded results are shown in grey. 
    Results for the 1D chain are shown in the left column, and 2D grid results are shown in the right column.
    The shading represents one standard deviation of the $\alpha$ distribution over qubits.
    Error bars on individual points represent uncertainty determined through bootstrap resampling.
    b)~Convergence of an example observable on qubit $(14,1)$ at $t=5$ as a function of the number of detectors kept for postselection $k$. 
    The shading represents uncertainty from bootstrap resampling.
    Results from the unencoded runs and from MPS are given by grey and black dashed horizontal lines.
    The vertical black dotted line shows the $k^*$ selected by the plateau finding algorithm.
    Acceptance rates for each $k$ are shown by the dotted grey line.
    }
    \label{fig:chain_grid_alpha_convergence}
\end{figure}

\noindent
Figure~\SubFigRef{fig:chain_grid_alpha_convergence}{} shows additional metrics comparing encoded to unencoded runs presented in Sec.~\ref{sec:results}.
As seen in Fig.~\ref{fig:improvement}, the encoding improvement in 1D is most pronounced at early and intermediate times, and decreases at late times.
In 2D, the improvement grows with time.
Figure~\SubFigRef{fig:chain_grid_alpha_convergence}{a}) shows $\alpha$ per logical qubit across all $t$.
For both 1D and 2D simulations, the encoded $\alpha$ values are on average higher than the unencoded, but are broadly distributed.
Several outliers are present in both encoded and unencoded results; 
these can be matched to locations on the quantum processor with faulty gates or qubits with high measurement error rate or low T1/T2 times.
Pauli twirling of the coherent and amplitude-damping/dephasing components of the noise could convert part of this structured error into a stochastic Pauli channel more amenable to detection and postselection.
Omitting these blocks in the analysis improves results considerably, consistent with most regions of the quantum processor having an operating error rate just below the pseudothreshold.
However, this exclusion is not done in the results reported in the main text.
Further, the per-qubit encoded points carry visibly larger bootstrap uncertainties than the unencoded points, which is an expected feature of shot loss from postselection and increased circuit depth due to encoding. The results obtained from the simulations in 2D have a significantly larger spread than those from simulations in 1D, and are generally noisier (lower $\alpha$). 
This again is a consequence of the deeper circuits required to implement a dense logical connectivity. 

Figure~\SubFigRef{fig:chain_grid_alpha_convergence}{b}) shows the effects of the plateau-finding algorithm explained in~\ref{sec:filtering} for qubit $(14,1)$ at $t=5$, 
as an example.
As detectors are added, the acceptance fraction (grey dotted line) falls steeply while the estimate settles onto a plateau in approaching the MPS value (black dashed line) and away from the biased unencoded result (grey dashed line).
The plateau-finding algorithm selects $k^*$ (blue dotted vertical line) at the onset of this stable region.
If such a value is not found, the algorithm selects $k^*$ to minimize the squared error, as explained in step five of the plateau-finding algorithm in~\ref{sec:det_selec}, and examples of this are shown in both 2D plots in Fig.~\SubFigRef{fig:chain_grid_alpha_convergence}{b}) and in the $g_z=1.5$ 1D plot.
Across all times and qubits, most show similar convergence behavior with the exception of especially faulty device regions.

\begin{figure}
    \centering
    \includegraphics[width=0.6\linewidth]{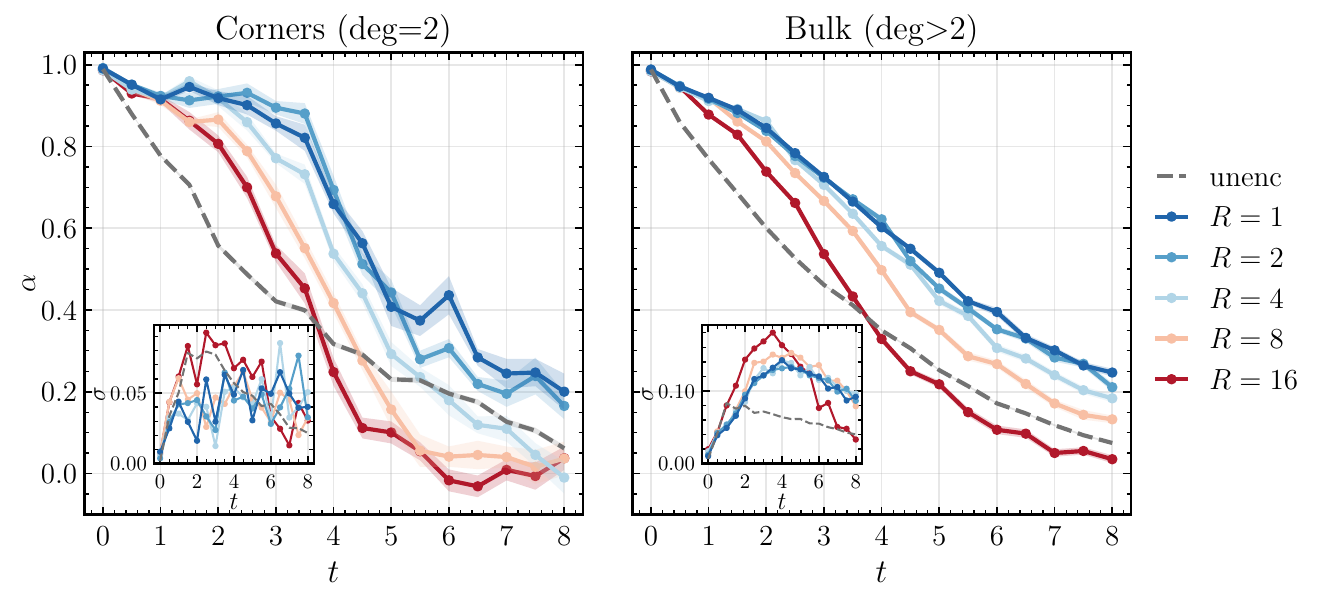}
    \caption{{\it Comparison of $\alpha$ for Corner and Bulk Qubits in 2D Simulations.}
    The signal-survival factor $\alpha$ is shown for encoded runs as a function of time $t$ for $R$ syndrome extraction rounds (colors) compared to unencoded runs (grey). 
    The left panel shows $\alpha$ for corners of the grid (logical qubits with two neighbors), and the right panel shows $\alpha$ for bulk qubits (those with more than two neighbors). 
    The insets show the error in the $\alpha$ fits, $\sigma$. 
    Results from the localized regime ($g_z=1.5$) are used.}
    \label{fig:corner_vs_bulk}
\end{figure}

Figure~\SubFigRef{fig:corner_vs_bulk}{} shows the difference in $\alpha$ between corner and bulk qubits in 2D grid simulations with $g_z=1.5$. 
Corners are defined as logical qubits with two connections, while bulk qubits have $\text{deg}>2$.
Since corners have fewer logical connections, fewer gates act on them and as a result they are less noisy.
In addition, the backwards lightcone on corner qubits is smaller, reducing the amount of errors that can affect observables there.
This is reflected in the larger $\alpha$ values in the left panel of Fig.~\SubFigRef{fig:corner_vs_bulk}{} compared with the right.
Further, the $\sigma$ values quantifying the spread in the $\alpha$ fit are smaller, indicating the corners more consistently reproduce the MPS expectations.

\subsection{Coupling Dependence}
\begin{figure}
    \centering
    \includegraphics[width=0.75\linewidth]{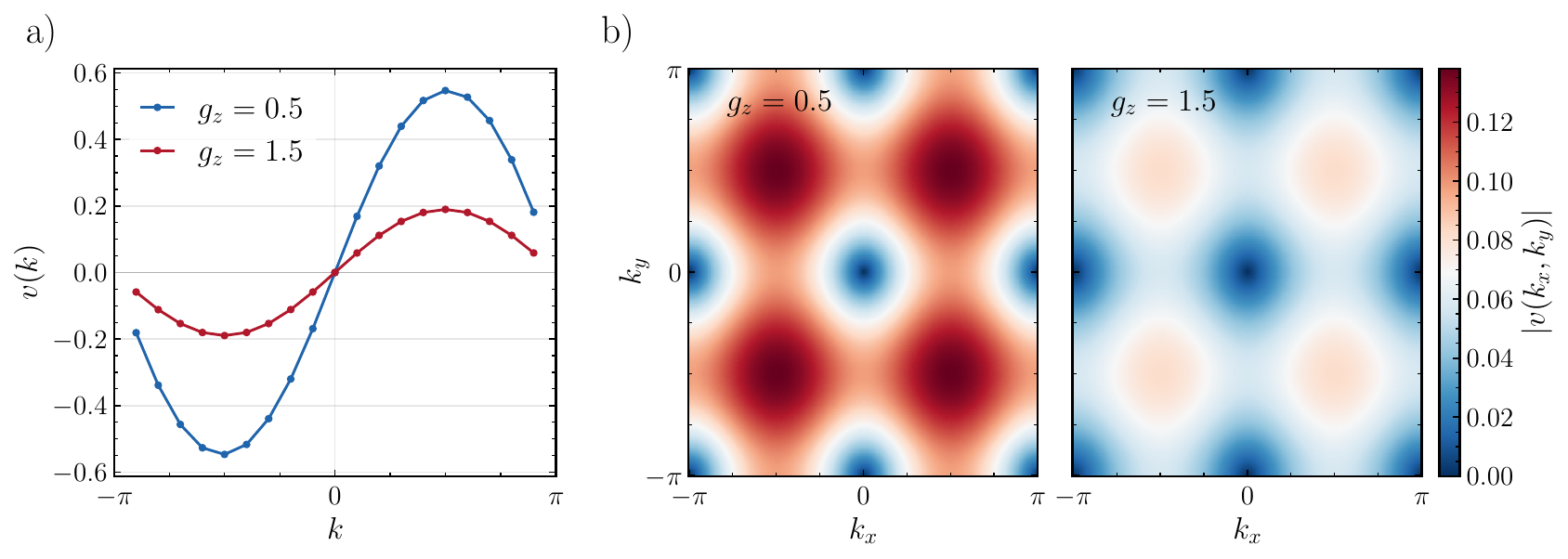}
    \caption{{\it Group Velocity in the 1D and 2D Ising Model.}
    a)~The group velocity $v(k)$ as a function of momentum $k$ for $g_z=0.5$ (blue) and $g_z=1.5$ (red) calculated in a 20-qubit 1D chain with periodic boundary conditions.
    b)~The group velocity $|v(k_x,k_y)|$ for a $5\times5$ 2D lattice with $g_z=0.5$ (left) and $g_z=1.5$ (right).
    Dispersion relations are calculated using exact diagonalization, and a spectral derivative is used to compute $v(k_x,k_y)$.
    }
    \label{fig:v_k_1d_2d}
\end{figure}

\noindent
As shown in Fig.~\SubFigRef{fig:improvement}{}, the encoding improvement is roughly independent of $g_z$ for 1D and 2D. 
However, $\alpha$ fits for both encoded and unencoded runs are consistently higher for $g_z=1.5$ than $0.5$, which is seen in Figs.~\SubFigRef{fig:chain_results}{e}) and~\SubFigRef{fig:grid_results}{d}). 
This difference originates in how far excitations (and errors) can propagate in each system.
As shown in Fig.~\SubFigRef{fig:v_k_1d_2d}{a}), the maximum group velocity $v(k) = dE(k)/dk$ is substantially larger at $g_z=0.5$ than at $g_z=1.5$ in 1D, where the stronger longitudinal field slows the propagation front.\footnote{These calculations are done using exact diagonalization with periodic boundary conditions. 
The limits of this method prevent determining the group velocity for our lattice geometry with missing links, so a smaller uniform lattice is used instead.}
A similar, but less dramatic difference exists in 2D, shown in Fig.~\SubFigRef{fig:v_k_1d_2d}{b}). 
The consequence of larger $v(k)$ for local observables is a larger backwards lightcone.
This implies more errors have the ability to affect a given $\langle \overline{Z}_n\rangle$ in the melting regime at fixed $t$. 
At a fixed number of syndrome extraction rounds $R$, errors can spread more under faster dynamics before they are caught by error detection.
This is the reason that $\alpha$ is consistently smaller in the melting regime (seen in Figs.~\SubFigRef{fig:chain_results}{e}) and~\SubFigRef{fig:grid_results}{d}), and degrades faster.
Further, the group velocity in 2D is significantly smaller than in 1D. 
The melting regime in 1D has by far the largest $v(k)$, $\sim2.5$ times larger than $g_z=1.5$ in 1D and $\sim5$ times larger than 2D $v(k_x,k_y)$. 
The fluctuations in $\alpha$ in Fig.~\SubFigRef{fig:chain_results}{e}) are attributed to this difference.
Since the difference in $v(k_x,k_y)$ between the regimes is only 0.04 in 2D, errors propagate at more similar speeds, which explains why the $\alpha$ plots in Fig.~\SubFigRef{fig:grid_results}{d}) are more similar than in 1D. 
This dependence on the backwards lightcone suggests that more frequent syndrome extraction is necessary for simulations of faster dynamics.
Unfortunately, the coherence time overhead is found to 
suppress most of the advantage of more frequent syndrome measurement in our simulations.

\subsection{Reset Dependence}
\begin{figure}
    \centering
    \includegraphics[width=0.75\linewidth]{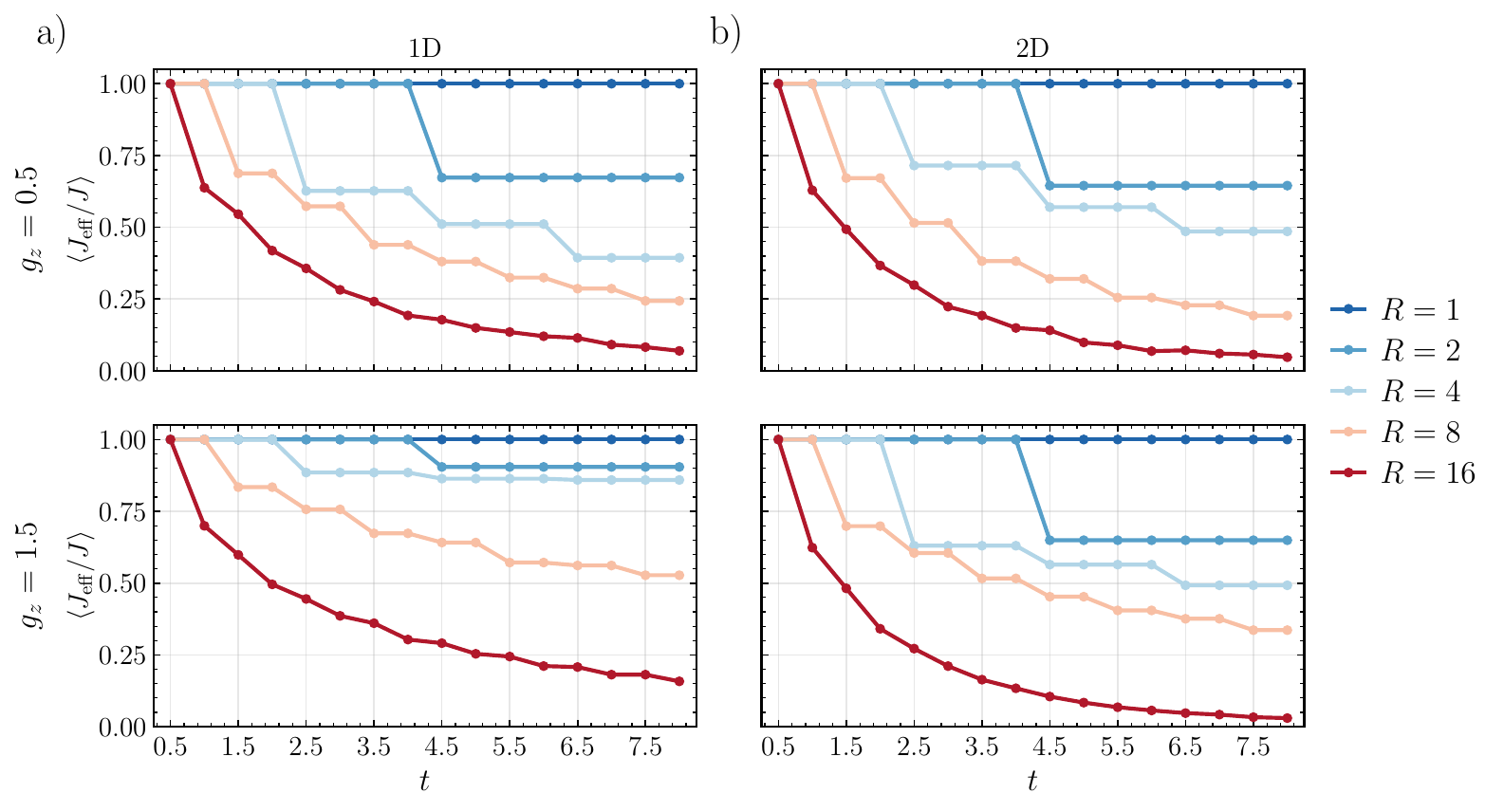}
    \caption{{\it Effective Inter-Block Coupling Under Residual Ancilla Sign Flips on} {\tt{ibm\_boston}}.
    a)~The effective inter-block coupling $\langle J_\text{eff}/J\rangle$ as a function of time $t$ averaged over all inter-block connections in the 1D simulations shown in Fig.~\SubFigRef{fig:chain_results}{} for $g_z=0.5$ (top) and $g_z=1.5$ (bottom).
    Different $R$ values indicate amount of syndrome measurements.
    b)~The same as a) but for 2D results presented in Fig.~\SubFigRef{fig:grid_results}{}.}
    \label{fig:j_eff_chain_grid}
\end{figure}

\noindent
As a result of omitting resets on ancilla qubits after syndrome measurements, an ancilla that fires remains in the state $|1\rangle$ until a subsequent error measurement flips it back to $|0\rangle$.\footnote{The absence of resets also means that leakage on an ancilla persists across all subsequent rounds, since there is no re-initialization to return a leaked ancilla to the computational subspace.}
Any inter-block $\overline R_{ZZ}(\theta)$ rotation that uses this ancilla in its implementation therefore applies the opposite rotation $\overline R_{ZZ}(-\theta)$.
Interspersing these opposite-angle inter-block rotations among the correct ones partially cancels the intended coupling $J$, producing an effective ``slowed-down" interaction $J_\text{eff} = J(1-2p_\text{flip})$, where $p_\text{flip}$ is the probability that the mediating ancilla is in $|1\rangle$ at the time of the gate.
The effect is amplified when two ancillas are used to implement the inter-block $\overline R_{ZZ}(\theta)$ rotation.
Figure~\SubFigRef{fig:j_eff_chain_grid}{} shows the average $J_\text{eff}$ over all inter-qubit couplings as a function of time for various amounts of syndrome extraction rounds $R$.
In 1D, the reduction in $J$ is smaller in the localized regime, where the stronger longitudinal field $g_z$ suppresses the spread of correlations along the chain.
This results in a lower average $p_\text{flip}$ for $g_z=1.5$.
The differences in $J_\text{eff}$ between regimes and dimensions is again attributed to the group velocity which governs the propagation speed of errors. 
Since errors propagate slower at $g_z=1.5$, the effects of incorrect rotations take longer to spread.
This slowdown could be further studied by running MPS simulations with the couplings $J_\text{eff}$, but this is not done in this work.

The effective reduction in $J$ could be removed entirely with the introduction of ancilla resets after syndrome measurements, and is the primary reason that high-$R$ simulations are found to perform worse than low-$R$ simulations.
Increasing syndrome measurement frequency adds more opportunities for stale ancillas to artificially inject errors into the simulation, and if the corresponding detection events are not selected for postselection (e.g., by ORP), they contaminate the results.
With ancilla resets, this problem is absent. 
However, ancilla resets consume a significant amount of coherence time, and as a result errors are nevertheless introduced.
We find that the implementation of resets does not significantly change the $R$-dependence of the results, with the exception of the highest values of $R$. 
Even for $R=16$, where the improvement from resets is largest, the $\alpha$ values do not exceed those of low $R$, indicating that the added coherence time injects more errors than the syndrome measurement rounds are able to detect. 
The implementation of faster resets on the hardware will likely shift this balance. 
This is another example of the tradeoff between the NISQ considerations of gate depth and coherence time and FT requirements. 

\subsection{Geometric Dependence in 1D Simulations}
\begin{figure}
    \centering
    \includegraphics[width=0.6\linewidth]{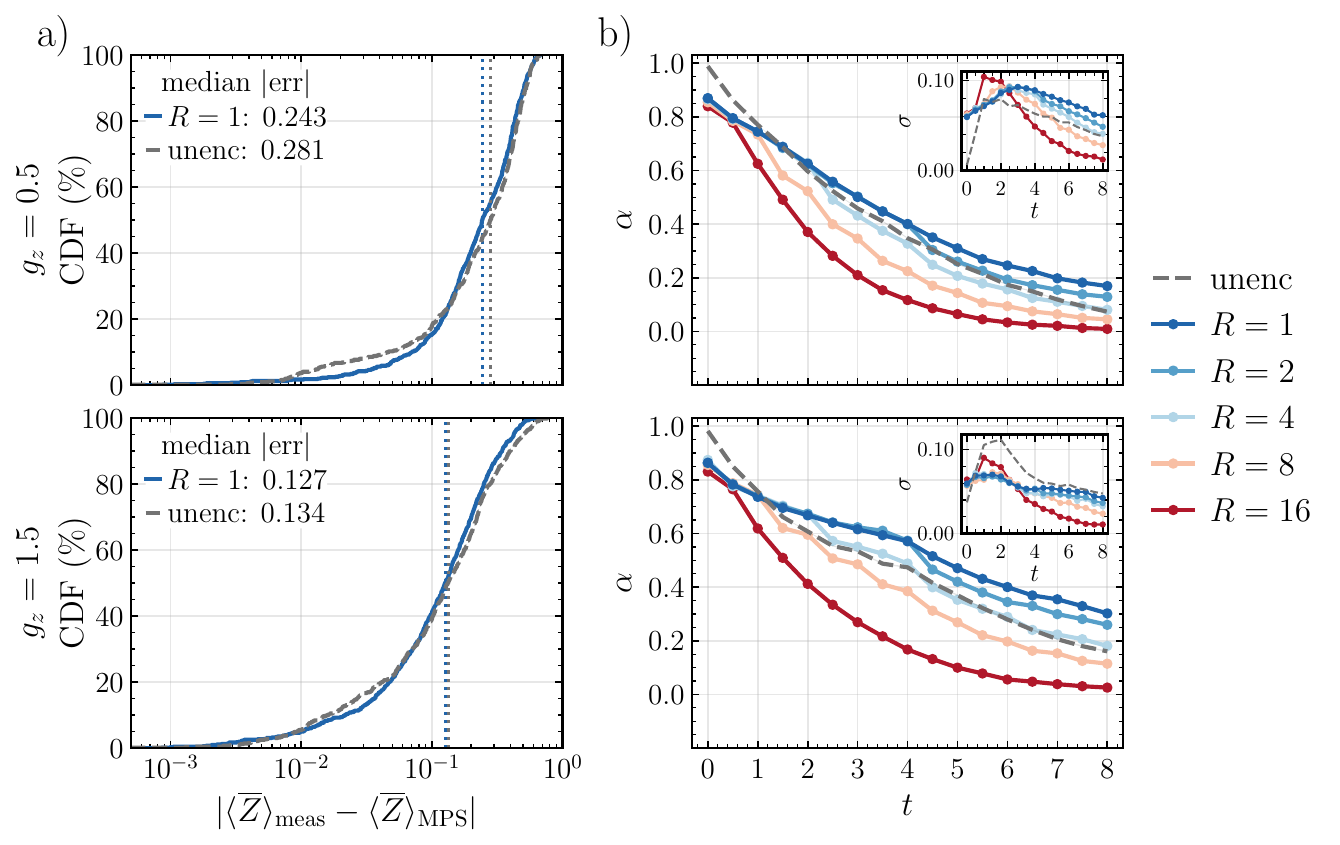}
    \caption{{\it Encoding Improvement Without Postselection in 2D Simulations.}
    a)~Cumulative distribution functions (CDFs) of the per-qubit, per-time absolute error $|\langle \overline{Z}\rangle_\text{meas} - \langle \overline{Z}\rangle_\text{MPS}|$ for both values of $g_z$.
    The median error and CDFs corresponding to $R=1$ syndrome measurements are shown compared to unencoded results (solid and dashed lines respectively).
    b)~The signal-survival factor $\alpha$ as a function of $t$ for various $R$ (colored lines) and each value of $g_z$, compared to unencoded results (dashed grey line).
    The insets show the residual error $\sigma$ in the fit. }
    \label{fig:grid_results_raw}
\end{figure}

\noindent
The encoded results presented in Fig.~\SubFigRef{fig:grid_results}{} show a large improvement over unencoded runs, increasing with time to a greater than $200\%$ improvement at late times.
This is partially due to the geometric gate depth overhead of embedding a square lattice onto heavy-hex connectivity. 
Unencoded circuits are 1.5 times deeper than encoded circuits at the latest time considered (see Table~\ref{tab:grid_circ_nums}). 
To isolate the encoding improvement from the geometric overhead, we compare the encoded results without any postselection to unencoded results in Fig.~\SubFigRef{fig:grid_results_raw}{}.
The CDF plots show that encoding without postselection has a small improvement at $g_z=0.5$ and a negligible improvement at $g_z=1.5$. 
The decay of the signal-survival factor $\alpha$ shows the performance of the encoded circuits over time without postselection.
At early times, encoded circuits without postselection perform worse than unencoded as a result of added circuit depth to perform syndrome measurements.
At intermediate and late times, low-$R$ encoded runs outperform unencoded due to the geometric overhead.
Comparing to Fig.~\SubFigRef{fig:grid_results}{d}), the maximum decrease in $\alpha$ from omitting postselection is $\sim 0.4$.
This indicates that much of the improvement seen in Fig.~\SubFigRef{fig:grid_results}{} is due to encoding as opposed to the geometric overhead in unencoded circuits.
Thus, removing errors that can potentially spread to $\mathcal O(t^2)$ qubits plays a large role in the improvement observed in 2D simulations.

\section{Classical Simulations of False-Vacuum Decay Dynamics}
\label{app:physics}
\noindent
Phase transitions play a key role in early-universe dynamics as well as in many condensed matter systems.
For high energy physics, a particularly important question is whether the Higgs field is in a true vacuum or in a false metastable vacuum.
The phase transition from a false vacuum to a true vacuum is known as false-vacuum decay and underlies a range of early-universe phenomena, from cosmological phase transitions and bubble nucleation to the possible fate of the electroweak vacuum~\cite{Coleman:1977py,Callan:1977pt}.
This section describes a simple 1D model for false-vacuum decay in spin systems as intuition for the oscillation and localization behavior observed in the simulations in the main text.
Classical MPS simulations carried out in the same regimes considered in Sec.~\ref{sec:results} are shown to support this intuition where analytical calculations are unavailable.

In false-vacuum decay phenomenology, the metastable ``false" vacuum decays to the stable and energetically favored ``true" vacuum by quantum tunneling~\cite{Coleman:1977py,Callan:1977pt}.
In the semiclassical regime, thermal or quantum fluctuations typically nucleate localized bubbles of true vacuum.
Bubbles exceeding a critical size then lower the energy through expansion, driving conversion of the system to the true vacuum.
The method in this work provides a lattice realization of false-vacuum decay through quench dynamics of a spin system~\cite{Lagnese:2021grb,Lagnese:2023xjg}.
Rather than directly studying bubble nucleation, we prepare a fixed, finite system in a ferromagnetic initial state with a true-vacuum bubble in the center and follow its unitary time evolution.
The presence of ``decay" in this setup is indicated by the melting of the true-vacuum bubble, and decay is absent when the bubble remains localized.

Section~\ref{sec:results} observes a difference in encoded quantum simulation performance between the $g_z=0.5$ (melting regime) and $g_z=1.5$ (localized regime). 
The melting or localization of true-vacuum bubbles in our simulations is due to domain-wall oscillations known as Bloch oscillations~\cite{Pomponio:2021ltz}.
The following analysis considers two domain walls in a 1D chain.
The Hamiltonian~\eqref{eq:h_ising} assigns a fixed energy per flipped spin, so the bubble's potential energy is linear in its size $r$
\begin{align}
    V(r) \ = \ -\chi r \ , \ \chi \ = \ 2m|g_z| \ , \ m = \left(1-\left(\frac{g_x}{|J|}\right)^2 \right)^{1/8} \ ,
\end{align}
where $m$ is the spontaneous magnetization at $g_z=0$~\cite{Pfeuty:1970qrn}.
Unitary time evolution under $\overline H$ conserves total energy, but converts the potential energy into the kinetic energy of expanding bubble walls.
As $V(r)$ changes from bubble expansion, the relative momentum of the bubble walls $k$ changes, as given by Hamilton's equations, 
\begin{align}
    \frac{dk}{dt} \ = \ -\frac{dV}{dr} \  = \ \chi \ .
\end{align}
Neglecting corrections due to nonzero $g_z$, the dispersion relation for a single domain wall is
\begin{align}
    E(k) \ = \ 2 |J| \sqrt{1 + g_x^2 + 2g_x \cos k} \ .
\end{align}
Since $E(k)$ is periodic, the group velocity $v(k) = dE(k)/dk$ changes sign as the momentum is increased in the Brillouin zone, resulting in oscillations of the bubble width
\begin{align}
    r(t) \ = \ \frac{4}{\chi} \left( E\left(k_0 + \frac{\chi}{2}t\right) - E(k_0)\right) \ .
\end{align}
The maximum amplitude of these oscillations is
\begin{align}
    r_\text{max} \ = \ \frac{4}{\chi} (E(\pi) - E(0)) \ = \ \frac{8|J| |g_x|}{m |g_z|}\ ,
    \label{eq:r_max}
\end{align}
and the period is 
\begin{align}
    T_B \ = \ \frac{4 \pi}{\chi} \ = \ \frac{2 \pi}{m |g_z|} \ .
\end{align}

\begin{figure}
    \centering
    \includegraphics[width=\linewidth]{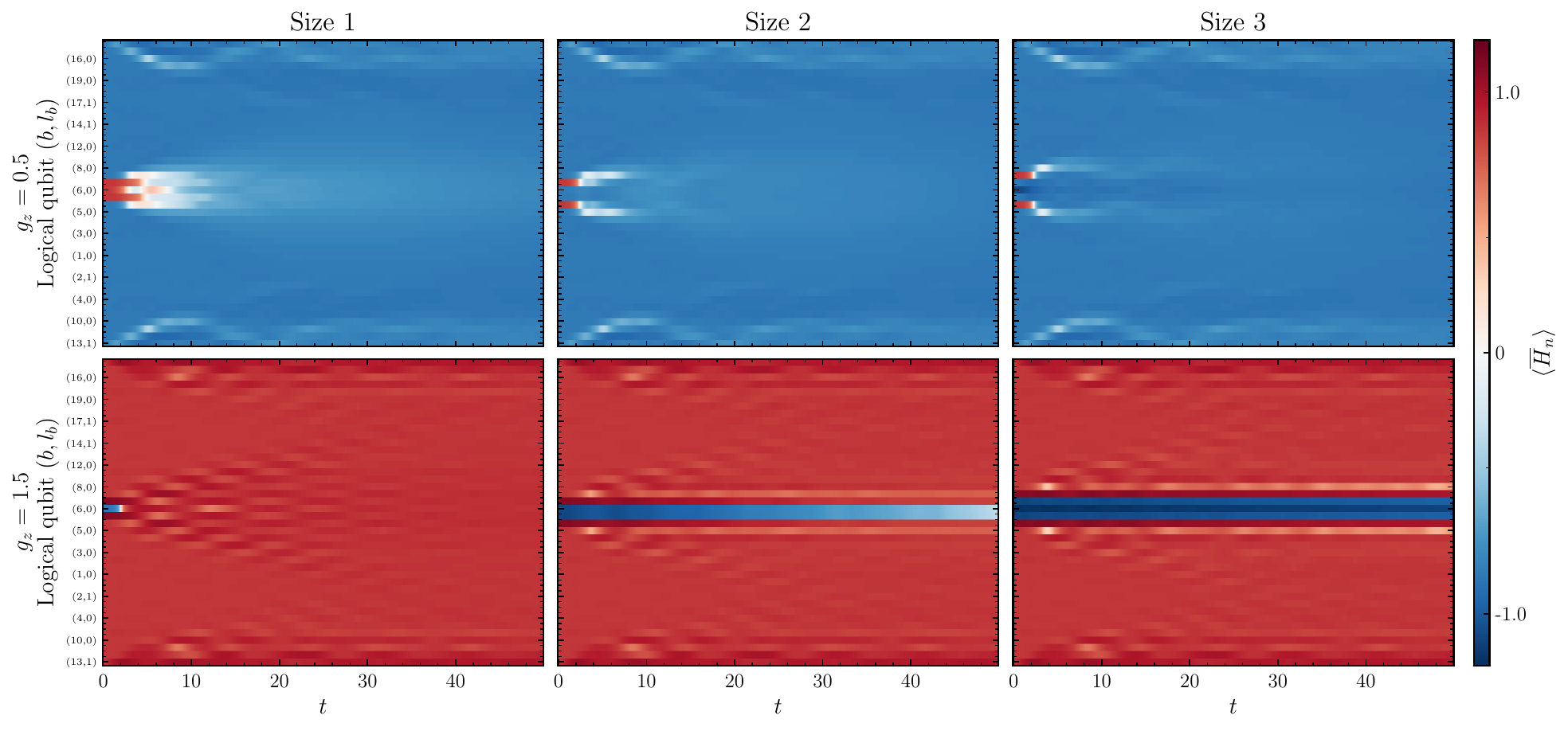}
    \caption{{\it Classical Simulations of False-Vacuum Decay Dynamics in 1D.}
    An initial true-vacuum bubble of size one, two, three (columns) at the center of the 42-qubit chain is evolved with $g_z=0.5$ (top row) and $g_z=1.5$ (bottom row).
    The color in the heatmap shows the energy density $\langle \overline{H}_n\rangle$ as a function of logical qubit position $n=(b,l_b)$ and time $t$.
    The time evolution uses $n_T=500$ Trotter steps of size $\delta t=0.1$ and is computed with MPS.}
    \label{fig:mps_1d_gz_comparison}
\end{figure}

Although strictly valid only in the analytically tractable $g_z\to 0$ limit, and not at the couplings considered in our simulations, this analysis provides intuition for how oscillations arise in the domain-wall initial state.
An equivalent treatment in the large-$J$ limit maps the system to free fermions in a linear potential~\cite{Balducci:2022zym,Balducci:2022kvd}, where the extent of oscillations is also found to be $r_\text{max}\sim |g_x|/|g_z|$.
Reference~\cite{Balducci:2022zym} finds a transition between the bubble-melting and localized regimes at $|g_z|/|g_x|\sim 1$ for large $J$.
Systems with $|g_z|/|g_x|< 1$ are found to thermalize, while $|g_z|/|g_x|> 1$ shows evidence of slower decay.
Results from our simulations at $J=1$ are consistent with these findings.
Figure~\SubFigRef{fig:mps_1d_gz_comparison}{} shows the energy density determined from MPS simulations through the relaxation dynamics following a quench under the Hamiltonian~\eqref{eq:h_ising} in 1D.
Initial bubble states ${|\psi(t=0)\rangle = \prod_{i\in B}\overline X_i |\overline 1\rangle^{\otimes N}}$ with bubble sizes $|B|=1,2,3$ are shown in the columns. 
The top row shows the bubble-melting regime ($g_z=0.5$).
Faint oscillations about the center of the bubble are seen for sizes two and three, consistent with Bloch oscillations with a period $T_B\sim12$ and $r_\text{max}\sim10$.
Since our simulations are far outside the small-$g_z$ or large-$J$ regime, we cannot compare these to predictions of $T_B$ or $r_\text{max}$ from the analysis above.
The bottom row ($g_z=1.5$) shows the bubble localized to all accessible times for sizes two and three.
This is consistent with Eq.~\eqref{eq:r_max}, where larger $g_z$ makes $r_\text{max}$ comparable to the bubble size and prevents oscillations. 
A size one bubble is seen to decay for both values of the coupling because the Bloch mechanism requires separation between the two bubble walls for the linear potential to act on, which does not exist for a single flipped spin.
This confirms the localization is not due to $Z$-field pinning, in which a strong longitudinal field suppresses interface-moving spin flips when $|g_z|\gg |g_x|$.

\begin{figure}
    \centering
    \includegraphics[width=0.75\linewidth]{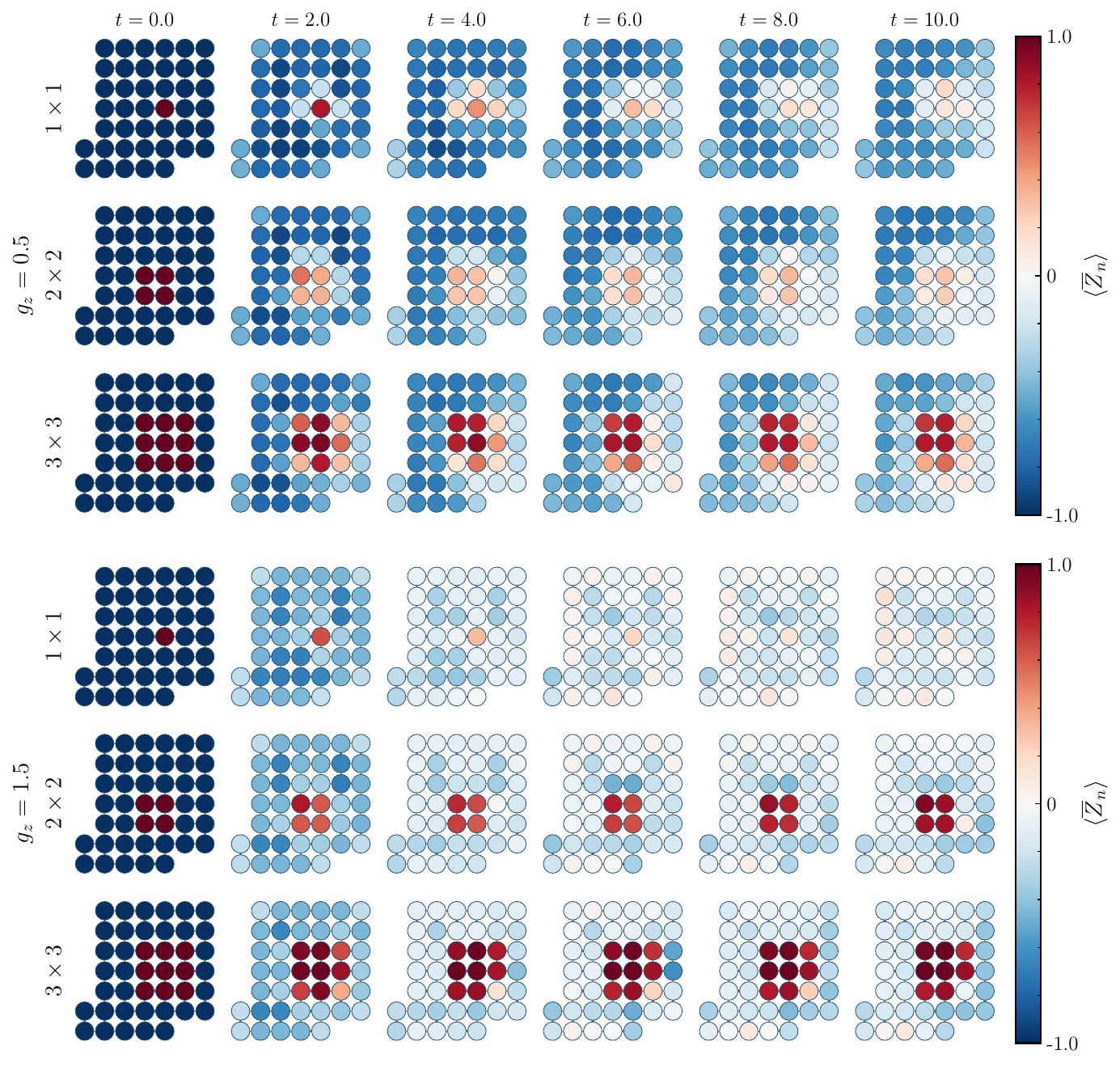}
    \caption{{\it Classical Simulations of False-Vacuum Decay Dynamics on a 2D Lattice with Missing Edges.}
    An initial true-vacuum bubble of size $1\times1$, $2\times2$, $3\times3$ (rows) at the center of the 42-qubit lattice is evolved with $g_z=0.5$ (top) and $g_z=1.5$ (bottom).
    The qubit colors show the magnetization $\langle \overline{Z}_n\rangle$ as a function of logical qubit position $n=(b,l_b)$ and time $t$ (columns).
    The time evolution uses $n_T=100$ Trotter steps of size $\delta t=0.1$ calculated with MPS.
    The logical connectivity grid corresponds to  Fig.~\SubFigRef{fig:chain_results}{a}) with all available edges used.}
    \label{fig:mps_2d_gz_comparison}
\end{figure}

We find similar behavior in 2D simulations. 
Figure~\SubFigRef{fig:mps_2d_gz_comparison}{} shows $\langle \overline{Z}_n\rangle$ determined from MPS simulations throughout the lattice for a selection of times.
The dynamics in 2D is found to be slower than in 1D, which is also evidenced by the group velocities shown in Fig.~\SubFigRef{fig:v_k_1d_2d}{}.
Further, the melting and localization dynamics is anisotropic due to the irregular geometry realized by the logical qubit connectivity. 
For $g_z=0.5$ sizes $2\times2$ and $3\times3$, the initial bubble is seen to spread more towards the right and bottom as a result of missing edges on the left and top boundaries.
These missing edges also artificially localize the bubble for $g_z=1.5$.
However, physics-driven localization (as opposed to localization driven by missing links) is still seen on the right and bottom, consistent with the dynamics seen in 1D simulations.
Additionally, the bottom right qubit of the $3\times3$ bubble for $g_z=1.5$ melts off as a result of a missing vertical edge that would connect it to the rest of the bubble.\footnote{The global quench of the background contributes a significant amount to the observed dynamics, which is seen in the relaxation dynamics shown in Figs.~\ref{fig:mps_1d_gz_comparison} and~\ref{fig:mps_2d_gz_comparison}.}

\begin{figure}
    \centering
    \includegraphics[width=0.75\linewidth]{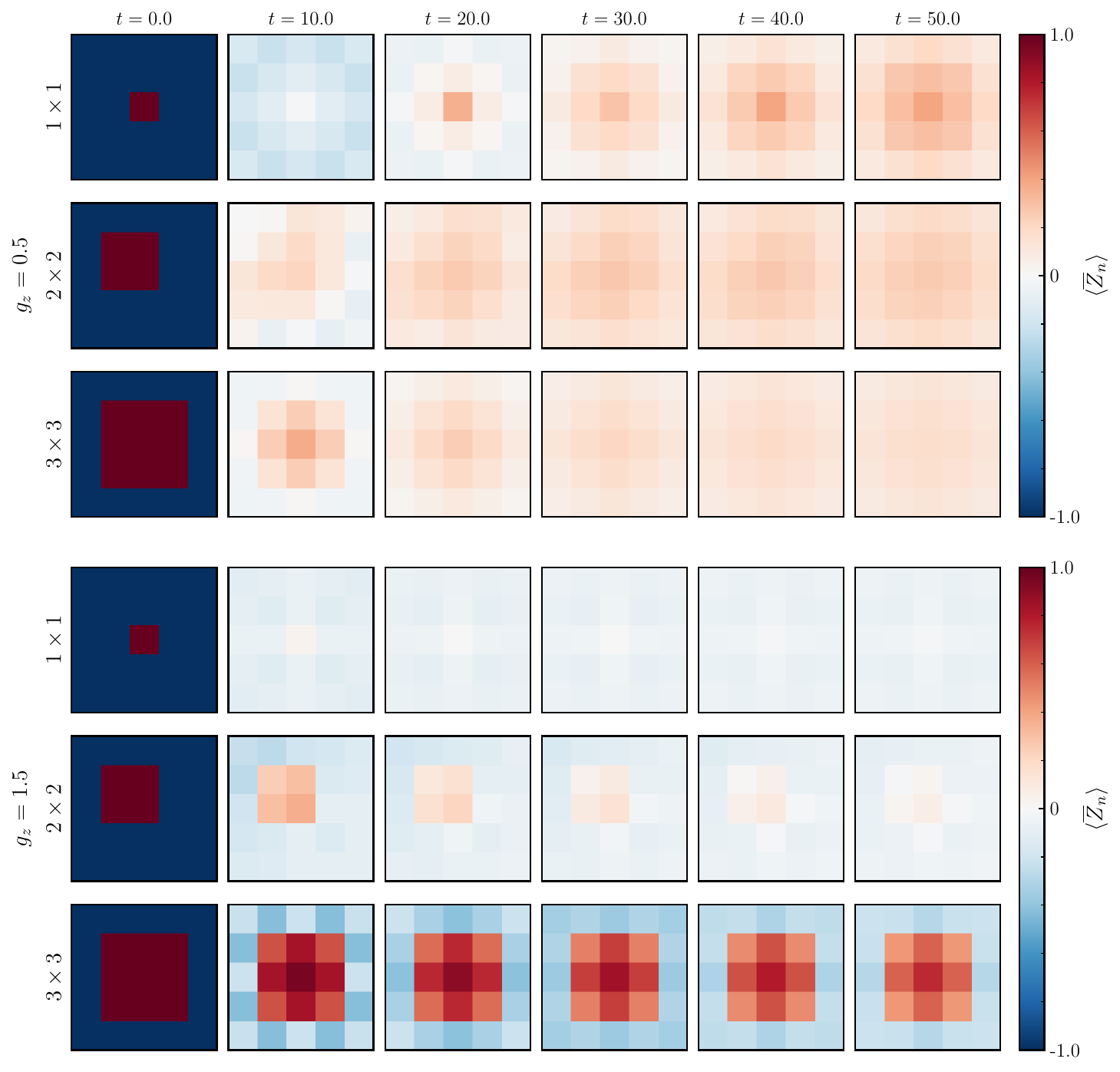}
    \caption{{\it Classical Simulations of False-Vacuum Decay Dynamics on a Uniform 2D Lattice.}
    An initial true-vacuum bubble of size $1\times1$, $2\times2$, $3\times3$ (rows) at the center of the $5\times5$ lattice is evolved with $g_z=0.5$ (top) and $g_z=1.5$ (bottom).
    The qubit colors show the magnetization $\langle \overline{Z}_n\rangle$ as a function of logical qubit position $n=(b,l_b)$ and time $t$ (columns).
    The time evolution uses $n_T=500$ Trotter steps of size $\delta t=0.1$ and is calculated with statevector simulations.}
    \label{fig:mps_2d_gz_uniform_comparison}
\end{figure}

To remove the contributions to the melting and localization dynamics from missing links, we compare the results of our realized grid to simulations on a uniform $5\times5$ lattice.
The comparison of $\langle \overline{Z}_n\rangle$ is shown for several selected times in Fig.~\SubFigRef{fig:mps_2d_gz_uniform_comparison}{}. 
This figure confirms that melting and localization consistent with Bloch oscillations are present in the absence of grid irregularities. 
The top row of the $g_z=0.5$ panel of Fig.~\SubFigRef{fig:mps_2d_gz_uniform_comparison}{} shows a recurrence for the $1\times1$ bubble.
This happens due to boundary effects and is absent in 1D simulations in Fig.~\SubFigRef{fig:mps_1d_gz_comparison}{}.
This is additionally verified by MPS simulations of a $7\times7$ lattice.
The $2\times2$ bubble is seen to melt for $g_z=1.5$.
This melting is slower than for $g_z=0.5$ and is also impacted anisotropically by the boundary.
This is consistent with the findings of Ref.~\cite{Balducci:2022kvd} where $|g_z|/|g_x|\gg1$ significantly slows down but does not fully prevent melting.

Together, these classical simulations show that the size-three bubbles (1D) and 
$3\times3$ bubbles (2D) studied in Sec.~\ref{sec:results} melt for 
$g_z=0.5$ and remain localized for $g_z=1.5$,
consistent with the Bloch oscillation picture. 
While the dynamics in 2D is significantly slower and the contrast between the melting and localized regimes is less pronounced, the same qualitative behavior holds, supporting the interpretation of the device results in the main text.
Although our couplings lie outside the small-$g_z$ and large-$J$ limits where the oscillation amplitude and period can be predicted analytically, the melting and localization observed here match the Bloch oscillation mechanism qualitatively.

\section{Other Layouts}
\label{app:other_layouts}
\begin{figure}
    \centering
    \includegraphics[width=\linewidth]{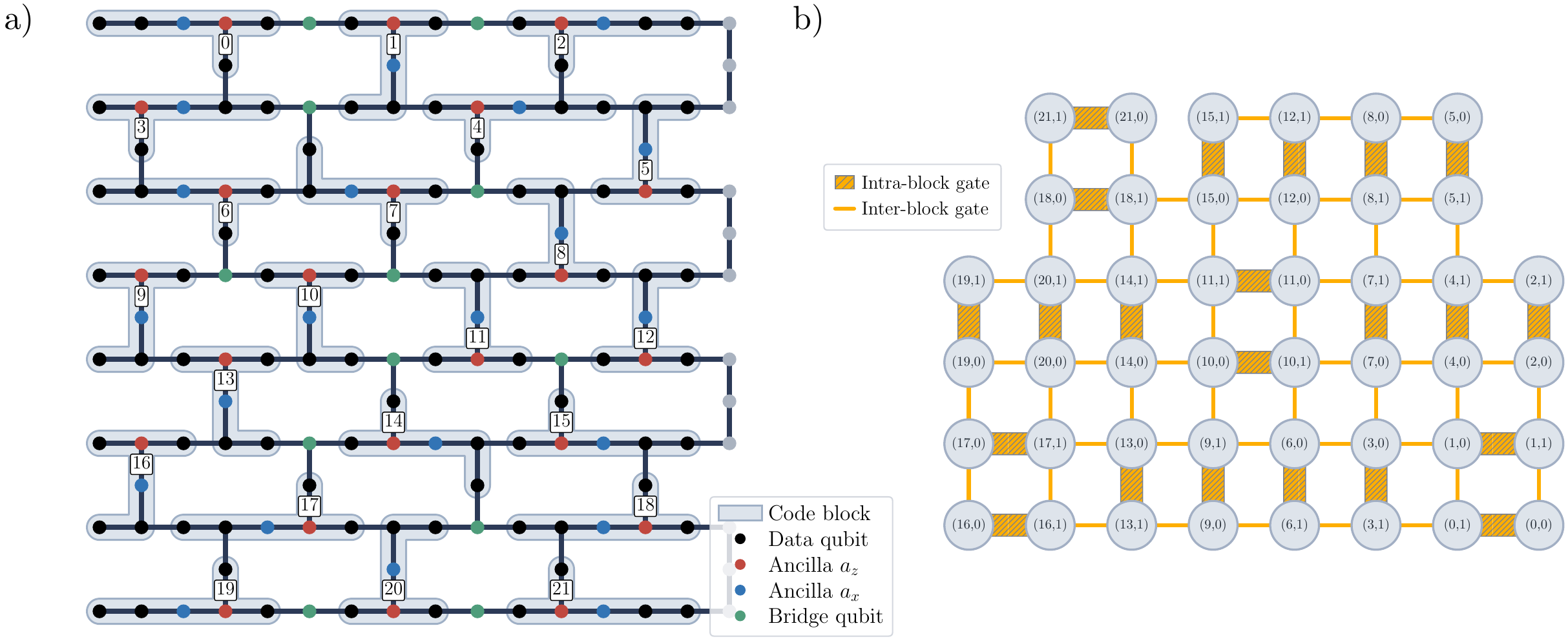}
    \caption{{\it Block Placement and Logical Connectivity Graph in the Absence of Faulty Qubits and Gates.}
    a)~An example of a block placement (grey) of 21 blocks assuming no faulty qubits and gates on a heavy-hex quantum computer.
    Within each block, the data qubits are shown in black, $Z$-ancillas are shown in red, $X$-ancillas are shown in blue, and bridge qubits are shown in green.
    b)~The resulting connectivity between logical qubits (grey) optimized for similarity to a square lattice.
    Thick orange hatched edges represent intra-block connections and thin orange edges show inter-block connections.}
    \label{fig:placement_grid_all_qubits}
\end{figure}

\noindent
As discussed in Sec.~\ref{sec:422_device}, 
there is significant freedom in $[[4,2,2]]$ code block placement onto the heavy-hex connectivity. 
In this work, we select a placement that avoids particularly faulty qubits and gates.
The block placement options are further expanded if this constraint is lifted.
While the maximum number of blocks is still 21 (corresponding to 42 logical qubits), the properties of the resulting placement may be more favorable. 
In particular, the resulting logical connectivity graph may be more similar to a regular square lattice.
Figure~\SubFigRef{fig:placement_grid_all_qubits}{a}) shows an example of one such placement.
This placement uses 144 active physical qubits, compared to 136 in Fig.~\SubFigRef{fig:overview}{b}), a result of more bridge qubits present. 
An example of the resulting logical connectivity graph possible with this placement is shown in Fig.~\SubFigRef{fig:placement_grid_all_qubits}{b}). 
This grid has 51 inter-block edges (72 edges total), whereas the grid used in the main text (Fig.~\SubFigRef{fig:chain_results}{a})) has 45 inter-block edges (66 edges total).
Importantly, this grid has fewer missing interior edges, which significantly contribute to the dynamics in our simulations. 
This suggests that more regular square lattice simulations are possible using the current generation of quantum computers with more uniform characteristics.

\begin{figure}
    \centering
    \includegraphics[width=0.475\linewidth]{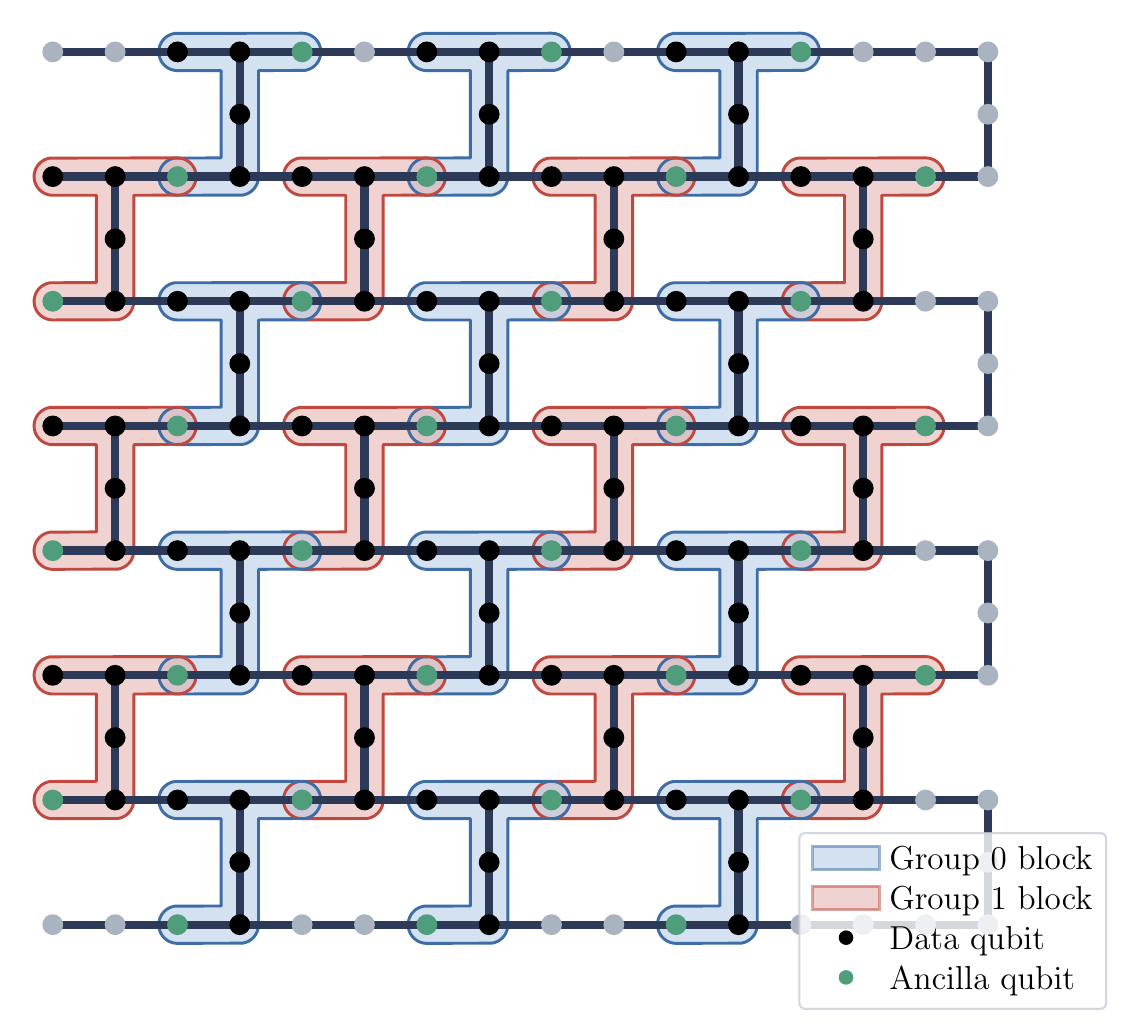}
    \caption{{\it Overlapping Block Placement.}
    Code blocks are split into two groups (red and blue) for serialized syndrome extraction.
    Data qubits within each block are shown in black and shared ancilla qubits are shown in green.}
    \label{fig:block_placement_overlapping}
\end{figure}

It is not possible to regularly tile a heavy-hex connectivity device with blocks that have the topology required for our syndrome extraction circuits (center of Fig.~\SubFigRef{fig:overview}{c})). 
Therefore bridge qubits are necessary, which contribute significantly to the depth of inter-block gates. 
A regular tiling with our block topology is possible if blocks are allowed to share ancillas. 
An example of such a placement is shown in Fig.~\SubFigRef{fig:block_placement_overlapping}{}.
This placement supports a higher encoding rate, using 126 physical qubits to encode 48 logical qubits with 24 overlapping blocks.
In addition, the logical connectivity graph possible with this placement is more dense, allowing for three-body gates between triples of neighboring blocks (with ancilla shuttling).
In an overlapping block placement, code blocks share ancilla qubits (shown in green) that are used only for syndrome extraction, and data qubits (black) are permanently tied to a given block.
This removes the overhead of bridge qubits.
Since data qubits of neighboring blocks are located physically closer, inter-block gates are considerably shallower, which is favorable for the total depth budget.
The tradeoff associated with overlapping layouts is that syndrome extraction must now be serialized as a result of ancilla sharing. 
Figure~\SubFigRef{fig:block_placement_overlapping}{} splits the blocks into two groups (shown in red and blue), corresponding to the serial order in which syndromes must be measured.
Ancilla shuttling is also required in this scenario, which introduces an overhead larger than that of switching between the compute and syndrome layouts in Fig.~\SubFigRef{fig:overview}{c}). 
Further, the conversion to detectors discussed in Sec.~\SubFigRef{sec:filtering_main_text}{} and ~\SubFigRef{sec:filtering}{} is modified due to ancilla sharing.
Syndrome measurement is by far the highest overhead in coherence time, which is doubled by serialization.
This work finds that the quality of local observables measured degrades significantly faster with overlapping code blocks compared to non-overlapping placements, and for this reason the non-overlapping placement is used.

\section{Tables of Results}
\label{app:tables}
\noindent
Tables~\ref{tab:chain_results_t_8},~\ref{tab:chain_cdf} and~\ref{tab:chain_alpha_sigma} show the data from {\tt ibm\_boston} corresponding to Fig.~\SubFigRef{fig:chain_results}{}. 
A 1D 42-logical qubit chain is run with $\delta t=0.5$ for $g_z=0.5$ and $g_z=1.5$. 
The initial state in each of these simulations is a size-three true-vacuum bubble product state.
Tables~\ref{tab:grid_results_t_4},~\ref{tab:grid_cdf} and~\ref{tab:grid_alpha_sigma} show the corresponding data for the 2D lattice simulations displayed in Fig.~\SubFigRef{fig:grid_results}{}.
The initial state on the 42-logical qubit lattice is a $3\times3$ true-vacuum bubble product state.
Table~\ref{tab:acceptance_rates} reports the acceptance rate of $R=1$ runs for 1D and 2D simulations for both values of $g_z$.

\begin{table}[h]
\renewcommand{\arraystretch}{1.1}
\begin{tabularx}{\linewidth}{|c||Y|Y|Y||Y|Y|Y|}
\hline
\rule{0pt}{10pt} & \multicolumn{6}{c|}{\large$\langle \overline{Z}_n\rangle $}\\\hline\hline
\rule{0pt}{10pt} \multirow{2}{*}{\makecell{Logical\\qubit $(b,l_b)$}} &  \multicolumn{3}{c||}{$g_z=0.5$}  &  \multicolumn{3}{c|}{$g_z=1.5$} \\\cline{2-7}
\rule{0pt}{10pt} & MPS & Unencoded & $R=1$ & MPS & Unencoded & $R=1$  \\
\hline\hline
$(13,1)$ & 0.217 & 0.100(6) & 0.098(19) & -0.568 & -0.486(5) & -0.329(27) \\
$(13,0)$ & 0.142 & 0.131(5) & 0.113(16) & -0.372 & -0.250(5) & -0.385(26) \\
$(10,1)$ & 0.084 & 0.090(5) & 0.081(17) & -0.138 & -0.066(5) & -0.244(36) \\
$(10,0)$ & -0.079 & -0.006(6) & 0.000(21) & -0.445 & -0.213(5) & -0.256(11) \\
$(9,0)$ & 0.003 & -0.006(6) & -0.019(21) & -0.381 & -0.263(5) & -0.249(11) \\
$(9,1)$ & 0.082 & 0.009(6) & 0.019(34) & -0.330 & -0.260(5) & -0.299(18) \\
$(4,0)$ & 0.080 & -0.043(6) & 0.024(32) & -0.289 & -0.263(6) & -0.248(22) \\
$(4,1)$ & 0.099 & 0.022(6) & 0.071(32) & -0.381 & -0.248(6) & -0.239(33) \\
$(2,0)$ & 0.115 & 0.055(6) & 0.065(18) & -0.407 & -0.283(5) & -0.269(20) \\
$(2,1)$ & 0.130 & 0.109(5) & 0.057(13) & -0.342 & -0.229(5) & -0.229(46) \\
$(0,0)$ & 0.175 & 0.121(6) & 0.067(46) & -0.379 & -0.297(5) & -0.202(31) \\
$(0,1)$ & 0.124 & 0.076(6) & 0.056(15) & -0.351 & -0.286(5) & -0.136(18) \\
$(1,0)$ & 0.071 & 0.044(6) & 0.009(32) & -0.317 & -0.223(5) & -0.132(31) \\
$(1,1)$ & 0.029 & 0.043(6) & 0.030(18) & -0.397 & -0.286(6) & -0.106(42) \\
$(3,1)$ & 0.068 & 0.115(6) & -0.020(17) & -0.347 & -0.166(6) & -0.152(35) \\
$(3,0)$ & 0.071 & 0.109(6) & 0.015(43) & -0.374 & -0.194(5) & -0.268(18) \\
$(7,1)$ & 0.029 & 0.071(5) & -0.010(29) & -0.228 & -0.151(6) & -0.223(11) \\
$(7,0)$ & -0.090 & -0.026(5) & -0.090(50) & -0.371 & -0.271(5) & -0.269(34) \\
$(5,0)$ & -0.011 & -0.028(6) & 0.029(17) & -0.348 & -0.245(5) & -0.237(40) \\
$(5,1)$ & 0.145 & 0.096(6) & 0.082(17) & -0.549 & -0.462(5) & -0.450(26) \\
$(6,1)$ & 0.206 & 0.217(6) & 0.163(41) & 0.623 & 0.549(5) & 0.555(21) \\
$(6,0)$ & 0.170 & 0.219(5) & 0.149(47) & 0.928 & 0.806(3) & 0.813(10) \\
$(11,0)$ & 0.245 & 0.223(5) & 0.211(30) & 0.809 & 0.677(4) & 0.595(12) \\
$(11,1)$ & 0.142 & 0.108(6) & 0.087(17) & -0.705 & -0.499(5) & -0.489(13) \\
$(8,0)$ & -0.071 & -0.058(6) & -0.108(30) & -0.304 & -0.231(6) & -0.179(10) \\
$(8,1)$ & -0.108 & -0.070(6) & -0.089(14) & -0.329 & -0.291(5) & -0.345(32) \\
$(12,1)$ & -0.001 & -0.005(6) & -0.010(28) & -0.252 & -0.206(6) & -0.137(32) \\
$(12,0)$ & 0.084 & 0.038(6) & 0.012(17) & -0.373 & -0.249(5) & -0.309(39) \\
$(15,1)$ & 0.095 & 0.024(6) & 0.083(20) & -0.275 & -0.152(5) & -0.191(13) \\
$(15,0)$ & 0.100 & 0.052(6) & 0.063(28) & -0.403 & -0.318(5) & -0.243(20) \\
$(14,1)$ & 0.101 & 0.096(5) & -0.002(11) & -0.403 & -0.242(6) & -0.237(30) \\
$(14,0)$ & 0.088 & 0.171(5) & -0.010(12) & -0.280 & -0.053(6) & -0.163(39) \\
$(17,0)$ & 0.080 & 0.104(5) & -0.012(11) & -0.333 & -0.199(5) & -0.252(19) \\
$(17,1)$ & 0.066 & 0.076(6) & 0.040(18) & -0.357 & -0.293(5) & -0.246(13) \\
$(20,0)$ & 0.054 & 0.059(5) & 0.032(10) & -0.425 & -0.335(5) & -0.258(26) \\
$(20,1)$ & 0.067 & 0.076(6) & 0.007(15) & -0.274 & -0.200(5) & -0.135(9) \\
$(19,0)$ & 0.088 & 0.067(6) & -0.016(40) & -0.323 & -0.229(5) & -0.288(22) \\
$(19,1)$ & 0.013 & 0.012(5) & -0.048(50) & -0.383 & -0.290(5) & -0.233(9) \\
$(16,1)$ & -0.077 & -0.074(6) & -0.024(20) & -0.443 & -0.278(5) & -0.245(26) \\
$(16,0)$ & 0.083 & 0.012(6) & 0.089(11) & -0.139 & -0.125(6) & -0.247(14) \\
$(18,0)$ & 0.142 & 0.120(6) & 0.099(11) & -0.372 & -0.203(5) & -0.489(35) \\
$(18,1)$ & 0.217 & 0.166(6) & 0.054(23) & -0.568 & -0.301(6) & -0.468(17) \\
 \hline
\end{tabularx}
\caption{{\it Magnetization in 1D Simulations at $t=8$.}
For each logical qubit $(b,l_b)$ (first column), $\langle \overline{Z}_n\rangle$ from MPS results (second and fifth columns) is compared to unencoded results (third and sixth columns) and encoded results with $R=1$ syndrome extraction rounds (fourth and seventh columns).
The left columns show $g_z=0.5$ and the right columns show $g_z=1.5$.
The uncertainty is determined through bootstrap resampling.}
\label{tab:chain_results_t_8}
\renewcommand{\arraystretch}{1.0}
\end{table}

\begin{table}[h]
\renewcommand{\arraystretch}{1.4}
\scriptsize
\begin{tabularx}{\linewidth}{|c||Y|Y|Y|Y|Y|Y||Y|Y|Y|Y|Y|Y|}
\hline
\rule{0pt}{10pt} & \multicolumn{12}{c|}{\large ${|\langle \overline{Z}\rangle_\text{meas} - \langle \overline{Z}\rangle_\text{MPS}|}$}\\\hline\hline
\rule{0pt}{10pt} \multirow{2}{*}{\makecell{$\%$}} &  \multicolumn{6}{c||}{$g_z=0.5$}  &  \multicolumn{6}{c|}{$g_z=1.5$} \\\cline{2-13}
\rule{0pt}{10pt}  & Unenc. & $R=1$ & $R=2$ & $R=4$  & $R=8$ & $R=16$ & Unenc. & $R=1$ & $R=2$ & $R=4$  & $R=8$ & $R=16$ \\
\hline\hline
0\% & 0.000(0) & 0.000(0) & 0.000(0) & 0.000(0) & 0.000(0) & 0.000(0) & 0.001(0) & 0.000(0) & 0.000(0) & 0.000(0) & 0.000(0) & 0.000(0) \\
10\% & 0.009(0) & 0.005(1) & 0.007(1) & 0.007(1) & 0.008(1) & 0.013(1) & 0.014(1) & 0.009(1) & 0.012(1) & 0.010(1) & 0.011(1) & 0.016(1) \\
20\% & 0.016(1) & 0.014(1) & 0.016(1) & 0.015(1) & 0.017(1) & 0.023(1) & 0.031(1) & 0.018(1) & 0.020(1) & 0.021(1) & 0.023(1) & 0.028(1) \\
30\% & 0.022(1) & 0.021(1) & 0.025(1) & 0.024(1) & 0.027(1) & 0.037(2) & 0.048(1) & 0.033(1) & 0.033(1) & 0.039(2) & 0.041(2) & 0.051(2) \\
40\% & 0.031(1) & 0.031(1) & 0.034(1) & 0.035(1) & 0.040(1) & 0.051(1) & 0.062(1) & 0.047(1) & 0.049(2) & 0.056(2) & 0.062(2) & 0.083(3) \\
50\% & 0.040(1) & 0.039(1) & 0.043(1) & 0.046(1) & 0.055(1) & 0.065(1) & 0.074(1) & 0.067(2) & 0.066(2) & 0.074(2) & 0.084(2) & 0.137(3) \\
60\% & 0.050(1) & 0.049(1) & 0.053(1) & 0.057(2) & 0.068(2) & 0.081(2) & 0.091(1) & 0.084(2) & 0.086(2) & 0.095(2) & 0.124(3) & 0.213(4) \\
70\% & 0.062(1) & 0.061(2) & 0.066(2) & 0.075(2) & 0.082(2) & 0.098(2) & 0.109(1) & 0.108(2) & 0.111(2) & 0.127(3) & 0.167(3) & 0.282(4) \\
80\% & 0.078(1) & 0.076(2) & 0.083(2) & 0.091(2) & 0.105(2) & 0.128(2) & 0.129(1) & 0.137(2) & 0.143(2) & 0.161(3) & 0.215(3) & 0.349(4) \\
90\% & 0.109(1) & 0.104(3) & 0.115(3) & 0.123(3) & 0.136(3) & 0.173(4) & 0.165(1) & 0.170(3) & 0.197(3) & 0.217(4) & 0.283(4) & 0.421(4) \\
100\% & 0.306(5) & 0.398(28) & 0.370(15) & 0.386(22) & 0.429(14) & 0.582(19) & 0.291(5) & 0.330(17) & 0.455(22) & 0.478(19) & 0.548(21) & 0.750(23) \\
 \hline
\end{tabularx}
\caption{{\it Absolute Error in 1D Simulations.}
The absolute error ${|\langle \overline{Z}\rangle_\text{meas} - \langle \overline{Z}\rangle_\text{MPS}|}$ by decile (first column) is compared between unencoded runs (second and seventh columns) and encoded runs for various numbers of syndrome extraction rounds $R$ (columns 3-6 and 8-11).
The left columns show $g_z=0.5$ and the right columns show $g_z=1.5$.
The uncertainty is determined through bootstrap resampling.}
\label{tab:chain_cdf}
\renewcommand{\arraystretch}{1.0}
\end{table}

\begin{table}[h]
\renewcommand{\arraystretch}{1.4}
\scriptsize
\begin{tabularx}{\linewidth}{|c||Y|Y|Y|Y|Y|Y||Y|Y|Y|Y|Y|Y|}
\hline
\rule{0pt}{10pt} & \multicolumn{12}{c|}{ \large Signal-survival factor $\alpha (\sigma)$}\\\hline\hline
\rule{0pt}{10pt} \multirow{2}{*}{\makecell{$t$}} &  \multicolumn{6}{c||}{$g_z=0.5$}  &  \multicolumn{6}{c|}{$g_z=1.5$} \\\cline{2-13}
\rule{0pt}{10pt}  & Unenc. & $R=1$ & $R=2$ & $R=4$  & $R=8$ & $R=16$ & Unenc. & $R=1$ & $R=2$ & $R=4$  & $R=8$ & $R=16$ \\
\hline\hline
0.0 & 0.983(19) & 0.989(12) & 0.987(9) & 0.987(12) & 0.988(9) & 0.986(11) & 0.983(19) & 0.988(10) & 0.987(11) & 0.987(10) & 0.989(10) & 0.984(13) \\
0.5 & 0.931(34) & 0.955(44) & 0.959(44) & 0.947(46) & 0.953(42) & 0.954(46) & 0.929(36) & 0.954(39) & 0.955(44) & 0.955(44) & 0.956(40) & 0.959(35) \\
1.0 & 0.905(45) & 0.948(44) & 0.938(43) & 0.941(44) & 0.937(45) & 0.916(89) & 0.898(47) & 0.944(38) & 0.942(42) & 0.949(43) & 0.943(42) & 0.924(66) \\
1.5 & 0.878(65) & 0.918(79) & 0.893(85) & 0.897(76) & 0.897(73) & 0.870(110) & 0.877(61) & 0.913(71) & 0.908(74) & 0.911(74) & 0.927(62) & 0.922(89) \\
2.0 & 0.870(67) & 0.918(66) & 0.887(69) & 0.883(65) & 0.851(83) & 0.828(136) & 0.888(58) & 0.927(68) & 0.930(66) & 0.915(67) & 0.939(70) & 0.912(106) \\
2.5 & 0.841(78) & 0.875(80) & 0.841(80) & 0.871(80) & 0.823(98) & 0.771(129) & 0.913(55) & 0.960(58) & 0.966(60) & 0.951(71) & 0.983(67) & 0.953(84) \\
3.0 & 0.820(67) & 0.829(90) & 0.802(80) & 0.768(96) & 0.724(99) & 0.694(116) & 0.908(45) & 0.922(76) & 0.945(85) & 0.904(69) & 0.927(72) & 0.839(91) \\
3.5 & 0.777(74) & 0.834(53) & 0.822(62) & 0.802(70) & 0.757(83) & 0.692(95) & 0.893(40) & 0.918(55) & 0.934(55) & 0.914(65) & 0.864(72) & 0.713(77) \\
4.0 & 0.777(62) & 0.923(42) & 0.805(44) & 0.767(44) & 0.721(60) & 0.642(74) & 0.864(37) & 0.886(79) & 0.874(71) & 0.857(74) & 0.798(81) & 0.636(97) \\
4.5 & 0.755(57) & 0.766(47) & 0.876(56) & 0.723(46) & 0.739(55) & 0.639(83) & 0.824(44) & 0.859(72) & 0.858(67) & 0.792(82) & 0.731(116) & 0.539(120) \\
5.0 & 0.842(45) & 0.725(46) & 0.699(56) & 0.682(48) & 0.569(53) & 0.346(58) & 0.801(50) & 0.800(71) & 0.816(88) & 0.784(98) & 0.703(127) & 0.480(140) \\
5.5 & 0.898(51) & 0.592(49) & 0.636(58) & 0.568(49) & 0.448(48) & 0.265(49) & 0.776(61) & 0.810(78) & 0.766(94) & 0.723(107) & 0.645(145) & 0.430(146) \\
6.0 & 0.845(43) & 0.593(52) & 0.591(58) & 0.449(42) & 0.378(46) & 0.127(42) & 0.757(59) & 0.771(77) & 0.729(94) & 0.696(102) & 0.598(129) & 0.371(152) \\
6.5 & 0.822(47) & 0.572(53) & 0.529(51) & 0.327(49) & 0.291(41) & 0.157(38) & 0.750(64) & 0.753(87) & 0.731(86) & 0.670(113) & 0.603(141) & 0.345(146) \\
7.0 & 0.760(44) & 0.542(63) & 0.426(50) & 0.223(56) & 0.152(40) & 0.039(33) & 0.741(60) & 0.757(74) & 0.715(84) & 0.686(100) & 0.583(123) & 0.285(135) \\
7.5 & 0.802(46) & 0.481(45) & 0.465(48) & 0.259(41) & 0.179(34) & -0.019(22) & 0.737(58) & 0.750(74) & 0.701(82) & 0.687(90) & 0.568(126) & 0.255(119) \\
8.0 & 0.779(42) & 0.552(38) & 0.420(39) & 0.253(33) & 0.110(40) & 0.123(30) & 0.735(56) & 0.728(81) & 0.679(79) & 0.667(90) & 0.534(117) & 0.246(115) \\
\hline
\end{tabularx}
\caption{{\it Signal-Survival Factor for 1D Simulations.}
The signal-survival factor $\alpha$ calculated as a function of time $t$ (first column) in unencoded results (second and seventh columns) is compared to encoded results for various numbers of syndrome extraction rounds $R$ (columns 3-6 and 8-11).
The left columns show $g_z=0.5$ and the right columns show $g_z=1.5$.
The uncertainty is the residual error $\sigma$ in the $\alpha$ fits.}
\label{tab:chain_alpha_sigma}
\renewcommand{\arraystretch}{1.0}
\end{table}

\begin{table}[h]
\renewcommand{\arraystretch}{1.1}
\begin{tabularx}{\linewidth}{|c||Y|Y|Y||Y|Y|Y|}
\hline
\rule{0pt}{10pt} & \multicolumn{6}{c|}{\large$\langle \overline{Z}_n\rangle $}\\\hline\hline
\rule{0pt}{10pt} \multirow{2}{*}{\makecell{Logical\\qubit $(b,l_b)$}} &  \multicolumn{3}{c||}{$g_z=0.5$}  &  \multicolumn{3}{c|}{$g_z=1.5$} \\\cline{2-7}
\rule{0pt}{10pt} & MPS & Unencoded & $R=1$ & MPS & Unencoded & $R=1$  \\
\hline\hline
$(13,1)$ & -0.558 & -0.113(6) & -0.257(16) & -0.093 & 0.094(6) & -0.147(34) \\
$(13,0)$ & -0.725 & -0.159(6) & -0.362(25) & -0.116 & 0.047(6) & -0.094(13) \\
$(10,1)$ & -0.694 & -0.134(6) & -0.357(9) & -0.111 & -0.003(5) & -0.123(13) \\
$(10,0)$ & -0.538 & -0.095(5) & -0.283(14) & -0.138 & -0.017(6) & -0.191(10) \\
$(9,0)$ & -0.524 & -0.139(5) & -0.069(24) & -0.375 & -0.142(5) & -0.127(17) \\
$(9,1)$ & -0.640 & -0.206(6) & -0.284(26) & -0.145 & -0.092(6) & 0.004(18) \\
$(4,0)$ & -0.603 & -0.253(5) & -0.388(17) & 0.005 & -0.078(5) & -0.102(13) \\
$(4,1)$ & -0.617 & -0.262(5) & -0.352(12) & -0.036 & -0.058(5) & -0.115(10) \\
$(2,0)$ & -0.706 & -0.210(5) & -0.418(17) & -0.134 & 0.017(6) & -0.223(28) \\
$(2,1)$ & -0.686 & -0.216(5) & -0.408(17) & -0.279 & -0.102(5) & -0.230(24) \\
$(0,0)$ & -0.627 & -0.271(6) & -0.180(44) & -0.087 & -0.110(6) & -0.119(14) \\
$(0,1)$ & -0.770 & -0.235(5) & -0.278(27) & -0.139 & -0.080(6) & -0.067(26) \\
$(1,0)$ & -0.732 & -0.205(5) & -0.103(24) & -0.137 & -0.025(5) & -0.085(30) \\
$(1,1)$ & -0.554 & -0.166(6) & -0.098(48) & -0.087 & -0.025(5) & -0.148(15) \\
$(3,1)$ & -0.432 & -0.089(6) & -0.228(29) & -0.165 & -0.075(5) & -0.153(32) \\
$(3,0)$ & -0.620 & -0.147(5) & -0.289(23) & -0.186 & -0.056(6) & -0.185(31) \\
$(7,1)$ & -0.457 & -0.037(6) & -0.298(32) & -0.043 & -0.081(6) & -0.115(30) \\
$(7,0)$ & -0.403 & -0.071(6) & -0.296(18) & -0.208 & -0.102(6) & -0.179(29) \\
$(5,0)$ & -0.249 & -0.075(6) & -0.199(19) & -0.283 & -0.097(5) & -0.288(10) \\
$(5,1)$ & 0.182 & 0.166(5) & 0.203(33) & 0.658 & 0.333(5) & 0.420(14) \\
$(6,1)$ & 0.597 & 0.201(5) & 0.579(22) & 0.695 & 0.242(5) & 0.652(29) \\
$(6,0)$ & 0.696 & 0.231(5) & 0.597(16) & 0.804 & 0.371(5) & 0.704(32) \\
$(11,0)$ & 0.762 & 0.343(5) & 0.642(12) & 0.845 & 0.479(5) & 0.656(15) \\
$(11,1)$ & 0.678 & 0.335(5) & 0.579(23) & 0.506 & 0.323(5) & 0.466(9) \\
$(8,0)$ & 0.330 & 0.149(5) & 0.285(23) & 0.486 & 0.267(5) & 0.442(15) \\
$(8,1)$ & -0.263 & -0.077(6) & -0.159(24) & -0.364 & -0.135(5) & -0.304(35) \\
$(12,1)$ & -0.199 & -0.129(6) & -0.182(15) & -0.268 & -0.286(5) & -0.234(18) \\
$(12,0)$ & -0.125 & -0.079(6) & -0.176(11) & -0.120 & -0.099(5) & -0.123(31) \\
$(15,1)$ & -0.219 & -0.083(5) & -0.147(49) & -0.215 & -0.125(6) & -0.216(32) \\
$(15,0)$ & 0.162 & 0.094(6) & 0.164(35) & 0.212 & 0.156(5) & 0.310(17) \\
$(14,1)$ & 0.520 & 0.248(5) & 0.493(12) & 0.751 & 0.350(5) & 0.655(16) \\
$(14,0)$ & 0.228 & 0.168(5) & 0.487(38) & 0.767 & 0.245(5) & 0.649(27) \\
$(17,0)$ & -0.412 & -0.095(6) & -0.288(27) & -0.212 & -0.007(6) & -0.172(13) \\
$(17,1)$ & -0.107 & -0.005(5) & -0.170(15) & -0.297 & -0.068(5) & -0.273(46) \\
$(20,0)$ & -0.289 & -0.065(6) & -0.215(11) & -0.127 & -0.144(6) & -0.013(39) \\
$(20,1)$ & -0.576 & -0.277(5) & -0.313(15) & -0.219 & -0.191(5) & -0.194(11) \\
$(19,0)$ & -0.679 & -0.374(5) & -0.434(10) & -0.102 & -0.081(6) & -0.078(39) \\
$(19,1)$ & -0.709 & -0.287(5) & -0.481(22) & -0.396 & -0.168(5) & -0.246(26) \\
$(16,1)$ & -0.742 & -0.351(5) & -0.618(25) & -0.365 & -0.256(6) & -0.249(20) \\
$(16,0)$ & -0.630 & -0.262(5) & -0.507(21) & -0.076 & -0.023(6) & -0.109(20) \\
$(18,0)$ & -0.282 & -0.092(5) & -0.227(25) & -0.261 & -0.230(6) & -0.298(10) \\
$(18,1)$ & -0.279 & -0.123(6) & -0.200(16) & -0.205 & -0.134(6) & -0.133(31) \\
 \hline
\end{tabularx}
\caption{{\it Magnetization in 2D Simulations at $t=4$.}
For each logical qubit $(b,l_b)$ (first column), $\langle \overline{Z}_n\rangle$ from MPS results (second and fifth columns) is compared to unencoded results (third and sixth columns) and encoded results with $R=1$ syndrome extraction rounds (fourth and seventh columns).
The left columns show $g_z=0.5$ and the right columns show $g_z=1.5$.
The uncertainty is determined through bootstrap resampling.}
\label{tab:grid_results_t_4}
\renewcommand{\arraystretch}{1.0}
\end{table}

\begin{table}[h]
\renewcommand{\arraystretch}{1.4}
\scriptsize
\begin{tabularx}{\linewidth}{|c||Y|Y|Y|Y|Y|Y||Y|Y|Y|Y|Y|Y|}
\hline
\rule{0pt}{10pt} & \multicolumn{12}{c|}{\large ${|\langle \overline{Z}\rangle_\text{meas} - \langle \overline{Z}\rangle_\text{MPS}|}$}\\\hline\hline
\rule{0pt}{10pt} \multirow{2}{*}{\makecell{$\%$}} &  \multicolumn{6}{c||}{$g_z=0.5$}  &  \multicolumn{6}{c|}{$g_z=1.5$} \\\cline{2-13}
\rule{0pt}{10pt}  & Unenc. & $R=1$ & $R=2$ & $R=4$  & $R=8$ & $R=16$ & Unenc. & $R=1$ & $R=2$ & $R=4$  & $R=8$ & $R=16$ \\
\hline\hline
0\% & 0.000(1) & 0.000(0) & 0.000(0) & 0.000(0) & 0.000(0) & 0.000(0) & 0.000(0) & 0.000(0) & 0.000(0) & 0.000(0) & 0.000(0) & 0.000(0) \\
10\% & 0.047(2) & 0.014(1) & 0.013(1) & 0.016(1) & 0.016(1) & 0.021(2) & 0.016(1) & 0.011(1) & 0.011(1) & 0.010(1) & 0.011(1) & 0.015(1) \\
20\% & 0.109(1) & 0.035(2) & 0.033(2) & 0.037(2) & 0.046(2) & 0.063(3) & 0.042(1) & 0.022(1) & 0.021(1) & 0.022(1) & 0.024(1) & 0.029(2) \\
30\% & 0.163(2) & 0.063(2) & 0.060(2) & 0.067(2) & 0.081(2) & 0.097(3) & 0.072(1) & 0.036(2) & 0.035(1) & 0.036(2) & 0.041(2) & 0.052(2) \\
40\% & 0.215(2) & 0.091(2) & 0.090(2) & 0.102(2) & 0.118(2) & 0.150(3) & 0.102(1) & 0.052(2) & 0.049(2) & 0.051(2) & 0.057(2) & 0.072(2) \\
50\% & 0.281(2) & 0.123(2) & 0.131(3) & 0.137(3) & 0.165(3) & 0.205(4) & 0.134(1) & 0.067(2) & 0.066(2) & 0.065(2) & 0.074(2) & 0.101(2) \\
60\% & 0.347(2) & 0.172(3) & 0.173(3) & 0.193(3) & 0.222(4) & 0.300(4) & 0.177(1) & 0.087(2) & 0.086(2) & 0.084(2) & 0.096(2) & 0.132(3) \\
70\% & 0.403(2) & 0.255(4) & 0.257(4) & 0.277(4) & 0.315(5) & 0.421(5) & 0.222(1) & 0.113(2) & 0.113(2) & 0.110(2) & 0.123(2) & 0.173(3) \\
80\% & 0.456(2) & 0.342(4) & 0.349(5) & 0.373(4) & 0.425(5) & 0.505(4) & 0.279(2) & 0.146(2) & 0.142(2) & 0.147(3) & 0.162(3) & 0.231(3) \\
90\% & 0.527(2) & 0.448(4) & 0.467(4) & 0.495(5) & 0.548(5) & 0.599(4) & 0.398(2) & 0.200(4) & 0.206(4) & 0.222(4) & 0.236(5) & 0.341(6) \\
100\% & 0.669(5) & 0.667(21) & 0.672(20) & 0.747(36) & 0.766(25) & 0.792(34) & 0.763(6) & 0.518(27) & 0.501(15) & 0.535(26) & 0.540(19) & 0.855(28) \\
 \hline
\end{tabularx}
\caption{{\it Absolute Error in 2D Simulations.}
The absolute error ${|\langle \overline{Z}\rangle_\text{meas} - \langle \overline{Z}\rangle_\text{MPS}|}$ by decile (first column) is compared between unencoded runs (second and seventh columns) and encoded runs for various numbers of syndrome extraction rounds $R$ (columns 3-6 and 8-11).
The left columns show $g_z=0.5$ and the right columns show $g_z=1.5$.
The uncertainty is determined through bootstrap resampling.}

\label{tab:grid_cdf}
\renewcommand{\arraystretch}{1.0}
\end{table}

\begin{table}[h]
\renewcommand{\arraystretch}{1.4}
\scriptsize
\begin{tabularx}{\linewidth}{|c||Y|Y|Y|Y|Y|Y||Y|Y|Y|Y|Y|Y|}
\hline
\rule{0pt}{10pt} & \multicolumn{12}{c|}{ \large Signal-survival factor $\alpha (\sigma)$}\\\hline\hline
\rule{0pt}{10pt} \multirow{2}{*}{\makecell{$t$}} &  \multicolumn{6}{c||}{$g_z=0.5$}  &  \multicolumn{6}{c|}{$g_z=1.5$} \\\cline{2-13}
\rule{0pt}{10pt}  & Unenc. & $R=1$ & $R=2$ & $R=4$  & $R=8$ & $R=16$ & Unenc. & $R=1$ & $R=2$ & $R=4$  & $R=8$ & $R=16$ \\
\hline\hline
0.0 & 0.988(6) & 0.988(10) & 0.988(12) & 0.984(16) & 0.986(12) & 0.984(18) & 0.982(37) & 0.987(13) & 0.985(15) & 0.987(11) & 0.985(13) & 0.981(16) \\
0.5 & 0.862(41) & 0.947(37) & 0.948(41) & 0.943(40) & 0.943(44) & 0.944(44) & 0.850(73) & 0.944(44) & 0.942(45) & 0.946(42) & 0.949(33) & 0.940(43) \\
1.0 & 0.770(79) & 0.918(48) & 0.917(52) & 0.912(48) & 0.918(51) & 0.883(78) & 0.757(106) & 0.907(54) & 0.909(48) & 0.918(48) & 0.913(54) & 0.879(75) \\
1.5 & 0.688(76) & 0.896(63) & 0.885(66) & 0.899(70) & 0.861(71) & 0.833(104) & 0.662(109) & 0.892(63) & 0.888(68) & 0.884(65) & 0.881(72) & 0.864(90) \\
2.0 & 0.597(79) & 0.851(84) & 0.845(91) & 0.867(81) & 0.817(95) & 0.744(135) & 0.607(112) & 0.879(69) & 0.866(70) & 0.858(68) & 0.860(78) & 0.810(104) \\
2.5 & 0.524(72) & 0.791(112) & 0.787(106) & 0.773(109) & 0.738(129) & 0.664(151) & 0.553(98) & 0.860(78) & 0.855(64) & 0.854(77) & 0.850(76) & 0.762(102) \\
3.0 & 0.459(73) & 0.734(115) & 0.735(115) & 0.710(122) & 0.667(131) & 0.537(159) & 0.534(85) & 0.854(66) & 0.849(68) & 0.843(78) & 0.836(76) & 0.734(102) \\
3.5 & 0.411(68) & 0.675(126) & 0.684(125) & 0.642(117) & 0.590(140) & 0.434(170) & 0.487(72) & 0.836(72) & 0.845(69) & 0.812(63) & 0.830(68) & 0.725(78) \\
4.0 & 0.348(63) & 0.606(133) & 0.627(123) & 0.555(133) & 0.492(137) & 0.323(153) & 0.474(66) & 0.834(68) & 0.831(74) & 0.814(71) & 0.778(70) & 0.620(89) \\
4.5 & 0.305(60) & 0.551(124) & 0.519(124) & 0.506(130) & 0.387(144) & 0.241(144) & 0.416(61) & 0.788(65) & 0.764(77) & 0.768(73) & 0.730(82) & 0.547(95) \\
5.0 & 0.251(60) & 0.485(121) & 0.452(121) & 0.412(115) & 0.337(139) & 0.209(127) & 0.370(60) & 0.722(64) & 0.703(62) & 0.725(69) & 0.645(86) & 0.417(83) \\
5.5 & 0.215(54) & 0.418(116) & 0.395(115) & 0.374(127) & 0.270(128) & 0.143(119) & 0.322(57) & 0.660(74) & 0.668(66) & 0.621(71) & 0.592(94) & 0.321(90) \\
6.0 & 0.174(53) & 0.399(113) & 0.349(110) & 0.296(109) & 0.248(130) & 0.097(74) & 0.280(58) & 0.618(74) & 0.609(68) & 0.591(67) & 0.558(84) & 0.192(64) \\
6.5 & 0.149(48) & 0.328(99) & 0.321(108) & 0.267(116) & 0.204(110) & 0.087(80) & 0.241(54) & 0.561(70) & 0.575(72) & 0.598(81) & 0.521(87) & 0.123(66) \\
7.0 & 0.119(45) & 0.296(100) & 0.276(95) & 0.229(100) & 0.160(108) & 0.047(47) & 0.206(52) & 0.535(74) & 0.543(69) & 0.531(81) & 0.479(84) & 0.117(43) \\
7.5 & 0.095(41) & 0.263(82) & 0.266(100) & 0.191(96) & 0.133(94) & 0.050(48) & 0.180(50) & 0.505(62) & 0.521(66) & 0.503(79) & 0.444(81) & 0.072(36) \\
8.0 & 0.074(38) & 0.243(87) & 0.207(81) & 0.167(95) & 0.124(75) & 0.035(33) & 0.161(48) & 0.481(64) & 0.482(52) & 0.471(67) & 0.405(78) & 0.055(36) \\
 \hline
\end{tabularx}
\caption{{\it Signal-Survival Factor for 2D Simulations.}
The signal-survival factor $\alpha$ calculated as a function of time $t$ (first column) in unencoded results (second and seventh columns) is compared to encoded results for various numbers of syndrome extraction rounds $R$ (columns 3-6 and 8-11).
The left columns show $g_z=0.5$ and the right columns show $g_z=1.5$.
The uncertainty is the residual error $\sigma$ in the $\alpha$ fits.}
\label{tab:grid_alpha_sigma}
\renewcommand{\arraystretch}{1.0}
\end{table}

\begin{table}[h]
\renewcommand{\arraystretch}{1.4}
\scriptsize
\begin{tabularx}{\linewidth}{|c||Y|Y||Y|Y|}
\hline
\rule{0pt}{10pt} & \multicolumn{4}{c|}{ \large Acceptance rate}\\\hline\hline
\rule{0pt}{10pt} \multirow{2}{*}{\makecell{$t$}} &  \multicolumn{2}{c||}{Chain}  &  \multicolumn{2}{c|}{Grid} \\\cline{2-5}
\rule{0pt}{10pt}  & $g_z=0.5$. & $g_z=1.5$ & $g_z=0.5$ & $g_z=1.5$ \\
\hline\hline
 $0.0$ & $0.341 \pm 0.339$ & $0.436 \pm 0.347$ & $0.414 \pm 0.355$ & $0.289 \pm 0.314$ \\
 $0.5$ & $0.372 \pm 0.309$ & $0.292 \pm 0.310$ & $0.312 \pm 0.273$ & $0.268 \pm 0.280$ \\
 $1.0$ & $0.307 \pm 0.291$ & $0.343 \pm 0.286$ & $0.339 \pm 0.299$ & $0.298 \pm 0.261$ \\
 $1.5$ & $0.324 \pm 0.300$ & $0.338 \pm 0.264$ & $0.303 \pm 0.228$ & $0.247 \pm 0.251$ \\
 $2.0$ & $0.308 \pm 0.264$ & $0.241 \pm 0.246$ & $0.229 \pm 0.201$ & $0.252 \pm 0.231$ \\
 $2.5$ & $0.297 \pm 0.261$ & $0.294 \pm 0.253$ & $0.226 \pm 0.196$ & $0.274 \pm 0.221$ \\
 $3.0$ & $0.295 \pm 0.281$ & $0.321 \pm 0.258$ & $0.208 \pm 0.188$ & $0.205 \pm 0.183$ \\
 $3.5$ & $0.260 \pm 0.227$ & $0.293 \pm 0.214$ & $0.200 \pm 0.167$ & $0.233 \pm 0.197$ \\
 $4.0$ & $0.235 \pm 0.201$ & $0.249 \pm 0.241$ & $0.176 \pm 0.147$ & $0.208 \pm 0.191$ \\
 $4.5$ & $0.226 \pm 0.209$ & $0.228 \pm 0.177$ & $0.255 \pm 0.179$ & $0.169 \pm 0.160$ \\
 $5.0$ & $0.254 \pm 0.230$ & $0.310 \pm 0.224$ & $0.196 \pm 0.162$ & $0.249 \pm 0.199$ \\
 $5.5$ & $0.210 \pm 0.151$ & $0.183 \pm 0.148$ & $0.255 \pm 0.185$ & $0.211 \pm 0.172$ \\
 $6.0$ & $0.246 \pm 0.196$ & $0.226 \pm 0.203$ & $0.200 \pm 0.177$ & $0.218 \pm 0.141$ \\
 $6.5$ & $0.210 \pm 0.188$ & $0.231 \pm 0.197$ & $0.229 \pm 0.178$ & $0.216 \pm 0.165$ \\
 $7.0$ & $0.209 \pm 0.195$ & $0.214 \pm 0.197$ & $0.139 \pm 0.081$ & $0.227 \pm 0.155$ \\
 $7.5$ & $0.228 \pm 0.190$ & $0.233 \pm 0.207$ & $0.173 \pm 0.136$ & $0.173 \pm 0.124$ \\
 $8.0$ & $0.197 \pm 0.161$ & $0.187 \pm 0.179$ & $0.201 \pm 0.158$ & $0.171 \pm 0.129$ \\
 \hline
\end{tabularx}
\caption{{\it Acceptance Rates of $R=1$ Encoded Simulations}.
The acceptance rates are shown for both 1D simulations (second and third columns) and 2D simulations (fourth and fifth columns) as a function of time $t$ (first column).
These encoded simulations use $R=1$ syndrome extraction rounds, corresponding to the data reported in the main text. 
The uncertainty represents the standard deviation over all qubits for a given $t$.
The uncertainty is the standard deviation over all qubits for each $t$.}
\label{tab:acceptance_rates}
\renewcommand{\arraystretch}{1.0}
\end{table}

\clearpage
\newpage
\bibliography{dev_422.bib}

\end{document}